\documentclass[fleqn,usenatbib]{mnras}

\usepackage{newtxtext,newtxmath}

\usepackage[T1]{fontenc}

\DeclareRobustCommand{\VAN}[3]{#2}
\let\VANthebibliography\thebibliography
\def\thebibliography{\DeclareRobustCommand{\VAN}[3]{##3}\VANthebibliography}

\usepackage{graphicx}	
\usepackage{amsmath}	
\usepackage{caption}
\usepackage{booktabs}   
\usepackage{natbib}     
\usepackage{pdflscape}  

\usepackage{{threeparttable}} 
\usepackage{lineno}      
\usepackage{orcidlink} 
\usepackage{enumitem}   

\title[Coronal Iron Lines Reveal Shock Emission in SNR 0540–69.3]{MUSE Observations Reveal Optical Coronal Iron Lines from Shock Emission in Supernova Remnant 0540–69.3}

\author[L. Tenhu et al.]{
L. Tenhu\,\orcidlink{0000-0002-7746-8512},$^{1,2}$\thanks{E-mail: lcetenhu@kth.se (LT)}
J. Larsson\,\orcidlink{0000-0003-0065-2933},$^{1,2}$ P. Lundqvist\,\orcidlink{0000-0002-3664-8082},$^{3}$ I. Saathoff\,\orcidlink{0000-0002-3068-7275},$^{1,2}$ J. D. Lyman\,\orcidlink{0000-0002-3464-0642} $^{4}$ and J. Sollerman\,\orcidlink{0000-0003-1546-6615} $^{3}$
\\
$^{1}$KTH Royal Institute of Technology, Department of Physics, SE-10691 Stockholm, Sweden \\
$^{2}$The Oskar Klein Centre for Cosmoparticle Physics, AlbaNova University Centre, SE-10691 Stockholm, Sweden \\
$^{3}$Department of Astronomy, Stockholm University, The Oskar Klein Centre, AlbaNova, SE-106 91 Stockholm, Sweden \\
$^{4}$Department of Physics, University of Warwick, Gibbet Hill Road, Coventry CV4 7AL, UK
}
\date{Accepted XXX. Received YYY; in original form ZZZ}

\pubyear{\the\year{}}

\begin{document}
\label{firstpage}
\pagerange{\pageref{firstpage}--\pageref{lastpage}}
\maketitle

\defcitealias{L21}{L21}
\defcitealias{Tenhu-2024}{T24}

\begin{abstract}

\noindent We investigate the optical shock emission from the Large Magellanic Cloud supernova remnant 0540--69.3 (SNR\,0540) using MUSE integral-field-unit data from the VLT. The observations cover the spectral range \mbox{4650--9300\,Å} and provide a \mbox{$1\times1$\,arcmin$^{2}$} field of view, encompassing nearly the entire remnant. We analyse the spatial and spectral properties of shock-related emission lines, and identify clumpy optical shock emission e.g. from \mbox{[\ion{S}{II}]\,$\lambda\lambda$6716,6731} doublet and the coronal \mbox{[\ion{Fe}{XIV}]\,$\lambda$5303} line (typically at radial velocities \mbox{$\lesssim|100|$\,km\,s$^{-1}$} and \mbox{$\lesssim|170|$\,km\,s$^{-1}$}, respectively). These features trace the blast-wave shell seen in previous X-ray studies. Post-shock electron density estimates, based on the \mbox{[\ion{S}{II}]}-line ratio, reveal spatial variation, with the highest densities \mbox{($\sim10^4$\,cm$^{-3}$)} in the bright knots in the west, and lower densities \mbox{($\sim3\times10^3$\,cm$^{-3}$)} in the east. The density in the north (southwest) appears significantly lower (higher) but remains unconstrained due to limited signal. We also estimate blast-wave shock velocities using the \mbox{[\ion{Fe}{XIV}]\,$\lambda$5303/[\ion{Fe}{XI}]\,$\lambda$7892} ratio, finding low velocities \mbox{($\sim400$\,km\,s$^{-1}$)}, consistent with previous studies. All these results support the scenario that the blast wave is interacting with the surrounding interstellar medium, particularly in the western regions. Additionally, we detect four unidentified emission lines, \mbox{$\sim$2000--3000\,km\,s$^{-1}$} south from the pulsar in transverse velocity, but their origin remains unclear. Possible explanations, including Fe lines from a high-velocity ejecta clump, all present challenges. Our findings highlight the complex nature of the circum- and interstellar medium surrounding SNR\,0540.

\end{abstract}


\section{Introduction}

When a supernova (SN) explodes, it ejects matter at high velocities, eventually generating an outward-propagating blast wave (\citealt{Vink-2020,Jerkstrand-2025} and references therein). This blast wave, or forward-shock, decelerates as it interacts with the surrounding medium. Consisting of circumstellar matter (CSM, matter lost by the progenitor pre-explosion), interstellar medium (ISM) or both, this material is swept-up by the shock and forms a shell behind it. This shell of hot, often X-ray-emitting gas is thought to define the edge of the supernova remnant (SNR) and contains information on the properties of the progenitor and the surrounding medium. This information helps to disentangle the progenitor--SN explosion--SNR connection, as well as to understand the shock conditions in SNRs.

The shell of hot gas produces line emission also at optical wavelengths (as proposed by \citealt{Shklovskii-1967} and first observed in Cygnys Loop and Vela SNRs by \citealt{Woodgate-1974,Woodgate-1975}). This optical emission from shock-heated gas manifests as forbidden coronal Fe emission (including {[\ion{Fe}{XIV}]~$\lambda$5303}, {[\ion{Fe}{XI}]~$\lambda$7892}, and {[\ion{Fe}{X}]~$\lambda$6375}). More commonly, SNR-shocks can be studied in the optical by observing other known shock tracing emission lines such as the \mbox{[\ion{S}{II}]\,$\lambda\lambda$6716,6731} doublet, numerous [\ion{Fe}{II}] and [\ion{Fe}{III}] lines, and, for example, H$\beta$ from shocked gas (line ratios of some of these are also used in identifying new SNRs, see e.g. \citealt{Kopsacheili-2020}). One benefit in studying shock emission in the optical is the better spatial and spectral resolution than in X-rays, which helps characterising the physical conditions of the shock-excited regions in SNRs.

To date, forbidden coronal Fe emission has been detected in several young SNRs e.g. in N49 \citep{Murdin-1978}, Puppis\,A \citep{Lucke-1979}, IC\,443 \citep{Sauvageot-1990}, N132D \citep{Winkler-1997}, and SN\,1987A \citep{Groningsson-2008,Larsson-2023}. Recent integral-field spectroscopic observations have investigated such emission in more detail e.g in N49 \citep{Dopita-2016} and N132D \citep{Dopita-2018} in the Large Magellanic Cloud (LMC) as well as \mbox{1E\,0102.2--7219} \citep{Vogt-2017} in the Small Magellanic Cloud (SMC). Notably, coronal Fe emission has even been observed from several young Type Ia SNRs (e.g. \citealt{Seitenzahl-2019,Das-2025}). In this work, we extend this sample of integral-field-spectroscopically studied SNR shocks in the Magellanic Clouds (MCs) by adding SNR\,0540-69.3 (SNR\,0540 hereafter), a young ({$\sim$1150\,yr}, \citealt{Reynolds-1985,L21,Lundqvist-2022}) and bright SNR in the LMC.

SNR\,0540 is an oxygen-rich SNR, most likely originating from a Type II SN explosion of a massive progenitor ($\gtrsim$15~M$_\odot$, \citealt{Chevalier-2006,Williams-2008}). It is a composite remnant, hosting a bright pulsar (PSR) and an associated pulsar-wind nebula (PWN) observed throughout the electromagnetic spectrum from the radio to X-rays (\citealt{Kirshner-1989,Manchester-1993,Gotthelf-2000,Morse-2006,Lundqvist-2020}). The PWN in SNR\,0540 powers the optical line emission mostly arising from a region \mbox{$\sim$4 arcsec} ({$\sim$1~pc} assuming the LMC distance {$\sim$50 kpc}, \citealt{Pietrzynski-2019}) away from the PSR. SNR 0540 also has a faint [\ion{O}{III}] shell extending to \mbox{$\sim$8 arcsec} from the PSR \citep{Morse-2006,L21}.

SNR\,0540 has often been referred to as the "Crab Twin" due to its similar age, sky orientation, and pulsar spin period among other properties. However, key differences exist (e.g. \citealt{Petre-2007,Tenhu-2024} and references therein), including, but not limited to, the absence of the forward-shock shell in the Crab, which in SNR\,0540 has been detected \mbox{$\sim$30 arcsec} ($\sim$7~pc) from the PSR both in X-rays and in the radio (e.g. \citealt{Seward-1994,Manchester-1993,Hwang-2001,Park-2010}). 

This thermally emitting forward-shock shell is not uniform. In fact, the radius of the shell varies significantly, ranging from \mbox{$\sim20$ arcsec} (\mbox{5 pc}) in the northwest (NW) to \mbox{$\sim40$ arcsec} in the east (E) and south (S). A prominent indentation of the shell in the NW, appearing as a bright, nearly diagonal filament, is clearly visible in both radio and X-ray observations \citep{Manchester-1993,Hwang-2001,Park-2010,Brantseg-2014}. X-ray observations by \citet{Gotthelf-2000,Hwang-2001,Park-2010} and \citet{Brantseg-2014} have revealed systematic spatial variation in the brightness, spectral hardness, electron density and temperature across the shell. These works observe that the western regions are characterised by bright and soft emission, with relatively low temperatures (\mbox{$kT\sim0.5$--1\,keV}) and higher electron densities (\mbox{$n_e\sim$10\,cm$^{-3}$}, \citealt{Brantseg-2014}). In contrast, the eastern parts of the shell exhibit fainter and harder emission (a hard component with \mbox{$E>3$\,keV}) with higher electron temperatures (by a factor \mbox{of $\sim3$}), as well as lower densities (up to an order of magnitude lower). These variations along with the observed nominal LMC element abundances (\citealt{Hwang-2001,Brantseg-2014}) suggest that the SN blast wave is interacting with a non-uniform CSM/ISM. 

Despite many similarities, the X-ray and radio observations of the SNR\,0540 shell reveal distinct differences. In X-rays, bright hard arcs appear on opposite sides of the PSR, positioned equidistantly in the east and west just outside the shell. These hard arcs, which are absent in the radio observations, emit non-thermal, likely synchrotron emission \citep{Hwang-2001,Park-2010,Brantseg-2014} and have been suggested to originate from shock-accelerated relativistic electrons in the strong SN shock. Additionally, while the hard arcs are seen only in X-rays, another difference between the two wavebands is a bright radio blob in the southeast (SE, \citealt{Manchester-1993,Brantseg-2014}) that has no observable counterpart in X-rays \citep{Hwang-2001,Park-2010}.

In the optical, the interaction between the SNR blast wave and the surrounding CSM/ISM has been studied with imaging and long-slit spectroscopy by \citet{Serafimovich-2005} and \citet{Lundqvist-2022}. These studies identify several filaments on the western side of SNR\,0540 that exhibit coronal Fe emission {[\ion{Fe}{XIV}]~$\lambda$5303}. The elevated temperatures measured for these knots, along with their spatial coincidence with the X-ray emission from the forward shock, suggest that they are ionized by SN-driven adiabatic shocks. Additionally, \citeauthor{Lundqvist-2022} report that these filaments also emit in Fe-lines with lower ionization states, indicating the presence of radiative shocks. To confirm this interaction scenario through different types of shocks, new observations extending the spectral coverage to the coronal iron lines {[\ion{Fe}{X}]~$\lambda$6374} and \mbox{[\ion{Fe}{XI}]~$\lambda$7892} are important. Furthermore, observing the \mbox{[\ion{S}{II}]\,$\lambda\lambda$6716,6731} doublet is essential for accurately estimating the densities in these regions.

As the forward shock decelerates due to interaction with the surrounding media, a secondary shock, known as the reverse shock, forms. For SNR\,0540, no clear detection of a reverse shock has been reported, although \citet{Brantseg-2014} observed reheating of material that could be powered by a reverse shock. Given the young age of SNR\,0540, its reverse shock is likely still very close to the blast wave and remains unresolved \citep{Hwang-2001}. 

In this work, we study the shock-interacting regions within SNR\,0540 by utilising the optical observations obtained with the Multi Unit Spectroscopic Explorer (MUSE) from the Very Large Telescope (VLT) in the wavelength range \mbox{4650--9300\,\AA}, along with archival Chandra X-ray data. The MUSE observations provide the first optical integral-field-spectroscopic data covering the entire SNR\,0540. This allows us to investigate the spatial and spectral variations of shock emission in the entire remnant for the first time in the optical. 

This paper belongs to a series started by \citet{L21} and followed by \citet{Tenhu-2024}, which we hereafter call \citetalias{L21} and \citetalias{Tenhu-2024}, respectively. While those works focused on investigating the emission from the PSR and the PWN, this study analyses the entire SNR using the same MUSE data, excluding the already studied central region. We divide this paper into the following sections. Section\,\ref{sec:observations} describes the observations and data processing techniques. This is followed by Section\,\ref{sec:methods} that outlines our approach to investigating the shock-related emission lines associated with the SNR. Section\,\ref{sec:results} details the results and analysis, which are discussed in Section\,\ref{sec:discussion}. A summary and the main conclusions are presented in Section\,\ref{sec:conclusion}. In Appendix\,\ref{appendix:r5-r8}, we provide supplementary shock emission results for completeness.

\section{Observations and Post-processing of Data} \label{sec:observations}

\subsection{Observations} \label{subsec:observations}

MUSE \citep{muse10} observations, performed in 2019 January and March \footnote{ESO programme 0102.D-0769, PI J. D. Lyman} utilising the wide-field mode adaptive optics (WFM-AO), provide us with suitable Integral Field Unit data to study the entire SNR\,0540 in the optical wavelengths. The spectral dimension of MUSE covers the wavelengths \mbox{4650--9300\,\AA} (with a gap between \mbox{5760--6010\,\AA} caused by the Na laser)\footnote{After correcting the MUSE data to the LMC rest velocity (Section\,\ref{subsec:post-processing}), the MUSE spectra covers the range \mbox{4595--9340\,\AA} with the Na-laser gap between \mbox{5748--6002\,\AA}.} with a spectral bin size of 1.25\,Å and resolution $R=$\,\mbox{1750--3750} (or \mbox{189--84\,km\,s$^{-1}$} in radial velocity). The field-of-view (FOV) spans ${329\times 333\,\mathrm{pixel}^2}$, which with a spatial resolution of \mbox{0.3--0.4\,arcsec} and \mbox{0.2\,arcsec} per spatial pixel (spaxel) sampling corresponds to \mbox{$65.8\times66.6$\,arcsec$^2$} or ${\sim16\times16\,\mathrm{pc}^2}$. We refer to \citetalias{L21} and \citetalias{Tenhu-2024} for more information on these observations and the data reduction.

We compare the optical observations to the most recent ancillary Chandra data (previously presented in \citealt{Park-2010} and \citealt{Brantseg-2014}). SNR 0540–69.3 was observed three times with the Advanced CCD Imaging Spectrometer (ACIS) on board the \textit{Chandra X-ray Observatory} \citep{Chandra} between 2006 February 15–18\footnote{ObsID 5549, 7270, and 7271} with a total exposure time of approximately 114\,ks, hence providing also the longest exposure X-ray observations of SNR\,0540. These observations were downloaded and reprocessed using CIAO 4.15.2  (version 2023 January) with CALDB$\_$Main 4.10.7. To co-add the images, the task \texttt{reproject$\_$obs} (version 2021 November 15) was used.

\subsection{Corrections for the World Coordinate Solution, Systemic Velocity, and Extinction} \label{subsec:post-processing}

We followed the approach of \citet{Dopita-2018} to set the world coordinate system solution (WCS) for both MUSE and Chandra data by using GAIA catalogues. We chose a 2D slice at ${\sim 7100}$\,Å (as this part of the spectrum does not have significant emission lines nor instrumental artefacts) of the MUSE data and used that to identify bright stars in the FOV. The identification process was performed with a peak finding algorithm provided by \texttt{photutils}\footnote{\url{https://photutils.readthedocs.io/en/stable/index.html}} \citep{photutils24}, where we required the identified stars to reach at least $70\%$ of the PSR's luminosity.

We used \texttt{astroquery}\footnote{\url{https://astroquery.readthedocs.io/en/latest/index.html}} \citep{astroquery19}, which uses the GAIA catalogue EDR3 \citep{gaia20}, to identify the PSR and get the precise sky coordinates. We then retrieved field star coordinates within \mbox{0.5\,arcmin} radius from the PSR via the EDR3 query. We chose 13 stars that we could match with bright stars found by the peak finding algorithm to compute the mean offset to set a new WCS for MUSE. The MUSE coordinates were shifted \mbox{$+00^{\mathrm{h}} 00^{\mathrm{m}} 00.15^{\mathrm{s}}$} in RA and \mbox{$-00^\circ00'0.8''$} in DEC.

We also fixed the WCS of the Chandra data by utilising the queried PSR coordinates. The maximum of the emission was first determined using the tool \texttt{dmstat}, which returns the current PSR coordinates in pixels. We then converted the desired PSR coordinates to the pixel coordinates using \texttt{dmcoords} and calculated the difference between the two to be \mbox{$-00^{\mathrm{h}} 00^{\mathrm{m}} 00.01^{\mathrm{s}}$} in RA and \mbox{$-00^\circ00'0.15''$} in DEC. The WCS was updated using this information with the task \texttt{wcs$\_$update}. Finally, different energy bands were created with \texttt{dmcopy} in the energy ranges of \mbox{0.5--1.2}, \mbox{1.2--2.5}, \mbox{0.5--2.5}, and \mbox{2.5--7.0\,keV}.

Following \citetalias{L21} and \citetalias{Tenhu-2024}, we corrected all MUSE spectra for the local systemic velocity by using the measured emission line shift of $277$\,km\,s$^{-1}$. In addition, we used ${R_V=3.1}$ and ${E(B-V)=0.27}$\,mag (based on the investigation of the Balmer decrement of the ISM emission in \citetalias{Tenhu-2024}), and the analytic formula from \citet{Cardelli-1989} to correct the MUSE spectra for extinction. We checked the reliability of this extinction correction by computing the ratio \mbox{[\ion{Fe}{VII}]\,$\lambda5721$/$\lambda6087$} in Appendix\,\ref{appendix:r5-r8}.

\section{Analysis of Interaction Regions} \label{sec:methods}

\begin{figure*}
	\includegraphics[width=\textwidth]{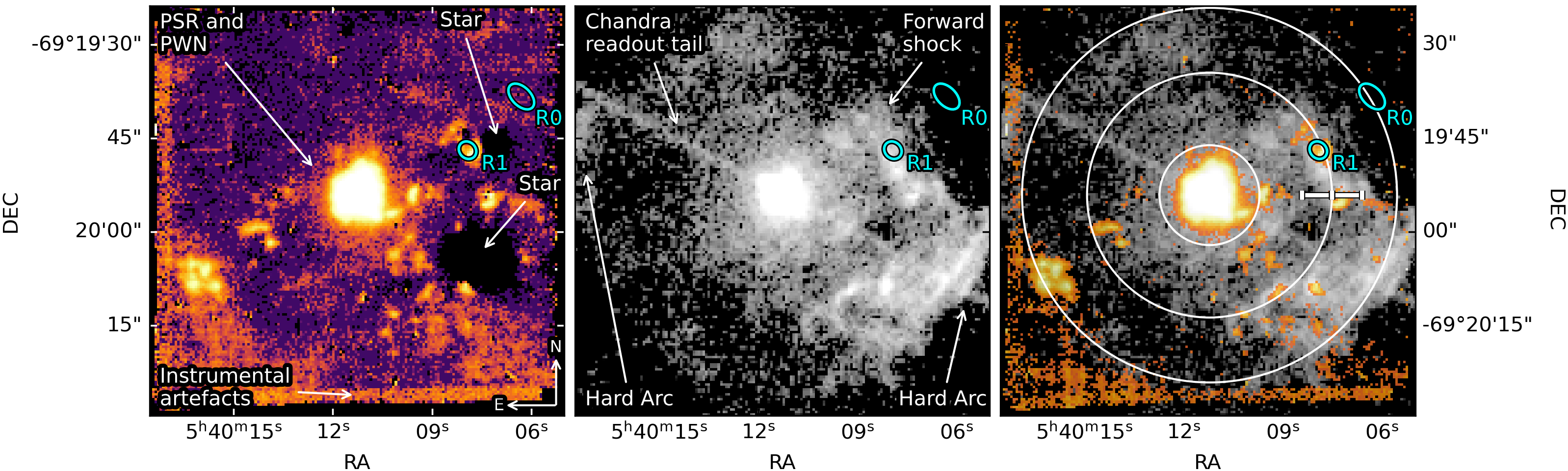}
    \caption{\textit{Left:} SNR\,0540 depicted by MUSE. The image is formed by spectrally integrating background- and continuum-subtracted fluxes from velocity slices \mbox{[$-450$, $-200$]}, \mbox{[$+200$, $+500$]}, and \mbox{[$+800$, $+1000$]}\,km\,s$^{-1}$ with respect to the shorter wavelength line in the \mbox{[\ion{S}{II}]\,$\lambda\lambda$6716,6731} doublet. \textit{Middle:} Chandra X-ray image of SNR\,0540 \mbox{(0.5--2.5\,keV)} in the same FOV as the MUSE image in the left panel. \textit{Right:} MUSE image superimposed on the Chandra image. The MUSE image is masked for clarity, so that only the brightest knots are visible. The white circles, centred at the PSR, have radii 8, 16, and \mbox{30\,arcsec} (\mbox{$\sim$1600}, \mbox{$\sim$4000}, and \mbox{$\sim$6000\,km\,s$^{-1}$}, respectively). The white horizontal line crossing the \mbox{16\,arcsec} circle shows the scale of 2000\,km\,s$^{-1}$ extending from \mbox{3000--5000\,km\,s$^{-1}$}. All velocities are computed assuming a remnant age of 1150\,yr and free expansion of the ejecta. The cyan ellipses denote regions from which extracted spectra exhibit emission from the unshocked ISM (R0) and unshocked ISM + shock-interactions (R1). In all panels, the colour scales are logarithmic intensities, where black denotes masked pixels.}
    \label{fig:intro}
\end{figure*}

We begin our analysis by providing an overview of the shock emission in SNR\,0540. Shock emission related to the remnant can be distinguished from unshocked emission by examining emission lines that are known shock-traces in SNRs (such as the \mbox{[\ion{S}{II}]\,$\lambda\lambda$6716,6731} doublet) and focusing on spectral regions at radial velocities significantly offset from zero, as unshocked emission typically occurs near zero velocity. Figure\,\ref{fig:intro} presents an overview of the shock interaction in SNR\,0540. The left panel shows spectrally integrated background-subtracted velocity slices of the \mbox{[\ion{S}{II}]\,$\lambda\lambda$6716,6731} doublet (for background subtraction details we refer to Section\,\ref{sec:emission-line-fits}). The velocity ranges are chosen so that the strong unshocked emission at near zero (rest) velocity for both lines in the doublet (\mbox{[$-200$, $+200$]} and \mbox{[$+500$, $+800$]\,km\,s$^{-1}$} with respect to the shorter wavelenght line in the doublet) are excludeded. Bright knots of shocked emission, clearly visible in the left panel, are predominantly located to the west (W), S, and southwest (SW) of the PWN.

The Chandra image shown in the middle panel of Figure\,\ref{fig:intro} is generated by summing all events within the energy range \mbox{0.5--2.5\,keV}. Prominent features include the X-ray-emitting forward shock, which dominates the western region of the SNR, and the hard arcs visible in the E and W. The right panel of Figure\,\ref{fig:intro} illustrates that the bright optically emitting knots of shocked-emission appear to overlap with the forward shock observed in X-rays, at least from the line-of-sight perspective. By assuming free expansion of the ejecta and an age of $1150$\,yr \citep{Kirshner-1989,L21,Lundqvist-2022}, we can compare the positions of the shock emission across the remnant by drawing circles at different transverse velocities. We observe that shock emission extends from $\sim1600$ to $\sim6000$\,km\,s$^{-1}$ in transverse velocity.

In addition to the SNR-related features, we note that both the MUSE and Chandra FOVs contain instrumental artefacts. For MUSE, these artefacts include stripes along the edges (in Figure\,\ref{fig:intro} they appear in the left and bottom edges) of the image, as well as incomplete subtraction of the stellar continuum from the brightest field stars. For these bright field stars, we mask the affected regions to varying spatial extents depending on the wavelength. The positions of these bright stars are indicated in the left panel of Figure\,\ref{fig:intro}. On the other hand, the Chandra FOV features a bright readout tail that overlaps the X-ray shell in the SW and in the E near the hard arc.

The rest of this section outlines the methods used in our spectral analysis. In Section\,\ref{sec:entire-spectrum} we describe the background subtraction process applied to the MUSE spectra, enabling the study of the entire spectral range simultaneously. Section\,\ref{sec:emission-line-fits} focuses on the detailed emission line analysis methods, including a localized background subtraction method (independent of the approach in Section\,\ref{sec:entire-spectrum}) and the fitting algorithm for a model comprising a combination of unshocked and shock components.

\subsection{Overview of the MUSE Spectra} \label{sec:entire-spectrum}

\begin{figure*}
    \includegraphics[width=\textwidth]{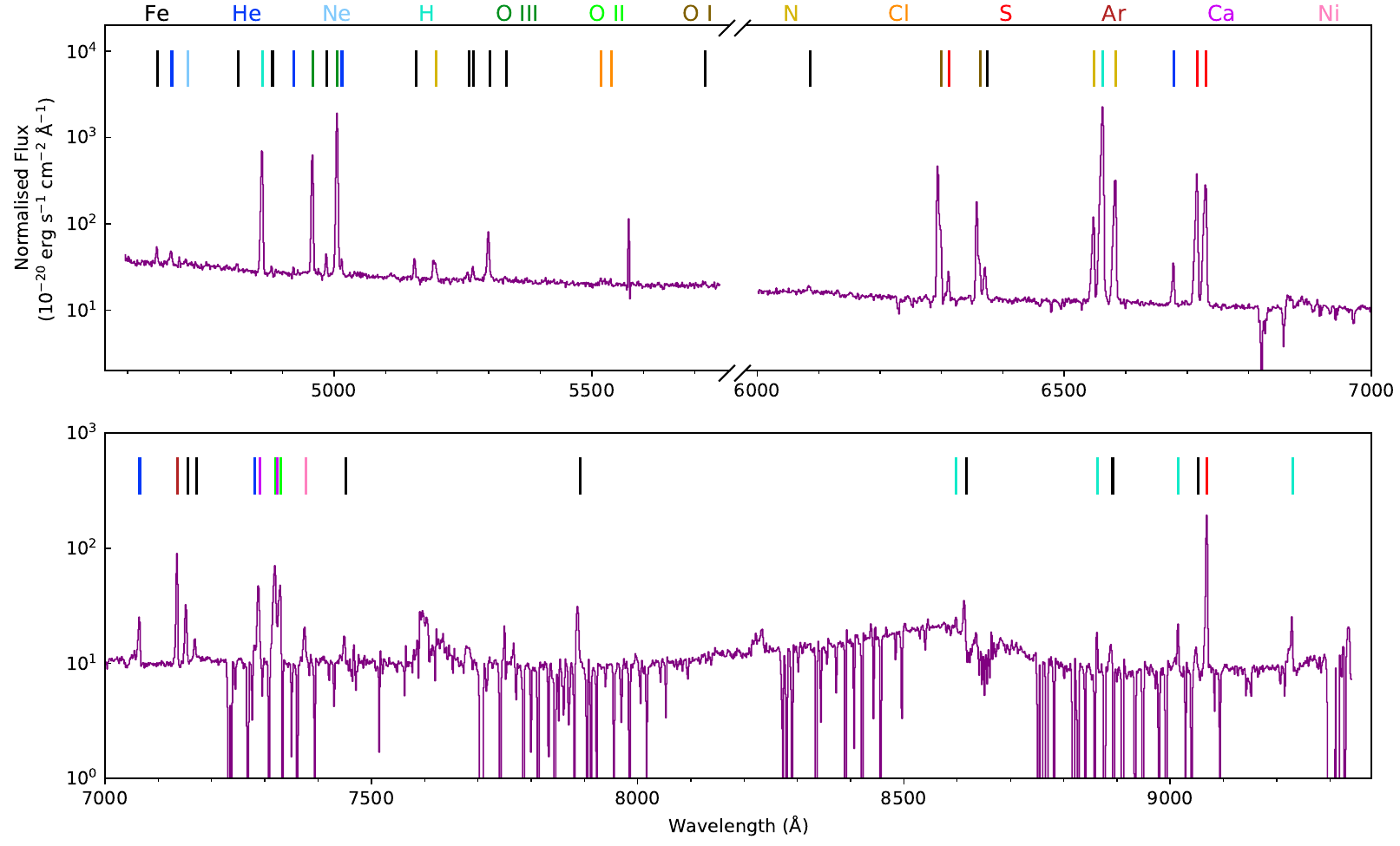}
    \caption{MUSE spectrum from a shock-dominated region R1 (with background). The fluxes are normalised with respect to the number of pixels in the extraction aperture (the location and size of the aperture is shown in Figure\,\ref{fig:intro}). Short vertical lines indicate the exact wavelengths of the brightest emission lines (also listed in Table\,\ref{tab:appendix_observed_emission_lines}), with different colors corresponding to different atomic or ionic species. The spectral axis is broken due to the gap in the spectrum between \mbox{5748–6002\,Å}.}
    \label{fig:methods-entire-spectra-log-nw}
\end{figure*}

Figure\,\ref{fig:methods-entire-spectra-log-nw} presents an example MUSE spectrum extracted from R1, one of the brightest shock emission knots in Figure\,\ref{fig:intro}, located to the NW from the PWN. This spectrum from R1 exhibits prominent unshocked ISM\footnote{Some of the emission attributed to the unshocked ISM may originate from unshocked CSM.} and shock-dominated emission. The brightest emission lines include the Balmer lines \mbox{H$\alpha$}, and \mbox{H$\beta$}, the doublets \mbox{[\ion{O}{III}]\,$\lambda\lambda$4959,5007}, \mbox{[\ion{O}{I}]\,$\lambda\lambda$6300,6364}, \mbox{[\ion{N}{II}]\,$\lambda\lambda$6548,6583}, and \mbox{[\ion{S}{II}]\,$\lambda\lambda$6716,6731}, as well as \mbox{[\ion{Ar}{III}]\,$\lambda$7136}, and \mbox{[\ion{S}{III}]\,$\lambda$9069}, which are all marked in Figure\,\ref{fig:methods-entire-spectra-log-nw} (for a more detailed look at the identified emission lines we refer to Section\,\ref{sec:results}). To gain a better overview of the fainter emission, we proceed by subtracting the background and continuum emission (i.e. all other components than the emission lines, referred to as background hereafter) from all spectra.

We model the background emission using the Locally Weighted Scatterplot Smoothing algorithm (LOWESS; \citealt{Cleveland-1979}). This method enables rapid inspection across the full MUSE spectral range from various spatial regions. However, while effective for providing a broad overview, this approach is less precise for examining weaker spectral features, as it may not fully account for e.g. stellar spectra or prominent instrumental artefacts. A more precise, emission-line-specific background subtraction method is detailed in Section\,\ref{sec:emission-line-fits}.

Prior to applying the LOWESS algorithm to the spectrum, we use the \texttt{astropy}\footnote{\url{https://docs.astropy.org/en/stable/index.html}} \citep{astropy13,astropy18,astropy22} sigmaclipping method on the spectra, to ensure that the LOWESS fit captures only the continuum, excluding the emission lines. Below, we outline the individual steps (following the approach of \citealt{Vogt-2017} for \mbox{SNR\,1E\,0102.2-7219}) in modelling the background emission:

\begin{enumerate}[label=\textup{(\roman*)}, leftmargin=*, align=right, widest=iii]
    \item Divide the entire spectrum into four separate blocks so that the edges of each block (4595, 5748, 6900, 8050, 9340\,Å) do not contain significant emission lines. Mask the wavelength range that contains the Na-laser gap \mbox{5748--6002\,Å}.
    \item For each individual spectral block, apply sigma clipping to only include features within $\sigma=[-1.5, 5]$.
    \item For each individual spectral block, fit the sigma-clipped data using the LOWESS smoother (provided by the Python module \texttt{statsmodels}\footnote{\url{https://www.statsmodels.org/devel/index.html}}, \citealt{statsmodels-2010}) with a window of ${15\%}$ ($\sim$170--195\,Å) of the block length.
    \item Use cubic spline interpolation (\texttt{scipy}\footnote{\url{https://docs.scipy.org/doc/scipy/index.html}}, \citealt{scipy}) to smoothly fill the gaps in the fit caused by sigma clipping.
\end{enumerate}

\begin{figure*}
    \includegraphics[width=\textwidth]{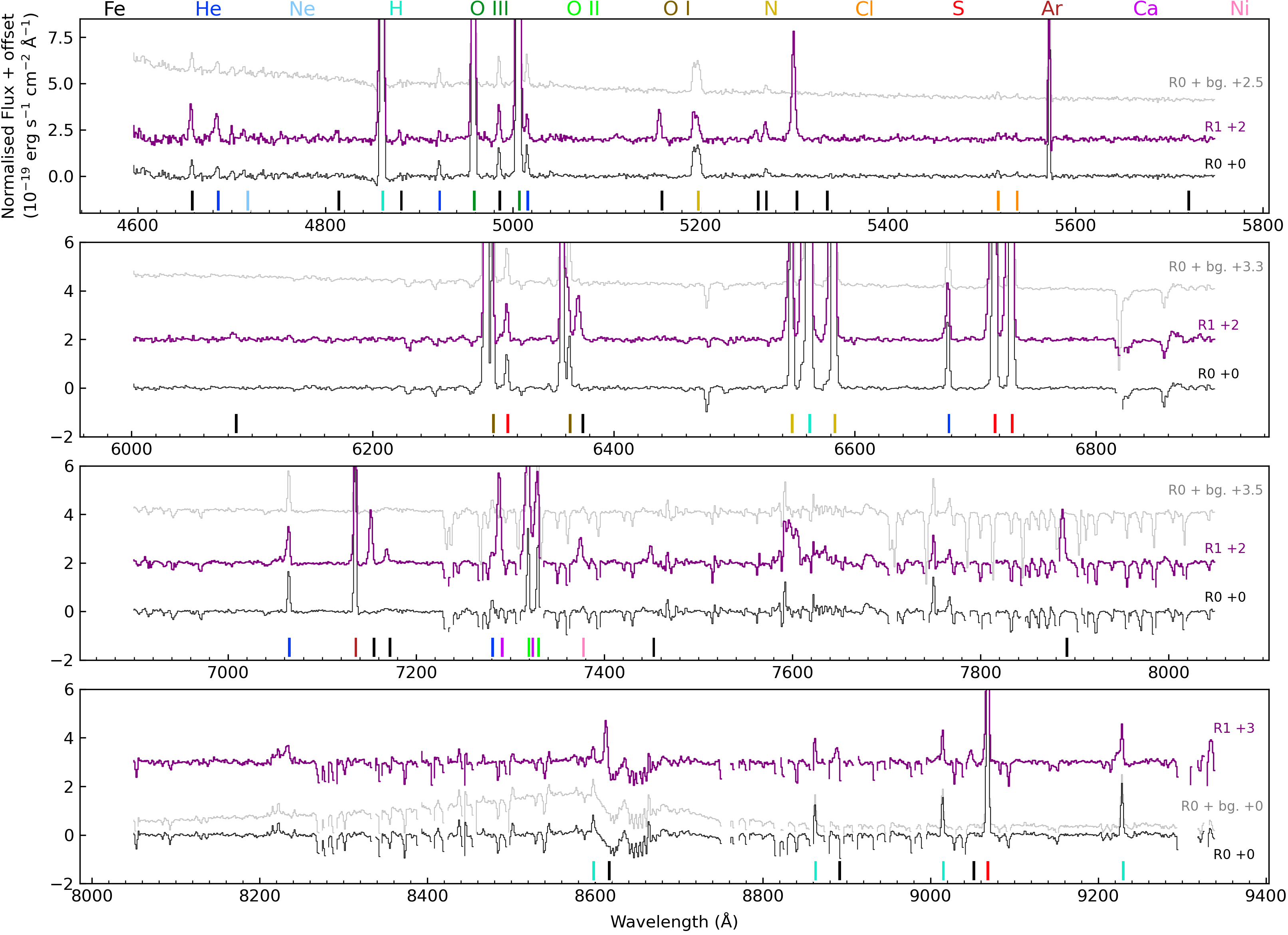}
    \caption{A closer look at the same MUSE spectrum (from shock-interacting region R1, in purple) as in Figure\,\ref{fig:methods-entire-spectra-log-nw} but now background-subtracted and compared with a spectrum extracted from an unshocked-ISM-dominated region (R0, in black, see Figure\,\ref{fig:intro}). We also show the same unshocked-ISM-dominated spectrum from R0 before background subtraction (grey). Different offsets and y-axis scales values are used for visual clarity. The short vertical lines (adopted from Figure\,\ref{fig:methods-entire-spectra-log-nw}) indicate the brightest identified emission lines, which are also listed in Table\,\ref{tab:appendix_observed_emission_lines}.}
    \label{fig:methods-entire-spectra-linear-ism-nw}
\end{figure*}

The resulting smooth background fit is then subtracted from the original spectra in each spaxel. Figure\,\ref{fig:methods-entire-spectra-linear-ism-nw} shows the background-subtracted spectra for two distinct spatial regions within the MUSE FOV: one dominated by unshocked ISM emission (R0) and another exhibiting significant emission originating also from shock-interaction (R1). The locations of these regions are shown in Figure\,\ref{fig:intro}. In Figure\,\ref{fig:methods-entire-spectra-linear-ism-nw} the background-subtracted spectrum from R1 reveal numerous fainter emission lines, particularly, emission lines that are entirely absent in the R0 background-subtracted spectrum, showcasing the shock-interaction origin of the emission from R1. Additionally, a comparison of the emission line profiles from both spectra indicates that the emission lines in the R1 spectrum are noticeably broader, consistent with the expectations for a shocked region. Some emission lines associated with shock interactions display complex, non-Gaussian structures, which will be addressed in the following section. While the broad shock-associated emission lines are clearly resolved, the unshocked ISM emission lines are not (the official resolving power of MUSE\footnote{Instrument description: \url{https://www.eso.org/sci/facilities/paranal/instruments/muse/inst.html}} varies across the spectrum from \mbox{$\sim175$\,km\,s$^{-1}$} for H$\beta$ to \mbox{$\sim85$\,km\,s$^{-1}$} for \mbox{[\ion{S}{III}]\,$\lambda$9069}). This limitation necessitates fixing the FWHM (full width at half maximum) for the unshocked ISM component fits, as presented in Section\,\ref{sec:emission-line-fits}. 

As seen in Figure\,\ref{fig:methods-entire-spectra-linear-ism-nw} and reported in \citetalias{Tenhu-2024}, the wavelength range \mbox{8000--9000\,Å} is affected by light contamination,\footnote{ESO Phase 3 Data Release Description for MUSE: \url{http://www.eso.org/rm/api/v1/public/releaseDescriptions/78} indicates that this contamination occurred in 2019 between February 1 and April 18. Our observations, conducted in the same year between January and March, exhibit a similar continuum feature.} manifesting as a pronounced continuum bump accompanied by significant features caused by poor atmospheric transmission. These features manifesting the poor atmospheric transmission are not confined just to the contaminated region but begin around 7200\,Å and persist through to 9000\,Å. Additional prominent features unrelated to unshocked ISM or shock emission include a complex noise peak around 7300\,Å, and narrow bright peaks (brighter than the actual emission lines) in association with the brightest \mbox{[\ion{O}{I}]} lines. Finally, the edges of several spectral blocks as well as the most drastic contamination feature at \mbox{$\sim$8600\,Å} suffer from incomplete background subtraction showcasing the limitations of the LOWESS approach. Consequently, we interpret all these wavelength regions with caution in our analysis.

\subsection{Analysis of Individual Emission Lines} \label{sec:emission-line-fits}

\begin{figure*}
    \includegraphics[width=\textwidth]{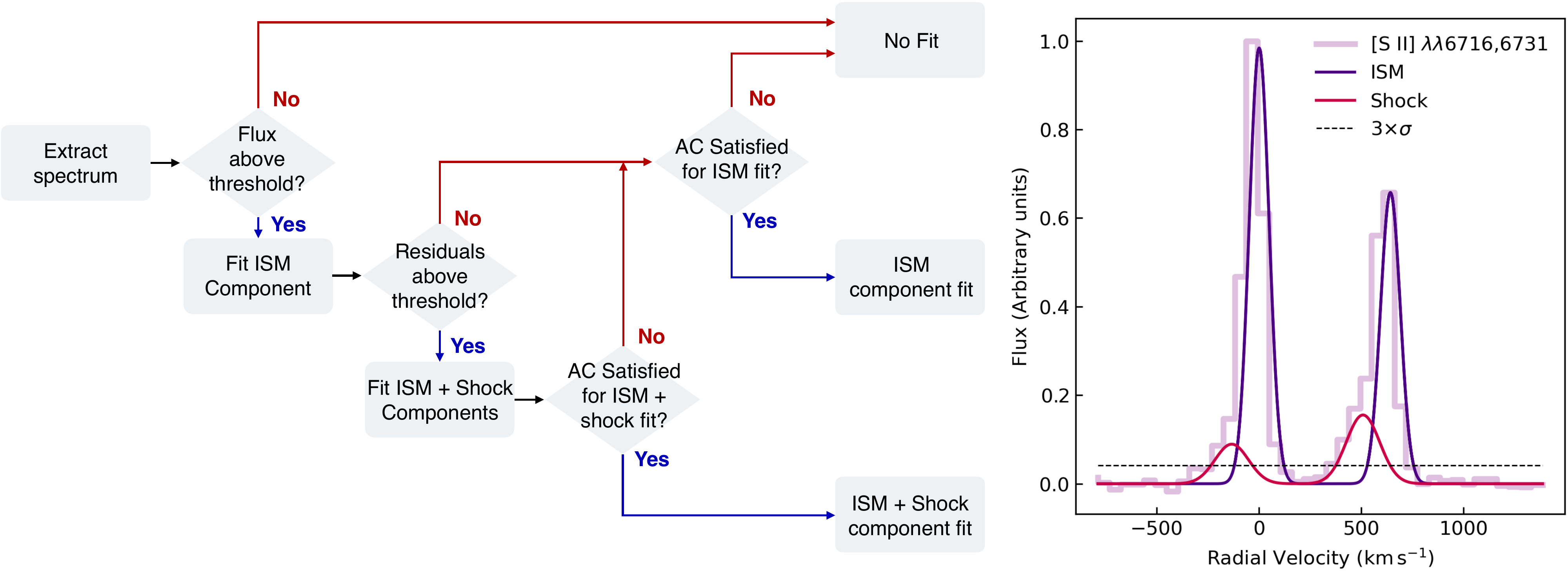}
    \caption{\textit{Left:} Schematic of the fit algorithm that accounts for two emission components: unshocked ISM and shock interaction. This process is repeated for each spaxel in the MUSE FOV. \textit{Right:} Example fit for the \mbox{[\ion{S}{II}]\,$\lambda\lambda$6716,6731} doublet. The background subtracted flux (MUSE, light purple line) is extracted from the central spaxel of region R1 (location shown in Figure\,\ref{fig:intro}), and is fitted with both unshocked ISM and shock components (dark purple and dark pink, respectively). The signal threshold for this emission line in this spaxel is presented with a black dashed line.}
    \label{fig:methods-fit-algorithm}
\end{figure*}

\begin{figure*}
    \includegraphics[width=\textwidth]{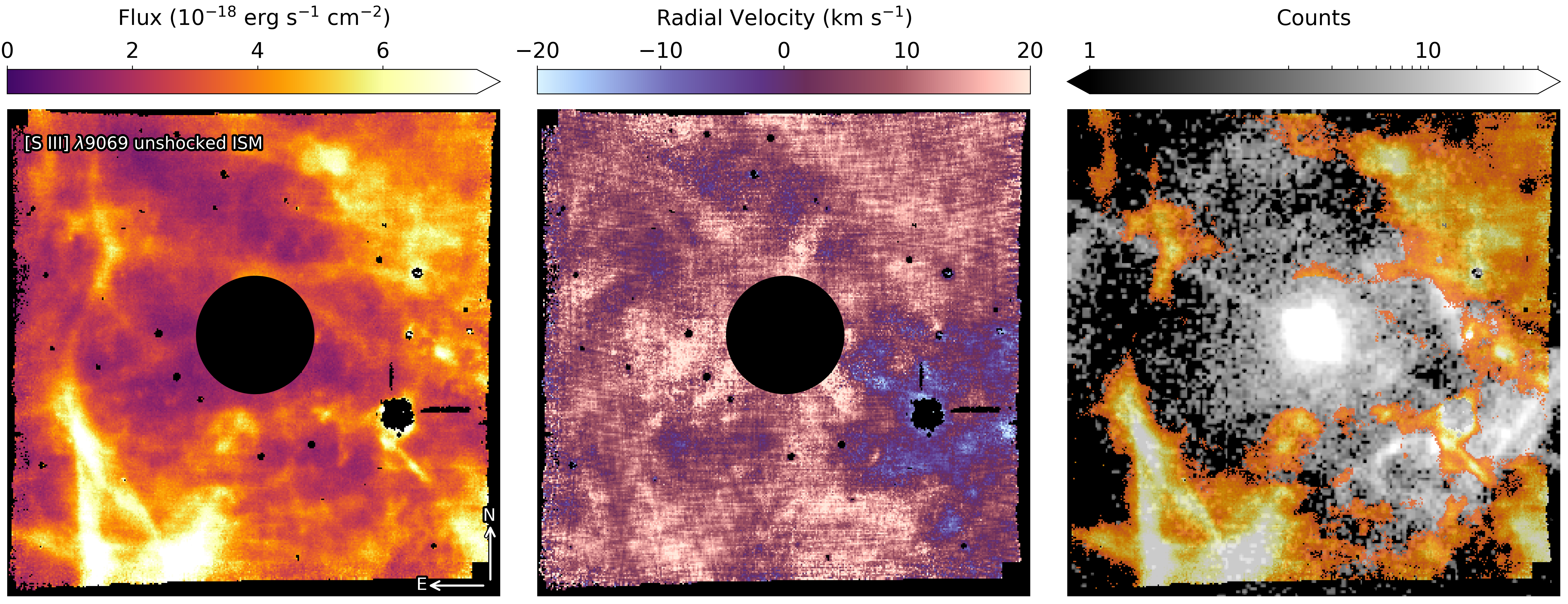}
    \caption{MUSE: Unshocked ISM component fit results of \mbox{[\ion{S}{III}]\,$\lambda$9069}. The fitted spectral range is 
    \mbox{[$-800$, $+1400$]\,km\,s$^{-1}$}. \textit{Left:} Spectrally integrated flux from the fitted unshocked ISM component. \textit{Middle:} Radial velocity of the fitted unshocked ISM component. \textit{Right:} Chandra X-ray image from \mbox{0.5--2.5~keV} reprojected to the MUSE FOV, with the brightest parts of the spectrally integrated fluxes obtained from the optical fits superposed. For all panels, black colour indicates masked spaxels, and the black circle masks the PWN and has a radius of \mbox{8\,arcsec}.
    \label{fig:methods-fit-results-ism}}
\end{figure*}

Following the spectral overview provided by the LOWESS algorithm, we now present how we obtain spatially resolved information by focusing on individual emission lines. While effective for summarising the full spectral range, the LOWESS algorithm is unsuitable for detailed emission line analysis. In this section, we summarise our approach for achieving a more accurate background subtraction for individual emission lines or line systems.

First, we restrict the spectral ranges so that we can assume the background emission to be modeled with a linear function. We then fit the continuum on both sides of the line and subtract the resulting best-fit model from the spectra in each spaxel. Subsequently, the spectral region is restricted to include only the relevant spectral range for emission line fitting, typically within \mbox{$\pm1500$\,km\,s$^{-1}$} of the emission line. Additionally, the signal threshold, which serves as one of the acceptance criteria (AC) for the following emission line fit algorithm, is defined as three times the standard deviation of the flux within the selected background regions. To optimise the amount of signal, larger spatial bins are used for weaker emission lines.

Next, we provide an overview of the emission line fitting algorithm also presented in the left panel of Figure\,\ref{fig:methods-fit-algorithm}. Here, we describe the process for a single MUSE spaxel, which is then applied across all spaxels in the MUSE FOV. The fitting algorithm accounts for emission from two components that represent the unshocked ISM and shocked emission. As the left panel of Figure\,\ref{fig:methods-fit-algorithm} illustrates, the process begins with spectral extraction, followed by a check for flux above the predefined signal threshold. If signal above this threshold is detected, the algorithm first fits the unshocked ISM component, as it is typically the dominant contributor. After this initial fit, the residuals are examined. If no residual signal remains above the signal threshold and the fit AC (defined below) are satisfied, the fit is accepted. On the other hand, if residual signal is detected above the signal threshold, the algorithm proceeds to a two-component fit, including both unshocked ISM and shock emission. The resulting fit is again evaluated against the AC before being accepted. If the algorithm proceeds to the No Fit option, the corresponding spaxel is masked for the emission line analysis. The algorithm uses \texttt{scipy}'s \texttt{curve\_fit} \citep{scipy} in all spectral fitting.

We now describe how the different emission components are modelled, and define the AC implemented in the fitting algorithm to ensure reproducibility of the results. The unshocked ISM line emission is modeled using a Gaussian function with three parameters: amplitude, radial velocity and FWHM. The radial velocity of the Gaussian model is restricted based on the typical velocities of \ion{H}{II} regions in the vicinity of SNR\,0540 in the LMC (up to $\pm 20\,\mathrm{km}\,\mathrm{s}^{-1}$, \citealt{Lah-2024}). As the unshocked ISM component in the emission lines remains unresolved, we fix the unshocked ISM FWHM to the spectral resolution at the relevant wavelength. We determine this resolution for each line by fitting only the unshocked ISM component within every spaxel in the MUSE FOV (when available) and taking the median of the fitted FWHMs. With this method, the (effective) resolving power for MUSE is \mbox{$\sim190$\,km\,s$^{-1}$} (an increase of $8.6\%$ or \mbox{$\sim15$\,km\,s$^{-1}$} compared to the reported official value for the instrument) for H$\beta$ and \mbox{$\sim92$\,km\,s$^{-1}$} (a corresponding increase $8.2\%$ or \mbox{$\sim5$\,km\,s$^{-1}$}) for \mbox{[\ion{S}{III}]\,$\lambda$9069}. If no significant unshocked ISM emission is available in most spaxels, we use the official spectral resolution to set the FWHM for the unshocked ISM component.

When needed, the shock component of the emission lines is also modelled using a Gaussian function, albeit with different constraints. The radial velocity is limited only by the spectral range used in the fit. This freedom is due to the shock component arising from a shock heated gas that generate emission lines with velocities distinctively higher (\mbox{$\gtrsim 100$\,km\,s$^{-1}$}) than that of the unshocked ISM. Finally, since the shock component of the emission lines is generally resolved and expected to be broadened mostly by a few hundred \mbox{km\,s$^{-1}$}, we set an upper constraint for the shock FWHM to \mbox{$500$\,km\,s$^{-1}$} for fit accuracy, while the lower bound is set to the (effective) spectral resolution (or if unavailable to the official MUSE spectral resolution). Additional constraints are set to emission line doublets, for example the \mbox{[\ion{S}{II}]\,$\lambda\lambda$6716,6731} doublet, where the distance between the peaks is fixed, and for the \mbox{[\ion{O}{III}]\,$\lambda\lambda$4959,5007} doublet, where the distance and the intensity ratio of the peaks are fixed.

As shown in the left panel of Figure\,\ref{fig:methods-fit-algorithm}, we require the fits to satisfy specific AC to ensure reproducible results. These criteria include a signal threshold, which the amplitude of a given component must exceed. Furthermore, the shock component must be spectrally resolved by $5\sigma$. This criterion aids in distinguishing the tail of the unresolved unshocked ISM line from the shock emission by guaranteeing that the shock Gaussian captures at least the brightest shock emission that is significantly different from the unshocked ISM component. Additionally, a Gaussian fit is only accepted if its reduced $\chi^2$ value is less than that of a corresponding linear model. 

We also assess the need for additional shock components. A second shock Gaussian is introduced to the model if significant residuals persist after fitting the unshocked ISM + shock component. However, the second shock component largely reflects the shock emission already captured by the first shock component. This highlights the fact that the shock emission in general seems to exhibit non-Gaussian features. Given the scope of this work, it suffices to focus only on the unshocked ISM component and the first shock component for the rest of this study.

The right panel of Figure\,\ref{fig:methods-fit-algorithm} presents an example fit for the \mbox{[\ion{S}{II}]\,$\lambda\lambda$6716,6731} doublet, where both unshocked ISM and shock components are included. The flux is obtained from the central spaxel of R1 (location of R1 is shown in Figure\,\ref{fig:intro}), with the corresponding signal threshold illustrating the significance of the shock component. Notably, the ratio of the peak heights differs between the unshocked ISM and shock components, as expected. 

Finally, Figure\,\ref{fig:methods-fit-results-ism} demonstrates the best spectral and spatial resolution achieved for the full MUSE FOV, where the unshocked ISM component of \mbox{[\ion{S}{III}]\,$\lambda$9069} is shown after fitting each spaxel in the FOV. The flux map (left panel) displays a circular morphology consistent with the X-rays (right panel). This shell-like structure most likely traces the boundary of the swept-up ISM/CSM. Furthermore, the low radial velocities (middle panel) provide additional support for the algorithm's effectiveness in accurately capturing the unshocked ISM component. The unshocked ISM appears to consist of a blueshifted component in the SW, while the rest of the unshocked ISM within the MUSE FOV exhibits patchy and predominantly redshifted radial velocities.

\section{Results} \label{sec:results}

In this section we present the emission line fit results for some of the brightest emission lines in the MUSE wavelength range in Section\,\ref{subsec:bright_lines} and Section\,\ref{subsec:results_coronal_fe_lines}. After this we focus on the unique properties of a region we refer to as R7, located in the southern part of the MUSE FOV, in Section\,\ref{subsec:regions_s}. All reported fit uncertainties are formal statistical uncertainties calculated from the fit covariance matrices and are reported as $1\sigma$.

\subsection{Spatial and Spectral Properties of the Shock Emission} \label{subsec:bright_lines}

\subsubsection{Spatial Properties}

\begin{figure*}
    \includegraphics[width=\textwidth]{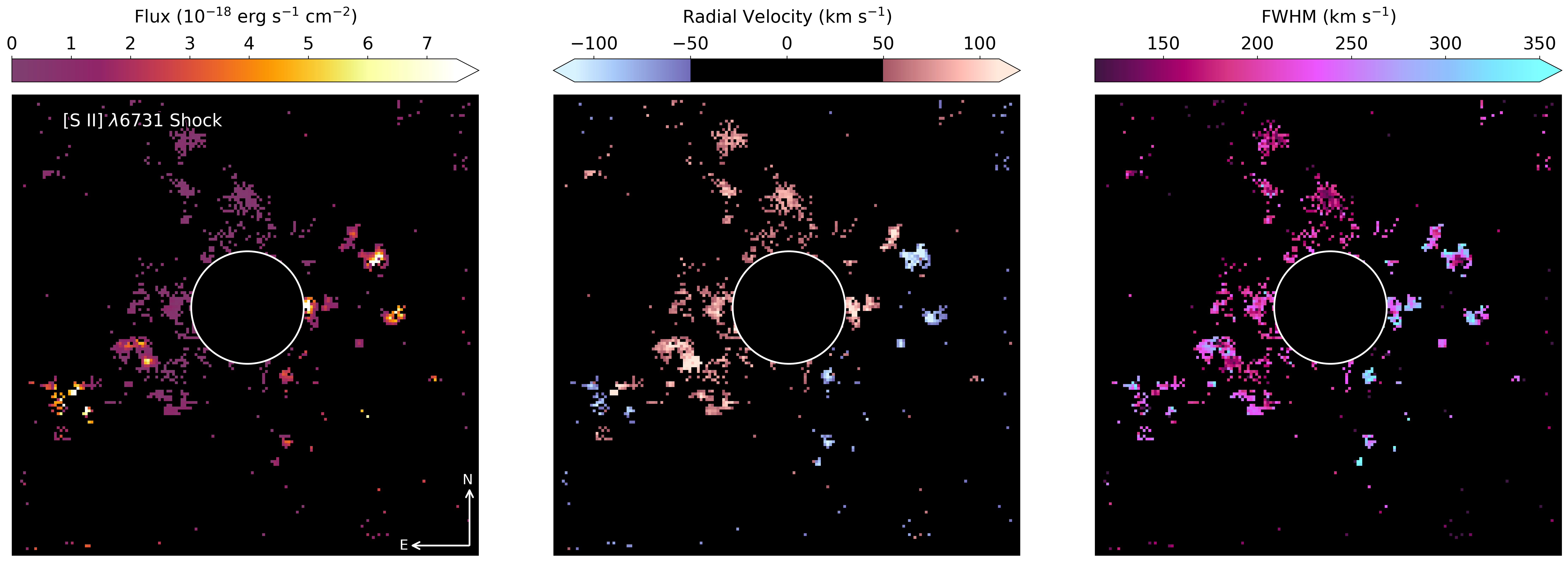}
    \includegraphics[width=\textwidth]{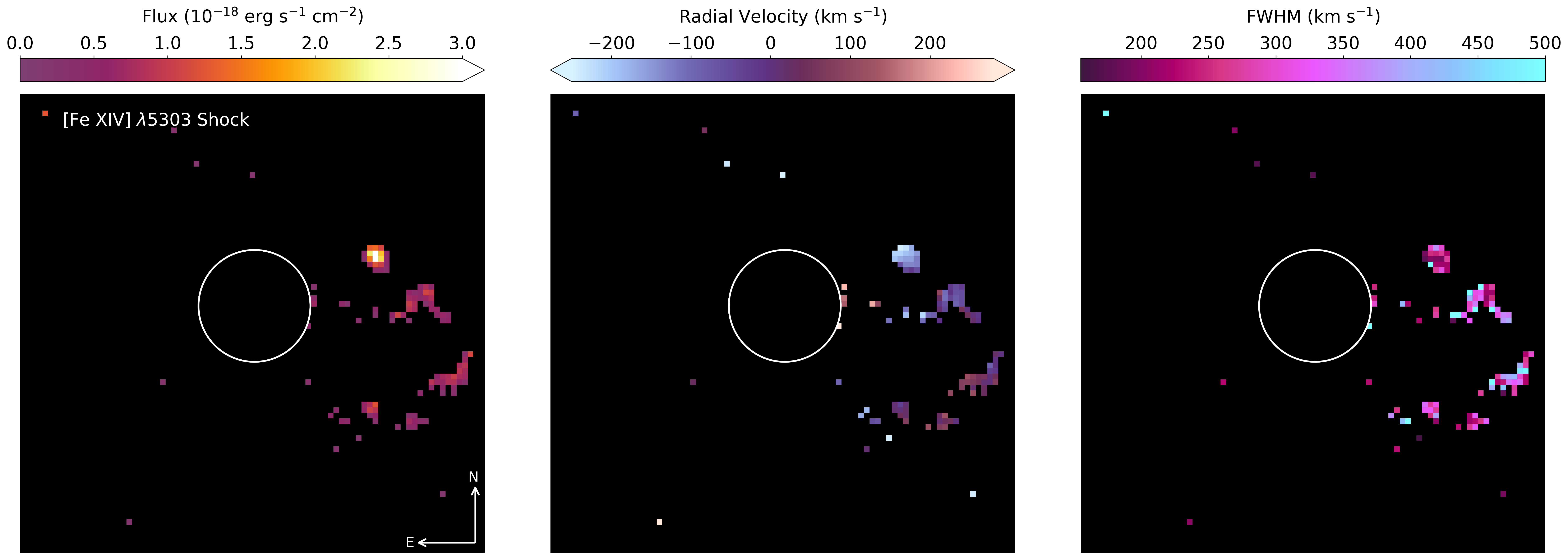}
    \caption{MUSE: shock component fit results. The PWN region within radius \mbox{8\,arcsec} (denoted with a white circle) is masked. \textit{Top Row:} \mbox{[\ion{S}{II}]\,$\lambda$6731} shock component. Spatial binning is $2\times2$ compared to the original MUSE spatial sampling. Pixels with radial velocities within \mbox{$\pm50$\,km\,s$^{-1}$} are masked due to limitations set by the spectral resolution in disentangling the shock and unshocked emission. \textit{Bottom Row:} \mbox{[\ion{Fe}{XIV}]\,$\lambda$5303} shock component. Spatial binning is $4\times4$ compared to the original MUSE spatial resolution. The colour scales differ between the panels.
    \label{fig:results-fit-results-shock}}
\end{figure*}

Because the \mbox{[\ion{S}{II}]\,$\lambda\lambda$6716,6731} doublet is a well-known shock tracer in SNRs and the MUSE spectra have good signal-to-noise ratio in the wavelength range covering said doublet (Figure\,\ref{fig:methods-entire-spectra-linear-ism-nw}), we begin our shock  emission analysis with this emission line. The top row of Figure\,\ref{fig:results-fit-results-shock} presents the properties of the shock component of \mbox{[\ion{S}{II}]\,$\lambda$6731} (the longer wavelength line in the \mbox{[\ion{S}{II}]\,$\lambda\lambda$6716,6731} doublet) obtained from the two-component fit algorithm outlined in Section\,\ref{sec:methods}. In the left panel, significant spatial variation in flux across the FOV is evident. The shock emission appears to consist of knots of varying luminosity, with the brightest knots located in the NW and SE. Larger, dimmer regions of shock emission are found in the N extending all the way to the SE. No shock emission from [\ion{S}{II}] is detected in the SW corner of the MUSE FOV, where the shock and unshocked ISM components of the [\ion{S}{II}] emission are most likely blended. This is presumably due to the lower radial velocities of the shock in this region (see Fe-line results presented in Section\,\ref{subsec:results-iron-lines}). The bright field star (Figure\,\ref{fig:intro}), which overpowers the much weaker shock emission across this large spatial region adds an additional uncertainty.

The middle panel of the top row in Figure\,\ref{fig:results-fit-results-shock} shows the radial velocity information provided by our fitting algorithm. We see that the majority of the knots show redshifted emission, in contrast to the brightest knots that are blueshifted and located far away from the PWN from the NW to the SE. In general, the radial velocities are typical for shocked regions ranging within \mbox{$\sim\pm$100\,km\,s$^{-1}$}. We have masked emission of the lowest radial velocities (within \mbox{$\sim\pm$50\,km\,s$^{-1}$}) due to limitations set by the spectral resolution, which prevents us from clearly disentangling the shock component with low radial velocities from the unshocked ISM.

Finally, the right panel of the top row of Figure\,\ref{fig:results-fit-results-shock} shows the FWHM of the shocked \mbox{[\ion{S}{II}]\,$\lambda$6731}. Similar spatial asymmetries are present here as in the radial velocities. The bright regions from the NW to the SE exhibit greater FWHMs than the dim and redshifted knots in the N and the western side of the remnant.

We also study other bright emission lines like \mbox{H$\beta$}, \mbox{[\ion{O}{III}]\,$\lambda\lambda$4959,5007}, and \mbox{[\ion{S}{III}]\,$\lambda$9069} (presented in Appendix\,\ref{subsec:appendix-additional-gaussian-results}), each of which show similar spatial morphology with the presented \mbox{[\ion{S}{II}]\,$\lambda$6731} line in all the three parameters: flux, radial velocity and FWHM. The lines at shorter wavelengths, \mbox{H$\beta$}, and \mbox{[\ion{O}{III}]\,$\lambda\lambda$4959,5007} suffer from the limitation set by the spectral resolution, and \mbox{[\ion{S}{III}]\,$\lambda$9069} is generally located in a noisy part of the MUSE spectrum, as discussed in Section\,\ref{sec:methods}. Consequently, the  \mbox{[\ion{S}{II}]\,$\lambda\lambda$6716,6731} doublet gives the best overview of the emission originating from shock-interaction of radiative shocks.

\begin{figure*}
    \includegraphics[width=\textwidth]{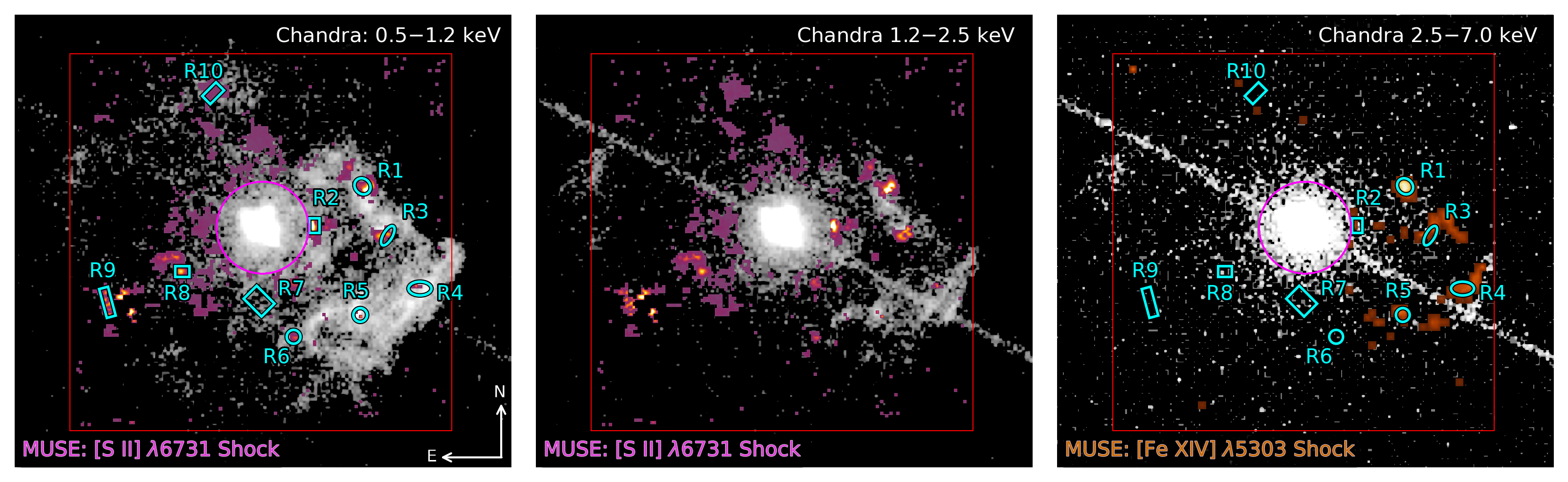}
    \caption{\textit{Left:} \mbox{[\ion{S}{II}]\,$\lambda$6731} shock component flux (from the top left panel in Figure\,\ref{fig:results-fit-results-shock}) superposed on Chandra X-ray image (\mbox{0.5--1.2\,keV}). Both images are reprojected to the same sky coordinates and the PWN region (within radius \mbox{8\,arcsec}) is indicated with a magenta circle. The cyan ellipses and rectangles denote extraction regions \mbox{R1--R10}. The red rectangle marks the MUSE FOV. \textit{Middle:} Same except Chandra image from \mbox{1.2--2.5\,keV}. \textit{Right:} \mbox{[\ion{Fe}{XIV}]\,$\lambda$5303} shock component flux (from the bottom left panel in Figure\,\ref{fig:results-fit-results-shock}) superposed on Chandra X-ray image (\mbox{2.5--7.0\,keV}), where the PWN and the Chandra read-out tail dominate the high-energy X-ray emission. As in the leftmost panel, the cyan ellipses and rectangles denote the regions \mbox{R1--R10}. The PWN is saturated in all panels to enhance the visibility of fainter structures.
    \label{fig:results-optical-shock-on-chandra}}
\end{figure*}

Figure\,\ref{fig:results-optical-shock-on-chandra} presents the optical shock component of \mbox{[\ion{S}{II}]\,$\lambda$6731} superimposed on Chandra X-ray images in different energy bands. The lower energy bands (\mbox{0.5--1.2\,keV}, and \mbox{1.2--2.5\,keV}) show significant overlap with the optical shock emission in regions other than the SW (where we do not have optical shock signal due to the bright field star) and in the SE, where we see optical shock emission but no emission from the X-rays. In X-rays, the straight diagonal feature spanning from the N to the W aligns with the brightest knots of optical shock emission in the two lower energy bands. The dimmest optical shock emission in the N also has a corresponding component in the X-rays. 

In hard X-rays (\mbox{2.5--7.0 keV}), the most prominent features (excluding the dominant PWN, and the Chandra readout tail) are the hard arcs located at the edges of the MUSE FOV in the northeast (NE) and SW. Due to these FOV edges being noisy (Figure\,\ref{fig:intro}), we cannot draw firm conclusions about  optical emission from these regions.

A similar overlap between the soft X-rays and the optical \mbox{[\ion{S}{II}]\,$\lambda$6731} shock emission can be observed between the optical and the radio, as the radio emission follows the X-ray morphology in the W (see the radio emission superposed on the X-rays in \citealt{Brantseg-2014}). In the N, we also see a rough correspondence between the optical and radio features. In contrast, the bright optical knots in the SE do not have a clearly observable X-ray counterpart but do have counterpart in the radio. However, it is important to note that our comparison is limited to the line-of-sight perspective, and the true three-dimensional origins of these components may differ. Based on the spatially resolved shock emission results, we define representative spatial regions (\mbox{R1--R10}, Figure\,\ref{fig:results-optical-shock-on-chandra}) in the MUSE FOV that we investigate in Section\,\ref{subsec:spectral-properties}.

\begin{figure}
    \includegraphics[width=\linewidth]{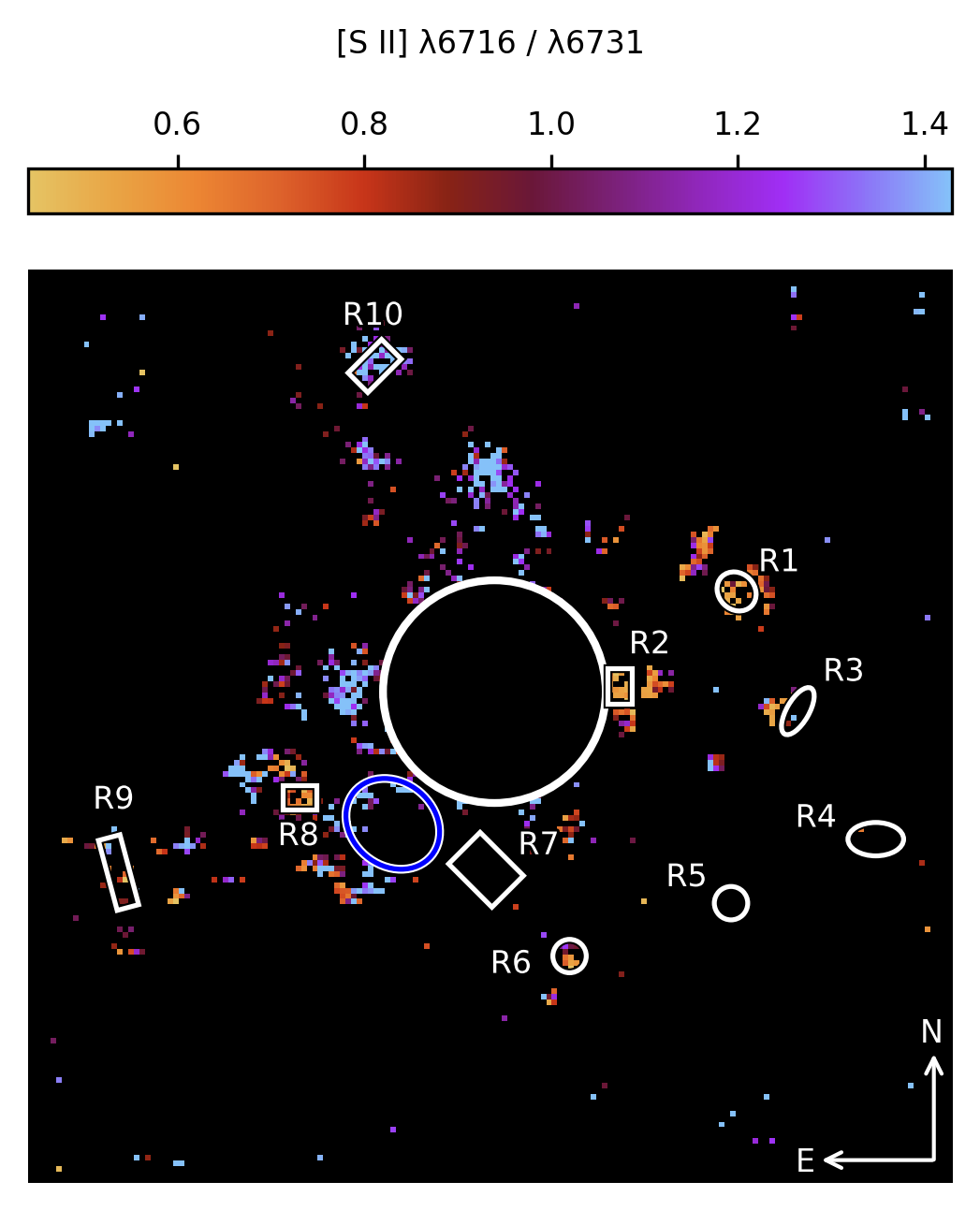}
    \caption{MUSE: \mbox{[\ion{S}{II}]\,$\lambda$6716/$\lambda$6731} shock-component ratio map derived from the emission line fits. The PWN is masked and indicated by a white circle with a radius of \mbox{8\,arcsec}. The labeled white ellipses and rectangles denote extraction regions \mbox{R1--R10}. The brightest parts of the H-blob in the SE, discovered in \citetalias{L21}, is marked with a blue ellipse.}
    \label{fig:results-s-ii-ratio}
\end{figure}

The \mbox{[\ion{S}{II}]\,$\lambda$6716 / [\ion{S}{II}]\,$\lambda$6731} ratio is a commonly used density diagnostic in SNRs. Figure\,\ref{fig:results-s-ii-ratio} presents this ratio for SNR\,0540 across the MUSE FOV. Spaxels with ratio values deviating more than \mbox{$3\sigma$} from the theoretically accepted values [0.44, 1.43] \citep{Osterbrock-2006} are masked (\mbox{$\sim20$ per cent} out of all spaxels that have signal in Figure\,\ref{fig:results-fit-results-shock}). These uncertainties are obtained by propagating the uncertainties of the fitted fluxes. Figure\,\ref{fig:results-s-ii-ratio} reveals clear spatial variations: we observe the lowest values in the brightest blueshifted knots from the NW to the SE, where the lower ratio values suggest higher densities. In contrast, the highest values are found in the dimmer knots located in the N and in the eastern side of the remnant, indicating relatively lower densities in these shock interacting regions.

\subsubsection{Spectral Properties}\label{subsec:spectral-properties}

\begin{figure*}
    \includegraphics[width=\textwidth]{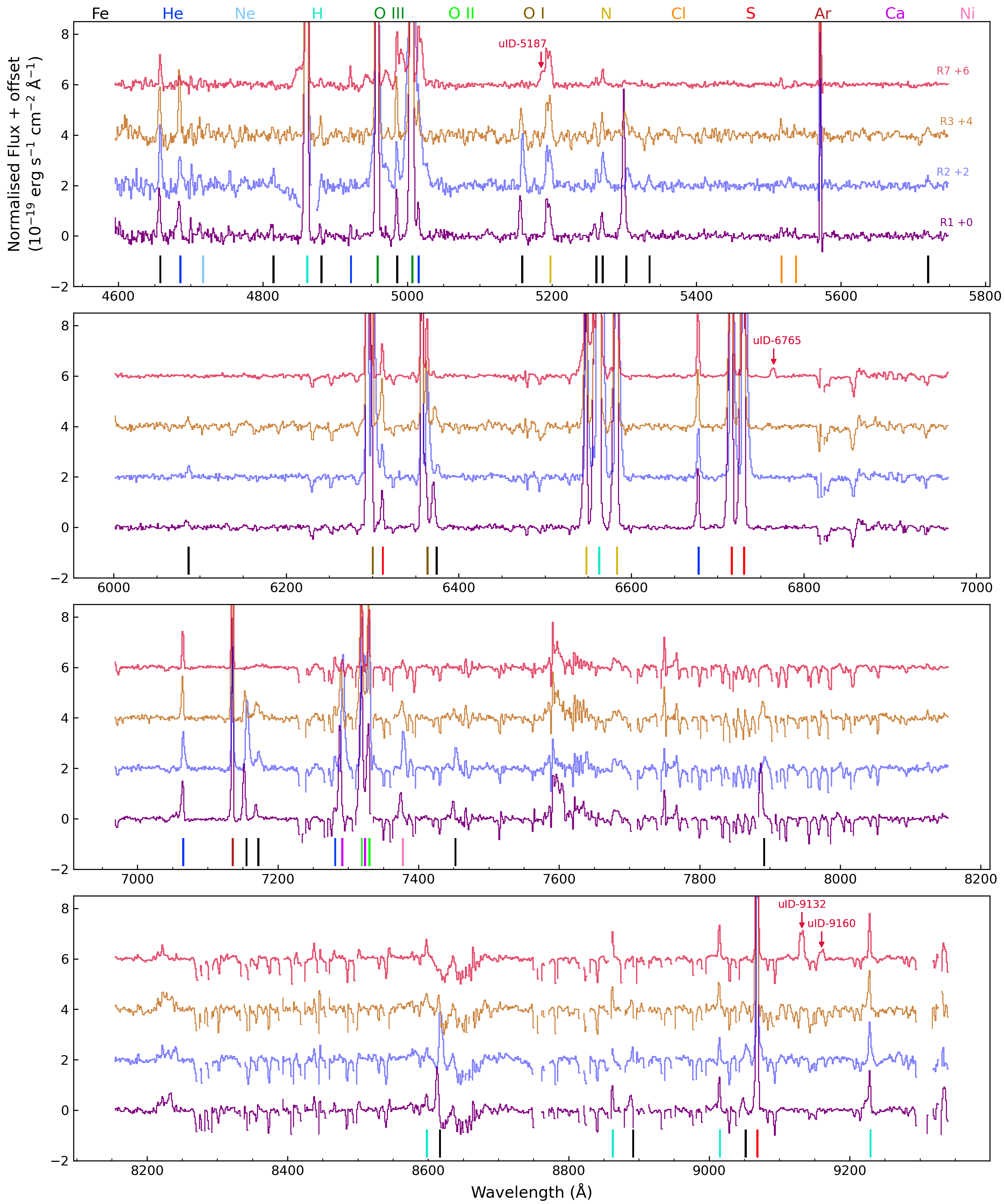}
    \caption{Background-subtracted MUSE spectra from shock-dominated regions R1 (purple), R2 (light blue), R3 (light brown), and R7 (red). The fluxes are normalised with the number of pixels in the extraction aperture (locations and sizes of the apertures are shown in Figure\,\ref{fig:results-optical-shock-on-chandra}) and an offset is added for visual clarity. Additionally, fluxes below $-1$ (in corresponding units) are masked for visual clarity before adding an offset. Short vertical lines at the bottom of each panel indicate the exact wavelengths of the brightest emission lines, with different colors corresponding to different atomic or ionic species. The brightest emission lines are also listed in Table\,\ref{tab:appendix_observed_emission_lines}. The red arrows point to unidentified emission lines in R7, see text.
    \label{fig:results-spectra-from-regions_nw_w_ww4}}
\end{figure*}

To study the shock emission in SNR\,0540 in more detail, we use the results from the previous section (\mbox{Figures\,\ref{fig:results-fit-results-shock},\,\ref{fig:results-optical-shock-on-chandra},\,and\,\ref{fig:appendix-gaussian-fit-results}}) to select a set of diverse spectral extraction regions across the full MUSE FOV. The regions were required to be bright, span different parts of the field with varying radial velocities, and be located away from field stars. The selected regions (\mbox{R1--R10}) thus reflect the range of shock emission characteristics in SNR\,0540. We present these regions in Figure\,\ref{fig:results-optical-shock-on-chandra}. We proceed by extracting entire MUSE wavelength range spectra from each of these regions and background-subtract them with the LOWESS algorithm presented in Section\,\ref{sec:methods}. The resulting spectra from \mbox{R1--R3}, and R7 can be found in Figure\,\ref{fig:results-spectra-from-regions_nw_w_ww4}, while those from \mbox{R4--R6}, and \mbox{R8--R10}, along with the brightest identified emission lines, are presented in \mbox{Figures\,\ref{fig:appendix-results-spectra-from-r4-5-6} and \ref{fig:appendix-results-spectra-from-r8-9-10}}, and Table\,\ref{tab:appendix_observed_emission_lines} in Appendix~\ref{appendix:r5-r8}, respectively. Table\,\ref{tab:appendix_observed_emission_lines} lists only emission lines that exceed three times the standard deviation measured from a corresponding background region.

These spectra reveal structure in shock-components of the emission lines (for example in the \mbox{[\ion{O}{III}]\,$\lambda\lambda$4959,5007} and \mbox{H$\alpha$ + \mbox{[\ion{N}{II}]\,$\lambda\lambda$6548,6583} systems}) showcasing the fact that the shock component of the emission is resolved by MUSE. The spectrum from region R2 (region closest to the PWN in line-of-sight perspective) exhibits wide oxygen tails, which most likely are due to the torus of oxygen-emitting ejecta (reported in \citetalias{L21}) overlapping with this region (also discussed in \citealt{Lundqvist-2022}).

\begin{figure*}
    \includegraphics[width=\textwidth]{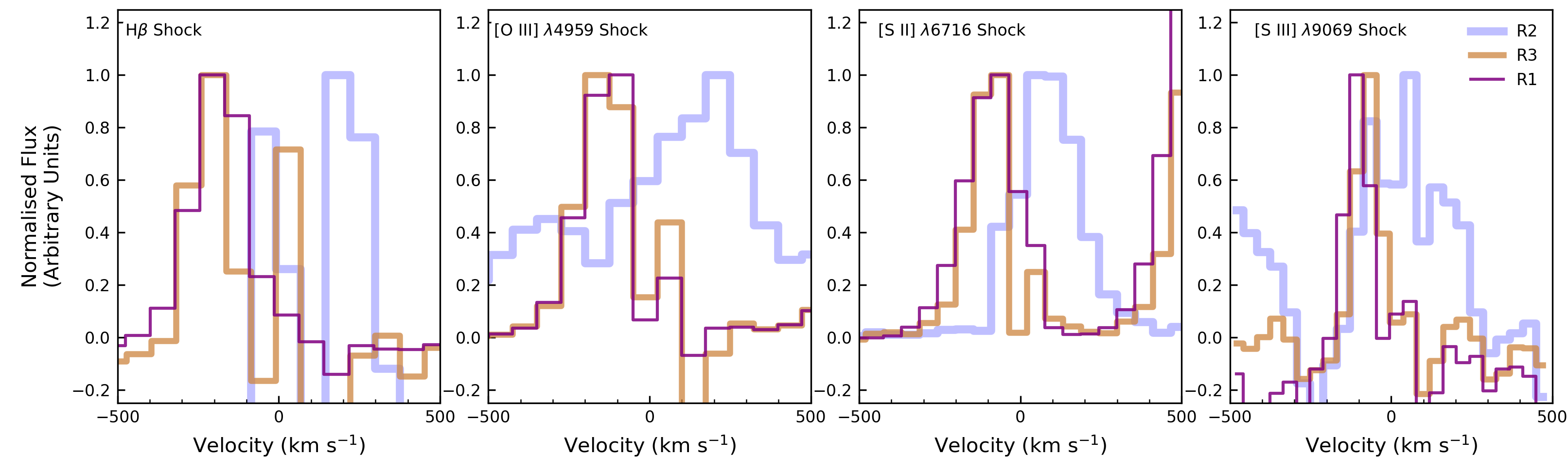}
    \caption{MUSE: Shock component velocity spectra of bright emission lines from regions R1 (purple), R2 (blue), and R3 (brown, locations of these regions can be found in Figure\,\ref{fig:results-optical-shock-on-chandra}). A strong unshocked ISM component has been subtracted from all lines before the fluxes are normalised with respect to the hight of the shock peak. Notably, as the wide tails suggest, the unshocked-ISM-subtracted [\ion{O}{III}] line from R2 (second panel) still contains emission from the O-halo around the PWN, see text.}
    \label{fig:results-vel-spectra-bright-emission-lines}
\end{figure*}

We also show line profiles of the shock components of the brightest emission lines in Figure\,\ref{fig:results-vel-spectra-bright-emission-lines}, where the shock component is revealed by subtracting the strong unshocked ISM component according to the fit results. As suggested in the radial velocity results (middle panel of the top row of Figure\,\ref{fig:results-fit-results-shock}) the three regions \mbox{R1--R3} (leftmost panel of Figure\,\ref{fig:results-optical-shock-on-chandra}) exhibit distinct radial velocity profiles, R1 being the most blueshifted across all emission lines and R2 the most redshifted. As discussed above, R2 shows prominent oxygen tails (second panel of Figure\,\ref{fig:results-vel-spectra-bright-emission-lines}) originating from the O-halo around the PWN, which the fit algorithm, optimised for fitting the narrow unshocked ISM component, fails to effectively subtract. Similar to R1, the line profiles of the the shock components of the brightest emission lines from R3 lie also on the blueshifted side in radial velocity, although at slightly lower velocities (in absolute value) as can be seen particularly in the rightmost panel of Figure\,\ref{fig:results-vel-spectra-bright-emission-lines} at the best spectral resolution. Finally, the velocity profiles of \mbox{[\ion{S}{II}]\,$\lambda$6716} from other regions with significant shock emission components are presented in Figure\,\ref{fig:appendix-results-vel-spectra-bright-emission-lines} in Appendix~\ref{appendix:r5-r8} for completeness.

We summarise all fit results, surface brightness ($S$), radial velocity ($v_\mathrm{r}$) and FWHM, in Table\,\ref{tab:appendix_fit_results_table_R1-R3} (for \mbox{R1--R3}), and Tables\,\ref{tab:appendix_fit_results_table_R4-R5} and \ref{tab:appendix_fit_results_table_R6-R8-R9} (for \mbox{R4 and R5}, and for \mbox{R6, R8, and R9}, respectively) in the Appendix~\ref{appendix:r5-r8}. The reported uncertainties reflect the formal statistical uncertainties from the emission line fits and do not account for systematic effects. One prominent source of systematic uncertainty arises from unresolved, strong unshocked ISM components. In this case, the limited spectral resolution makes it challenging to deblend the unshocked ISM and shock emission. To account for this, we also report the spectral resolution ($R$) alongside the other fit parameters.

Results from R7 are discussed separately in Section\,\ref{subsec:regions_s}. While the signal from \mbox{R10} confirms the presence of shock emission, it is insufficient to yield statistically significant fit results, and therefore fit results for \mbox{R10} are not listed.

\begin{figure*}
    \includegraphics[width=\textwidth]{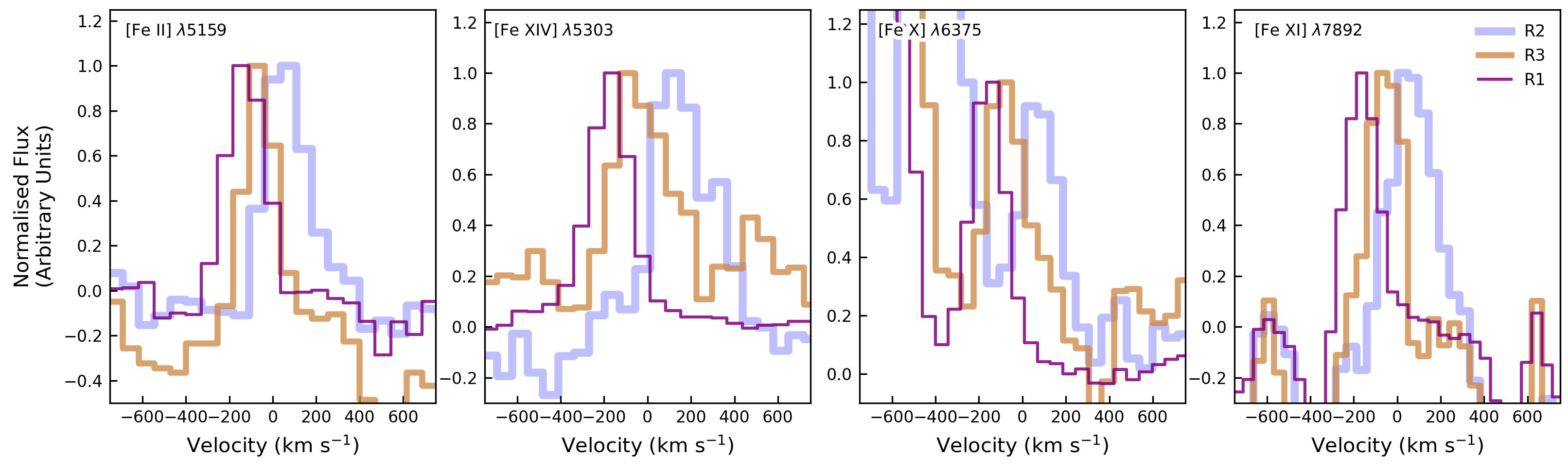}
    \caption{MUSE: Velocity spectra of Fe lines from regions R1 (purple), R2 (blue), and R3 (brown, locations of these regions can be found in Figure\,\ref{fig:results-optical-shock-on-chandra}). The fluxes are normalised with respect to the hight of the shock peak. Note that the panels have slightly different scales on the flux axes and are arranged in order of increasing wavelength.}
    \label{fig:results-vel-spectra-iron}
\end{figure*}

\subsection{Coronal Fe Lines} \label{subsec:results_coronal_fe_lines} \label{subsec:results-iron-lines}

Figure\,\ref{fig:results-spectra-from-regions_nw_w_ww4} displays bright Fe lines (indicated by short black vertical lines on the wavelength axis) for spectra from \mbox{R1--R3} (as defined in Figure\,\ref{fig:results-optical-shock-on-chandra}). The Fe lines are prominent in the \mbox{R1--R3} spectra (particularly in R1), where we can identify several coronal Fe lines, such as \mbox{[\ion{Fe}{XIV}]\,$\lambda$5303}, \mbox{[\ion{Fe}{X}]\,$\lambda$6375}, and \mbox{[\ion{Fe}{XI}]\,$\lambda$7892}. The presence of the coronal Fe lines further supports the notion that these regions are undergoing shock interaction.

Based on the brightness of \mbox{[\ion{Fe}{XIV}]\,$\lambda$5303} in the R1 spectrum, we proceed by fitting the entire MUSE FOV for this line and present the results in the bottom panel of Figure\,\ref{fig:results-fit-results-shock}. Similarly to \mbox{[\ion{S}{II}]\,$\lambda$6731}, the knot associated with R1 is the brightest also in \mbox{[\ion{Fe}{XIV}]\,$\lambda$5303} (R1 knot hereafter), as shown in the left panel of the bottom row. Instead of knots distributed throughout the MUSE FOV, the \mbox{[\ion{Fe}{XIV}]\,$\lambda$5303} emission is concentrated to a ring-like feature in the W and SW and show nearly no emission in other parts of the MUSE FOV. 

The radial velocity results for \mbox{[\ion{Fe}{XIV}]\,$\lambda$5303}, shown in the bottom middle panel of Figure\,\ref{fig:results-fit-results-shock}, indicate that the R1 knot is significantly blueshifted, consistent with the fit results obtained for \mbox{[\ion{S}{II}]\,$\lambda$6731}. The R1 knot has a clear gradient in the \mbox{[\ion{Fe}{XIV}]\,$\lambda$5303} radial velocity; the northeastern part of the knot having the highest blueshifted velocities (even up to \mbox{$\sim-200$\,km\,s$^{-1}$}), which then gradually decrease to radial velocities close to the rest velocity in the southwestern part of the knot. The rest of the Fe ring exhibits lower radial velocities and is even redshifted in the SW (\mbox{$\sim+50$\,km\,s$^{-1}$}). We note that such low velocity emission is masked for \mbox{[\ion{S}{II}]\,$\lambda$6731} in the top row of \mbox{Figure\,\ref{fig:results-fit-results-shock}} due to the limitations set by the spectral resolution in distinguishing the unshocked ISM component from the shock component, which likely explains why no significant shock emission is seen for \mbox{[\ion{S}{II}]\,$\lambda$6731} in the SW.

In the FWHM panel (bottom right of Figure\,\ref{fig:results-fit-results-shock}), the highest FWHM values of \mbox{[\ion{Fe}{XIV}]\,$\lambda$5303} emission are located in the W, similarly to \mbox{[\ion{S}{II}]\,$\lambda$6731}. In particular, the knot associated to R3 has the highest FWHM values (approaching \mbox{+500\,km\,s$^{-1}$}). However, these FWHM results do not reveal similar structure in the individual knots as is seen in the radial velocity, presumably due the larger spatial binning, and to the lower spectral resolution of MUSE in the \mbox{[\ion{Fe}{XIV}]\,$\lambda$5303} wavelength range.

Figure\,\ref{fig:results-spectra-from-regions_nw_w_ww4} illustrates that the other coronal Fe lines, \mbox{[\ion{Fe}{XI}]\,$\lambda$7892}, and \mbox{[\ion{Fe}{X}]\,$\lambda$6375}, also show significant emission in the spectrum extracted from R1. However, unlike \mbox{[\ion{Fe}{XIV}]\,$\lambda$5303}, these lines do not exhibit significant emission elsewhere in the MUSE FOV, not even with increased spatial binning. Only few isolated pixels outside the R1 knot exhibit any signal, which is why spatially resolved images of \mbox{[\ion{Fe}{XI}]\,$\lambda$7892}, and \mbox{[\ion{Fe}{X}]\,$\lambda$6375} are omitted. This difference is likely attributable to \mbox{[\ion{Fe}{XI}]\,$\lambda$7892}, and \mbox{[\ion{Fe}{X}]\,$\lambda$6375} lines being intrinsically lower in brightness, generally resulting in a weaker signal that can only be captured by extracting larger (targeted) spatial regions like \mbox{R1--R3}. Consequently, we focus the following spatial analysis on \mbox{[\ion{Fe}{XIV}]\,$\lambda$5303}.

The optical \mbox{[\ion{Fe}{XIV}]\,$\lambda$5303} emission appears to coincide with the region of brightest emission in the soft X-rays (soft X-rays in the leftmost panel and optical \mbox{[\ion{Fe}{XIV}]\,$\lambda$5303} shock emission in the rightmost panel of Figure\,\ref{fig:results-optical-shock-on-chandra}). However, the X-ray emission does not form a ring, but rather an extended region with sharp edges, combining a diagonal filament in the NW to a wider, nearly perpendicular feature in the SW. The differences between the optical and X-ray emission morphologies in the SW are likely due to the bright field star that is located approximately at the centre of the Fe ring and is masked in the optical results (see left panel of Figure\,\ref{fig:intro}), which prevents us from fitting the entire SW region in the optical. Additionally, we observe more shocked Fe emission in the SW because, unlike \mbox{[\ion{S}{II}]\,$\lambda$6731}, there are no bright unshocked ISM lines near \mbox{[\ion{Fe}{XIV}]\,$\lambda$5303} that could blend with and obscure the shock emission. Interestingly, parts of the optical Fe ring in the SW appear to overlap with the southwestern hard arc seen in the hard X-rays (rightmost panel of Figure\,\ref{fig:results-optical-shock-on-chandra}).

As with other bright emission lines, we construct velocity profiles for the (coronal) Fe lines. Figure\,\ref{fig:results-vel-spectra-iron} presents these profiles for four different Fe lines: \mbox{[\ion{Fe}{II}]\,$\lambda$5159}, \mbox{[\ion{Fe}{XIV}]\,$\lambda$5303}, \mbox{[\ion{Fe}{X}]\,$\lambda$6375}, and \mbox{[\ion{Fe}{XI}]\,$\lambda$7892}. The radial velocities are consistent across all regions, with R1 exhibiting the strongest blueshift and R2 the most pronounced redshift. R2 also exhibits the strongest \mbox{[\ion{Fe}{II}]\,$\lambda$5159} emission. The \mbox{[\ion{Fe}{X}]\,$\lambda$6375} line, located close to the longer wavelength line of the \mbox{[\ion{O}{I}]\,$\lambda\lambda$6300,6364} doublet, is clearly resolved and separable from the oxygen-line system (unlike in \citealt{Vogt-2017}, where \mbox{[\ion{Fe}{X}]\,$\lambda$6375} was reported to be mixed with [\ion{O}{I}]). Velocity profiles of Fe lines in other regions are presented in Figure\,\ref{fig:appendix-results-vel-spectra-coronal-fe-lines} in Appendix\,\ref{appendix:r5-r8}.

\subsection{Unique Emission Lines in Region R7} \label{subsec:regions_s}

Unlike other regions studied above, the spectrum from region R7 (red line in Figure\,\ref{fig:results-spectra-from-regions_nw_w_ww4}, R7 location shown in Figure\,\ref{fig:results-optical-shock-on-chandra}) seems to possess different spectral properties without any evidence of shock emission. First, the spectrum contains few Fe lines. Second, the brightest emission lines have narrow, unshocked-ISM-like profiles, best seen in the \mbox{[\ion{S}{II}]\,$\lambda\lambda$6716,6731} doublet. An exception to this is the broadening observed in the Balmer lines \mbox{H$\alpha$}, \mbox{H$\beta$}, and the \mbox{[\ion{O}{III}]\,$\lambda\lambda$4959,5007} doublet. This broadening likely results from the overlap of region R7 with the hydrogen-blob located SE of the PWN (\citetalias{L21}) and the PWN-associated oxygen-halo, which extends to the location of R7 (in line-of-sight). 

The R7 spectrum contains several emission lines that are not present in the spectra of other regions and do not directly match any known emission lines typically associated with SNRs. The most prominent examples are features at 6765 (on the red side of \mbox{[\ion{S}{II}]\,$\lambda\lambda$6716,6731} doublet), 9132, and 9160\,Å (both on the red side of \mbox{[\ion{S}{III}]\,$\lambda$9069}). An additional feature is blended to the blue side of the \mbox{[\ion{N}{I}]\,$\lambda\lambda$5199,5202} doublet. The identification and origin of these four unique features are discussed in Section\,\ref{subsec:discussion-R4} and the remainder of this section focuses on the spectral and spatial properties of the three unidentified lines which we call \mbox{uID-6765}, \mbox{uID-9132}, and \mbox{uID-9160} hereafter. We refer to the fourth feature as \mbox{uID-5187}, but since it is blended to the \mbox{[\ion{N}{I}]} doublet, we omit it from further analysis for simplicity.

\begin{figure}
    \includegraphics[width=\linewidth]{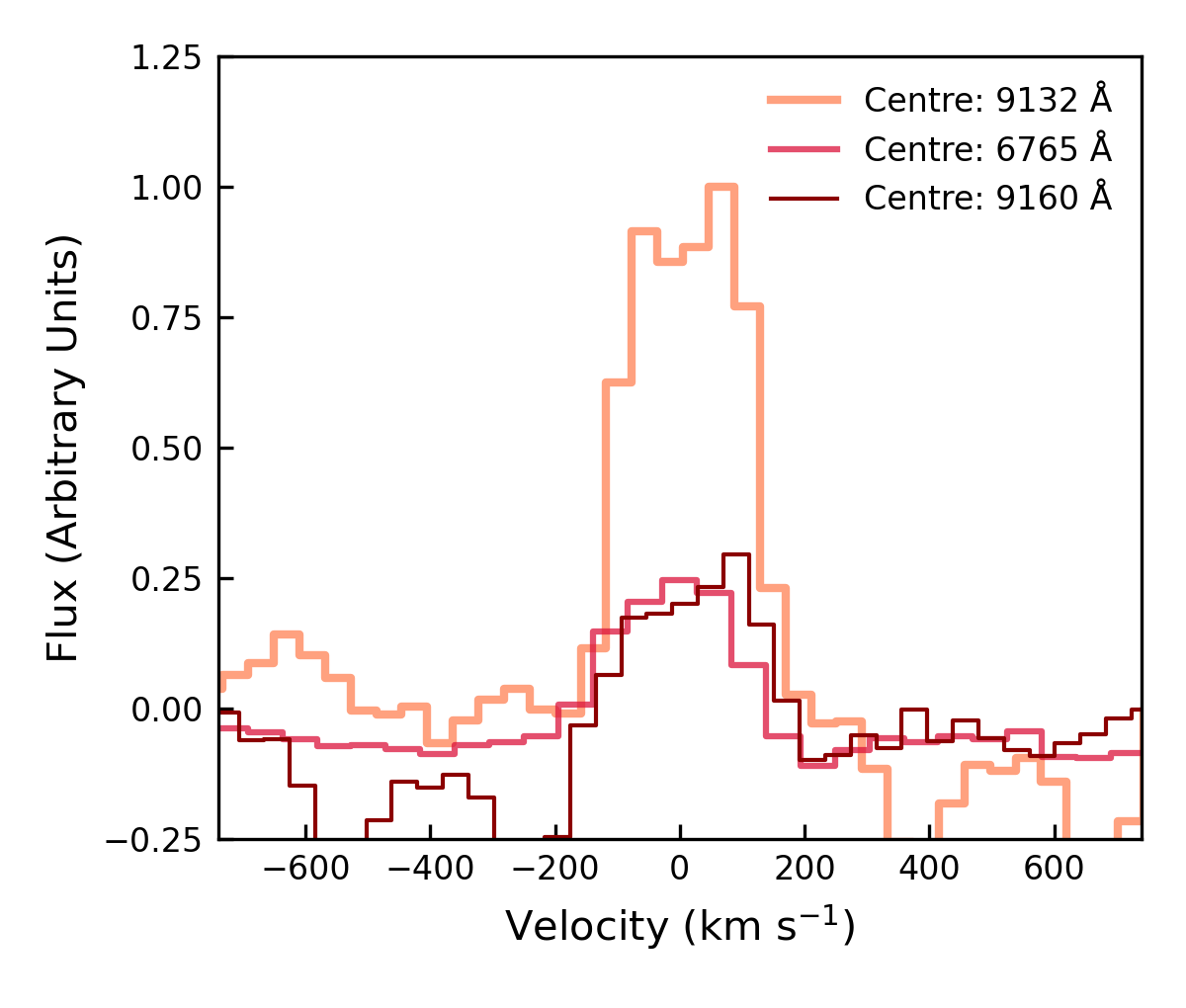}
    \caption{MUSE: Line profiles taken with respect to the rest wavelengths 9132 (light orange), 6765 (red), and 9160\,Å (dark red) from region R7. These profiles correspond to the unidentified features \mbox{uID-9132}, \mbox{uID-6765}, and \mbox{uID-9160}. The spectra are normalised to the peak flux of \mbox{uID-9132} within the displayed spectral range.}
    \label{fig:results-vel-spectra-unknown}
\end{figure}

We use the wavelengths 6765, 9132, and 9160\,Å to artificially set the rest velocities of the three unique lines, allowing us to construct their velocity profiles, as shown in Figure\,\ref{fig:results-vel-spectra-unknown}. We observe that all these unidentified lines exhibit roughly flat profiles with comparable widths, spanning approximately [$-200$, +200]\,km~s$^{-1}$. Among them, \mbox{uID-9132} is the brightest, while the other two have about one fourth of its brightness, and \mbox{uID-9160} is the most affected by noise. 

\begin{figure*}
    \includegraphics[width=\textwidth]{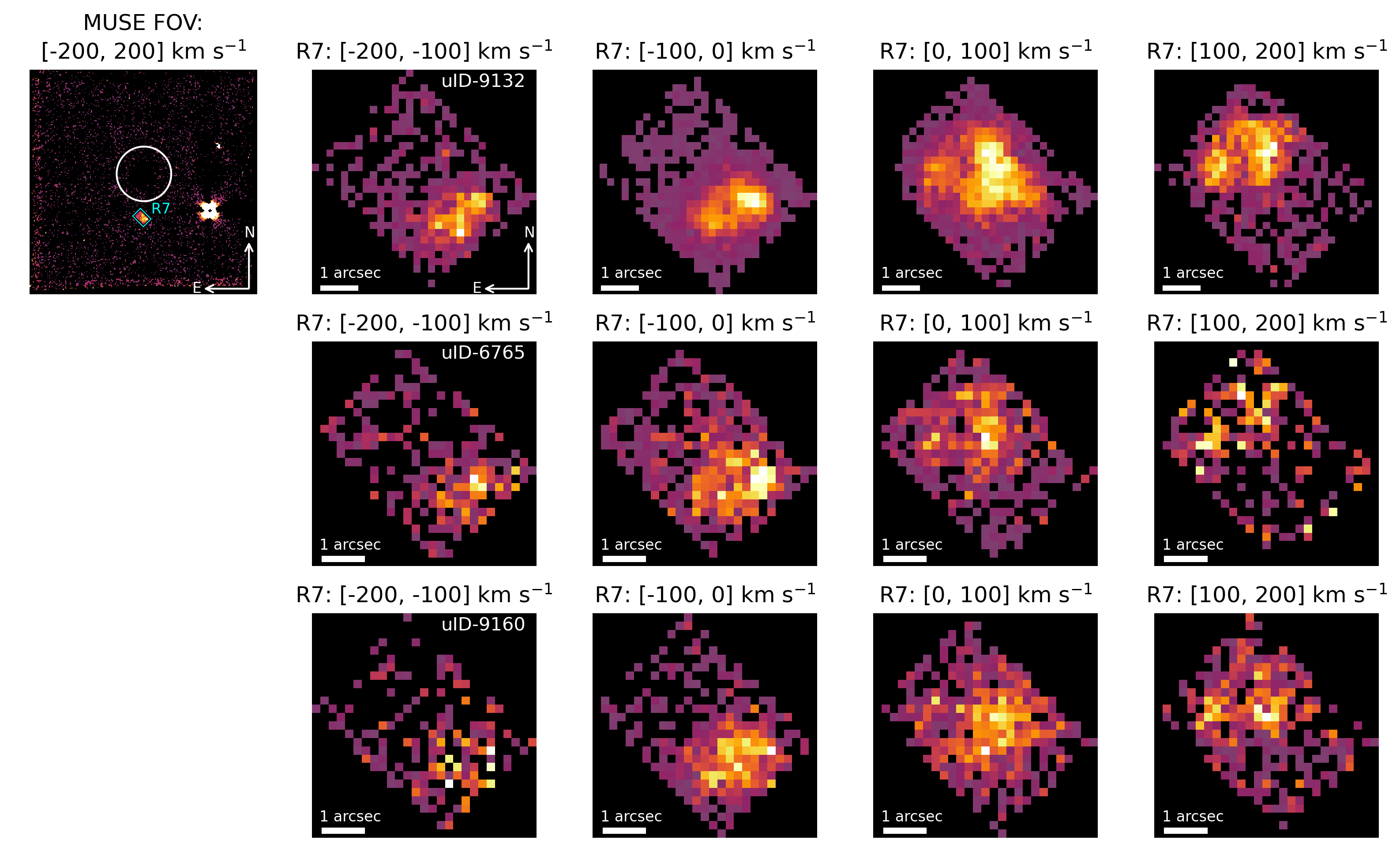}
    \caption{\textit{Top:} Spatial distribution of the brightest unidentified line {uID-9132} within \mbox{$\pm200$\,km\,s$^{-1}$} of 9132\,Å across the entire MUSE FOV. The bright spots not confined in R7 are incompletely subtracted field stars, see text. The PWN region is indicated by a white circle with a radius of \mbox{8\,arcsec}. The four rightmost panels on the top row show velocity slices taken from R7 with respect to the artificial rest wavelength 9132\,Å. All five panels share the same intensity scale.
    \textit{Middle:} Same as above but with respect to the artificial rest wavelength 6765 Å. \textit{Bottom:} Same as above but with respect to the artificial rest wavelength 9160 Å.}
    \label{fig:results-vel-slices-unknown}
\end{figure*}

Figure\,\ref{fig:results-vel-slices-unknown} illustrates how the emission from these three features, \mbox{uID-9132}, \mbox{uID-6765} and \mbox{uID-9160} are spatially distributed. The top left panel shows that we observe emission exclusively in R7 across the entire MUSE FOV within \mbox{$\pm200$\,km\,s$^{-1}$} of 9132\,Å. The two additional bright spots correspond to incompletely subtracted field stars (Figure\,\ref{fig:intro}).

Examining \mbox{uID-9132}, \mbox{uID-6765}, and \mbox{uID-9160} spatially at different radial velocities reveals a spatial shift in the brightest emission from the SW (at radial velocities \mbox{[$-200$, $-100$]\,km\,s$^{-1}$}) to the NE (\mbox{[+100, +200]\,km\,s$^{-1}$}) within R7. This radial-velocity-dependent spatial variation is presented in Figure\,\ref{fig:results-vel-slices-unknown}.

\section{Discussion} \label{sec:discussion}

Emission lines arising from shock interaction offer a way to extract information on physical properties of SNRs by taking ratios from different atomic or ionic species. For instance, such ratios can provide insights into the post-shock electron temperatures and densities as well as blast-wave shock velocities (e.g \citealt{Osterbrock-2006,Vogt-2017}). In this section, we discuss how the first optical integral-field-spectroscopic data covering the entire SNR\,0540 contribute to the knowledge of shock-interacting regions in this and other remnants. In Section\,\ref{subsec:discussion-shock-conditions}, we focus on investigating the physical conditions in the shocked regions by estimating post-shock electron temperatures and densities by measuring the \mbox{[\ion{S}{III}]\,$\lambda$9069/$\lambda$6312} and \mbox{[\ion{S}{II}]\,$\lambda$6716/[\ion{S}{II}]\,$\lambda$6731} ratios. We follow up on investigating the blast-wave shock velocities by examining the \mbox{[\ion{Fe}{XIV}]~$\lambda5303$/[\ion{Fe}{XI}]~$\lambda7892$} ratio. In Section\,\ref{subsec:discussion-R4}, we turn our attention to the unidentified emission lines observed in R7 and discuss their possible origins. Finally, we formulate an overall picture of shock emission in SNR\,0540 and place our findings in the context of other SNRs (Section\,\ref{subsec:discussion-overall_picture}).

\subsection{Physical Conditions in the Shocks}\label{subsec:discussion-shock-conditions}

\begin{figure}
    \includegraphics[width=\linewidth]{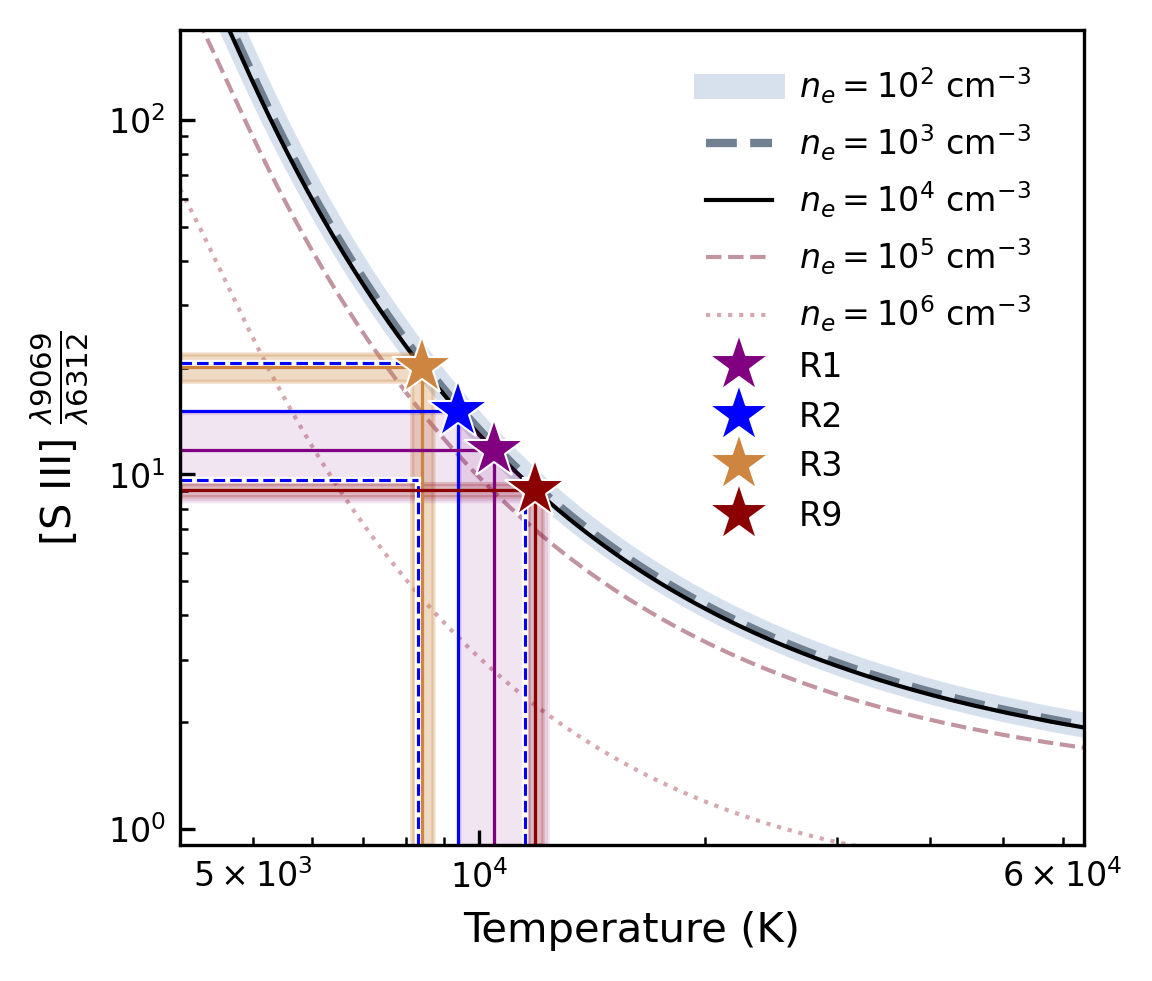}
    \caption{\mbox{[\ion{S}{III}]~$\lambda9069$/$\lambda6312$} ratio as a function of the post-shock electron temperature (all curves). The different curves correspond to different electron densities ($n_e$, see figure legend). The \mbox{[\ion{S}{III}]}-line ratio values for regions R1 (purple star), R2 (blue star), R3 (brown star), and R9 (dark red star) are shown with their ${1\sigma}$ uncertainties (shaded regions). For R2, only the edges of the uncertainty range are marked with blue dashed lines for visual clarity. }
    \label{fig:discussion-s-iii-ratio-shock-temp}
\end{figure}

\subsubsection{Post-Shock Electron Temperature} \label{subsubsec:discussion-temperature}

Certain optical emission line ratios, e.g. \mbox{[\ion{N}{II}]\,$\lambda\lambda$6548,6583/$\lambda$5755} and \mbox{[\ion{O}{III}]\,$\lambda\lambda$4959,5007/$\lambda$4363} are sensitive to the post-shock electron temperature since the excitation rates of the atomic transitions, and hence the line intensities, have a strong temperature dependence \citep{Osterbrock-2006}. Many such emission line ratios are unfortunately outside of the spectral range of MUSE, or fall within the instrument's spectral gap (after the LMC systemic velocity correction in Section\,\ref{sec:observations}, the spectral gap lies between \mbox{5748–6002\,Å}). As a result, we are limited to using the \mbox{[\ion{S}{III}]\,$\lambda$9069/$\lambda$6312} ratio. We model this ratio using the same atomic data as in \cite{Groningsson-2008} for SN\,1987A, with the exception of updated transition probabilities from the CHIANTI database \citep{DelZanna21}. The resulting models for five post-shock electron densities, $n_e$, between \mbox{$10^2$--$10^6$\,cm$^{-3}$} are shown in Figure\,\ref{fig:discussion-s-iii-ratio-shock-temp}. These models indicate that this ratio is insensitive for values \mbox{$n_e \lesssim 10^4$ cm$^{-3}$}.

To measure the \mbox{[\ion{S}{III}]} line ratio across different regions in SNR\,0540, we must first deblend the [\ion{S}{III}]\,$\lambda$6312 line from the nearby \mbox{[\ion{O}{I}]\,$\lambda$6300}. This step is essential in regions R2, R8, and R10, where the redshifted shock component of \mbox{[\ion{O}{I}]\,$\lambda$6300} blends with the much weaker \mbox{[\ion{S}{III}]\,$\lambda$6312} emission (\mbox{Figures\,\ref{fig:results-spectra-from-regions_nw_w_ww4}, \ref{fig:appendix-results-spectra-from-r4-5-6}, and \ref{fig:appendix-results-spectra-from-r8-9-10}}). Additionally, the \mbox{[\ion{O}{I}]\,$\lambda$6300} emission is affected by an instrumental artefact (a distinct, narrow peak on the blue side of the line, similar in shape across the whole MUSE FOV), which can be modeled using a gaussian with FWHM fixed to the spectral resolution. After subtracting this feature, we perform simultaneous fits for the unshocked ISM and shock components of both \mbox{[\ion{O}{I}]\,$\lambda$6300} and \mbox{[\ion{S}{III}]\,$\lambda$6312}. In contrast, for regions R1, R3, R6, R7 and R9, such deblending was not required due to either insignificant or blueshifted \mbox{[\ion{O}{I}]\,$\lambda$6300} shock components. We show the resulting \mbox{[\ion{S}{III}]-ratio} values for \mbox{R1--R3} and R9 in Figure\,\ref{fig:discussion-s-iii-ratio-shock-temp}, which we also summarise in Table\,\ref{tab:discussion_observed_ratio-density}. Other regions lacked sufficient shock emission signal for a robust measurement. 

Assuming \mbox{$n_e\lesssim 10^4$\,cm$^{-3}$}, we derive post-shock electron temperatures for \mbox{R1--R3} and R9, as shown in Figure\,\ref{fig:discussion-s-iii-ratio-shock-temp}. The values are in the range $\sim $ 0.8--1.2 $\times 10^4$\,K, as listed in Table\,\ref{tab:discussion_observed_ratio-density}. Comparing our results to previous studies, R1 overlaps with the filament F4 in \cite{Lundqvist-2022}, where an [\ion{O}{III}] temperature of $\sim2\times10^4$\,K was measured, consistent with our measurement. However, as the contributions of the shocked gas were not fully isolated, \cite{Lundqvist-2022} caution that this value might be a lower limit. For example, in their F1 (overlapping with our R2), a more realistic [\ion{O}{III}] shock temperature was argued to be $\gtrsim 5.5\times10^4$\,K. Assuming isobaric conditions for a radiative shock, \cite{Lundqvist-2022} inferred $n_e \sim 10^3$\,cm$^{-3}$ for the [\ion{O}{III}] emitting shocked gas. Further cooling and compression of the shocked gas would lower temperatures and increase densities for the shocked [\ion{S}{III}] gas, and perhaps even more so for [\ion{S}{II}]. Therefore, our choice of \mbox{$n_e\lesssim10^4$\,cm$^{-3}$} for the modeled [\ion{S}{III}] line ratio and the resulting temperatures of the order \mbox{$T\sim10^4$\,K} seem reasonable.

\subsubsection{Post-Shock Electron Density} \label{subsubsec:discussion-density}

\begin{table*}
    \begin{threeparttable}
    \caption{Observed \mbox{[\ion{S}{III}]\,$\lambda9069$/$\lambda6312$} shock-component ratios (\mbox{[\ion{S}{III}]-ratio}) with corresponding post-shock temperature estimates (T), and observed \mbox{[\ion{S}{II}] $\lambda$6716/$\lambda$6731} shock-component ratios (\mbox{[\ion{S}{II}]-ratio}) with corresponding post-shock electron density estimates ($n_e$) in SNR\,0540 (the locations for \mbox{R1--R10} are shown in \mbox{Figures\,\ref{fig:results-optical-shock-on-chandra}\,and\,\ref{fig:results-s-ii-ratio}}). The electron density estimates assume a temperature of $T=10^4$\,K. Listed are also the measured \mbox{[\ion{Fe}{XIV}] $\lambda$5303/[\ion{Fe}{XI}] $\lambda$7892} ratios (Fe-ratio) and derived blast-wave shock velocities ($v_\mathrm{s}$). The shock velocities are determined with a model presented in \citet{Vogt-2017}.}
    \begin{tabular}{lcccccc}
\toprule
           Region & [\ion{S}{III}]-ratio & $T$ & [\ion{S}{II}]-ratio           & $n_e$ & Fe-ratio & $v_\mathrm{s}$\\
                  &   & $(10^4\mathrm{K})$ &                & $(10^3\times\mathrm{cm}^{-3})$ & ($\log_{10}$) &(km\,s$^{-1}$)\\
\midrule
          R1$^\mathrm{a}$ & $12\pm4$ & $1.0^{+0.2}_{-0.1}$ & $0.609\pm0.001$ & $6.51\pm0.03$ & $1.60\pm0.08$ & $440^{+30}_{-20}$ \\
          R2$^\mathrm{b}$ & $15\pm6$ & $0.9^{+0.2}_{-0.1}$ &   $0.594\pm0.004$ & $7.1\pm0.2$ & $1.4\pm0.2$ & $400^{+30}_{-20}$ \\
          R3 & $20\pm2$ & $0.84^{+0.03}_{-0.03}$ &  $0.749\pm0.006$ & $3.52\pm0.08$ & $1.5\pm0.1$ & $410^{+30}_{-20}$ \\
          R4$^\mathrm{c}$ & -- & -- &  $0.536\pm0.004$ & $10.1\pm0.3$ & $1.9\pm0.3$ & $450$ \\
          R5$^\mathrm{d}$ & -- & -- &  $0.44\pm0.04$ & 100 & $1.3\pm0.1$ & $390^{+20}_{-10}$ \\
          R6 & -- & -- &  $0.82\pm0.04$ & $2.7\pm0.4$ & -- & -- \\
          R7 & -- & --  &  --  &  --  & -- & -- \\
          R8 & -- & -- &  $0.638\pm0.007$ & $5.6\pm0.2$ & -- & -- \\
          R9 & $9.0\pm0.4$ & $1.19^{+0.03}_{-0.03}$ &  $0.812\pm0.002$ & $2.78\pm0.02$ & -- & -- \\
          R10$^\mathrm{e}$ & -- & --  &  --  &  --  & -- & -- \\
\bottomrule
\end{tabular}
    \label{tab:discussion_observed_ratio-density}
    \begin{tablenotes}
        \footnotesize
        \item Note: a) The upper uncertainty for the blast-wave shock velocity ($v_\mathrm{s}$) is estimated, see text. \\
        b) In order to compute the \mbox{[\ion{S}{III}]-ratio}, the \mbox{[\ion{S}{III}]\,$\lambda6312$} line is deblended from the strong [\ion{O}{I}]\,$\lambda6300$ line, see text. \\
        c) The measured Fe-ratio exceeds the upper limit $\sim1.65$ reported in \citet{Vogt-2017} (but is within 1$\sigma$) and so only a lower bound to the derived velocity is reported. \\
        d) The measured \mbox{[\ion{S}{II}]-ratio} is at the lower limit $\sim0.44$ \citep{Osterbrock-2006} and so only a lower bound to the derived density is reported. \\
        e) The measured \mbox{[\ion{S}{II}]-ratio} exceeds the upper limit $\sim1.44$
        \citep{Osterbrock-2006} by more than 3$\sigma$ and is omitted.
    \end{tablenotes}
    \end{threeparttable}
\end{table*}

The ratio \mbox{[\ion{S}{II}]\,$\lambda$6716/$\lambda$6731} serves as a diagnostic for estimating post-shock electron densities, $n_e$ \citep{Osterbrock-2006}. Motivated by the results from the previous section, we adopt a uniform gas temperature of $T=10^4$\,K for all regions R1--R10. Using this assumption, we derive $n_e$ estimates for these regions (excluding R7, where no significantly broadened \mbox{[\ion{S}{II}]} lines, which we associate to shocked emission, are detected). The measured \mbox{[\ion{S}{II}]-ratios} for the entire remnant are mapped in Figure\,\ref{fig:results-s-ii-ratio}, and ratios with uncertainties for regions \mbox{R1--R10} with the corresponding $n_e$ are listed in Table\,\ref{tab:discussion_observed_ratio-density}. The general trend observed in Figure\,\ref{fig:results-s-ii-ratio}, lower \mbox{[\ion{S}{II}]}-doublet ratios indicating higher values of $n_e$ in the NW and W, and higher ratios indicating lower values of $n_e$ in the E and N, is confirmed in Table\,\ref{tab:discussion_observed_ratio-density}. The highest measured post-shock electron density is found in R5 (\mbox{$n_e\sim100\times10^3$\,cm$^{-3}$}), where the \mbox{[\ion{S}{II}]\,$\lambda$6716/$\lambda$6731} ratio is equal to the theoretical lower limit and thus provides only a lower bound for the highest density. The highest post-shock density measurements that are within the theoretical \mbox{[\ion{S}{II}]\,$\lambda$6716/$\lambda$6731} ratio range are found in regions R1, R2 (\mbox{$n_e\sim7\times10^3$\,cm$^{-3}$}), and R4 (\mbox{$n_e\sim10\times10^3$\,cm$^{-3}$}). Medium density values are found in R3 (\mbox{$n_e\sim4\times10^3$\,cm$^{-3}$}), and the lowest values in R6 and R9 (\mbox{$n_e\sim3\times10^3$\,cm$^{-3}$}).

Previous estimates for $n_e$ using \mbox{[\ion{S}{II}]} have been limited to the central regions of SNR\,0540 (e.g. \citealt{Sandin-2013, Lundqvist-2022}) and no direct comparison with our results can therefore be made (although \citeauthor{Lundqvist-2022} estimated $n_e$ to be of the order of \mbox{$10^3$\,cm$^{-3}$} from [\ion{O}{III}] emission for four filaments across SNR\,0540). However, density estimates for gas close to the blast wave shell in SNR\,0540 have been performed using X-rays by \citet{Hwang-2001,Park-2010}, and \citet{Brantseg-2014}, based on fits with plane parallel shock models. As discussed in Section\,\ref{subsubsec:discussion-shock-velocity}, the $n_e$ values are expected to differ between the X-ray-- and \mbox{[\ion{S}{II}]}--emitting gas, as they trace gases at different temperatures. Interestingly, however, we find that the $n_e$ estimates from both the X-rays and from the optical observations in this work show rather similar spatial trends. The X-ray studies consistently report that the highest densities reside in the bright, diagonal filament in the W (e.g. \citeauthor{Brantseg-2014} estimates\footnote{$n_e$ is derived from the emission measure of the thermal plasma and the estimated volume of the emitting region, incorporating a volume filling factor $f$, which accounts for the fraction of the volume actually occupied by the emitting gas. By definition, $f$ ranges between 0 and 1. For further details on these estimates, we refer to \citet{Brantseg-2014}.} \mbox{$n_e=7\pm1$\,cm$^{-3}\,f^{-1/2}$} for their region F), which includes our R5, where we obtain the highest post-shock density \mbox{$n_e=100\times10^3$\,cm$^{-3}$} (lower bound). However, the spectrum from R5 has a substantially higher noise level compared to other regions (Figure\,\ref{fig:appendix-results-spectra-from-r4-5-6}), which likely affects the measured \mbox{[\ion{S}{II}]\,$\lambda$6716/$\lambda$6731} ratio and, consequently, the estimated post-shock density. Our high post-shock density estimates in R4, \mbox{$n_e=\left(10.1\pm0.03\right)\times10^3$\,cm$^{-3}$}, (which roughly overlaps with region D in \citeauthor{Brantseg-2014}, who report \mbox{$n_e=4.4^{+4.4}_{-0.8}$\,cm$^{-3}\,f^{-1/2}$}) follow the X-ray trend of higher densities in the SW.

In the bright filament in the NW, \citeauthor{Brantseg-2014} estimates a slightly lower density \mbox{$n_e=4.0^{+0.6}_{-0.3}$\,cm$^{-3}\,f^{-1/2}$} for their region B (compared to their region D), which also includes our R1, where we estimate \mbox{$n_e=\left(6.51\pm0.03\right)\times10^3$\,cm$^{-3}$} (slightly lower than the corresponding value from R4). The X-ray measurements further indicate that densities decrease by a factor of \mbox{$\sim2$--3} close to our R3 (\mbox{$n_e=2.0^{+0.4}_{-0.3}$\,cm$^{-3}\,f^{-1/2}$}), a level of decrease similar to our optical result from R3, \mbox{$n_e=\left(3.52\pm0.08\right)\times10^3$\,cm$^{-3}$}. No corresponding X-ray investigation exists for R2, the region closest to the PWN (in the line-of-sight perspective), due to significant photon pileup. Similarly, R8, where we observe higher densities of \mbox{$\left(5.6\pm0.2\right)\times10^3$\,cm$^{-3}$}, does not have a corresponding X-ray measurement, likely due to the lower X-ray fluxes near this region.

We observe the highest \mbox{[\ion{S}{II}]}-doublet ratio values -- and consequently the lowest electron densities of \mbox{$\sim3\times10^3$\,cm$^{-3}$} -- for both R6 and R9. While R9 is roughly consistent with the trend seen in the X-ray estimates (\citealt{Park-2010} measure the entire eastern half, which overall shows lower densities by a factor of \mbox{$\gtrsim2$} compared to the NW or SW for all their models), R6 appears to deviate from this pattern. R6 shows indications of lower densities (a factor of \mbox{$\sim2$} lower than in R1, although with large uncertainties) compared to the trend in X-rays (approximately similar with the NW). This discrepancy between the trends in the optical and X-ray density estimates for regions near R6 is possibly due to the relatively low optical \mbox{[\ion{S}{II}]} signal from this region compared to others analysed in this work (Tables~\ref{tab:appendix_fit_results_table_R1-R3}, \ref{tab:appendix_fit_results_table_R4-R5}, and~\ref{tab:appendix_fit_results_table_R6-R8-R9} in Appendix\,\ref{appendix:r5-r8}). The weak \mbox{[\ion{S}{II}]} flux makes it challenging to fully deblend the shock component from the strong unshocked ISM component, affecting the optical ratio and density estimates.

For R10, the estimated \mbox{[\ion{S}{II}]}-doublet ratio exceeds the physical upper limit, most likely due to blending of the shock-component with the strong unshocked ISM component. This prevents us from drawing further conclusions about the density of this region, apart from that it is significantly lower than in any other region studied in this work. The X-ray estimations by \citeauthor{Brantseg-2014} (\citeyear{Brantseg-2014}; \mbox{$n_e\,^\mathrm{b}=0.5^{+0.3}_{-0.1}$\,cm$^{-3}\,f^{-1/2}$}) support this interpretation, as they find densities in the N to be an order of magnitude lower than in other measured regions. A notable caveat in comparing our optically measured post-shock densities to the corresponding X-ray estimates is that the spatial regions used, even if partly overlapping, differ in spatial extent. This difference affects the derived values and introduces uncertainties in the comparison.

\subsubsection{Blast-Wave Shock Velocity}\label{subsubsec:discussion-shock-velocity}

\begin{figure}
    \includegraphics[width=\linewidth]{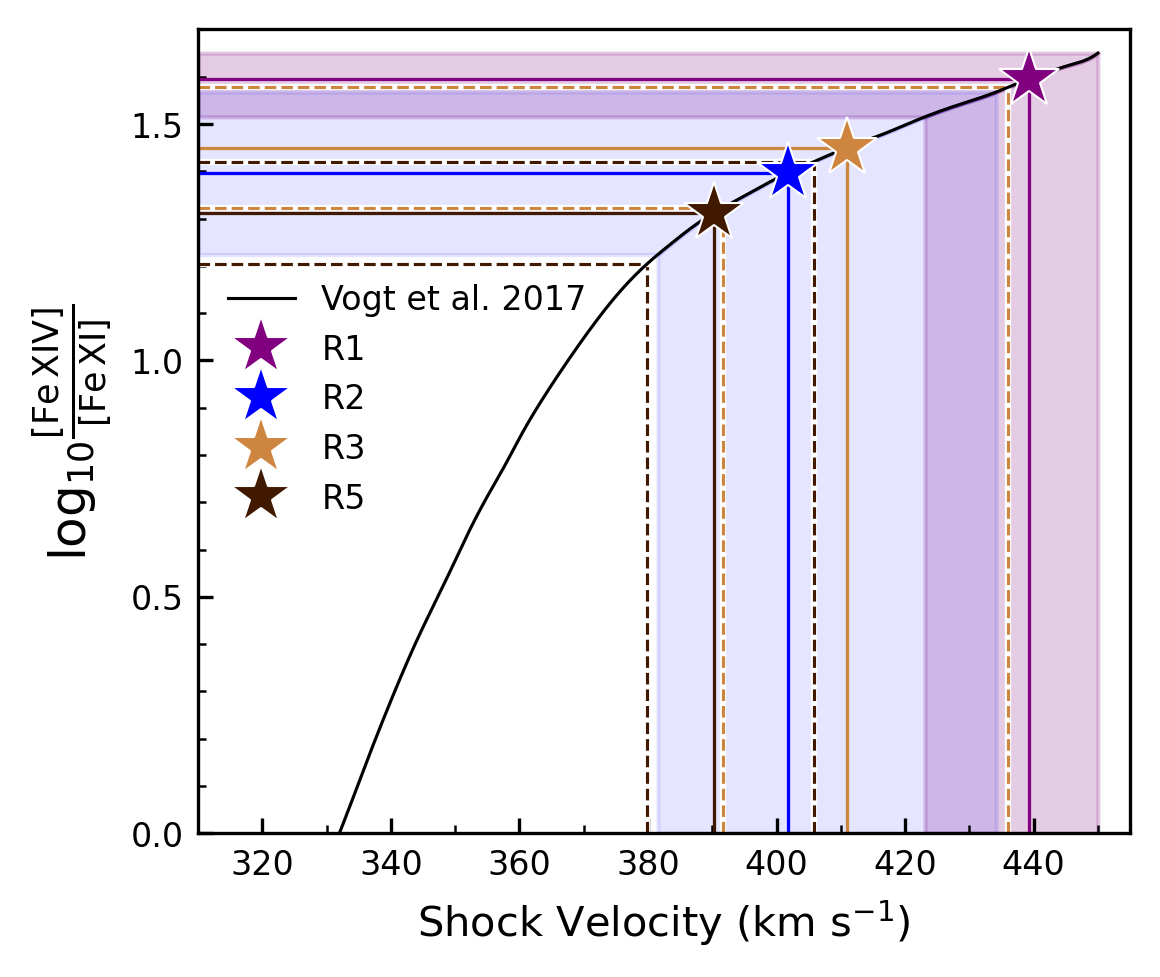}
    \caption{\mbox{[\ion{Fe}{XIV}]~$\lambda5303$/[\ion{Fe}{XI}]~$\lambda7892$} ratio as a function of blast wave wave velocity (black solid curve) from \citet{Vogt-2017}. The coronal Fe-line ratio values for regions R1 (purple star), R2 (blue star), R3 (light brown star), and R5 (dark brown star) are shown with their ${1\sigma}$ uncertainties (shaded regions). For R2 and R5, only the edges of the uncertainty range are marked with brown dashed lines for visual clarity. The upper edge of the $1\sigma$ uncertainty for R1 is limited by the available range presented by \citet{Vogt-2017} and is likely higher.}
    \label{fig:discussion-fe-ratio-shock-vel}
\end{figure}

We observe many different ionization states for Fe lines across SNR\,0540. The brightest regions \mbox{R1--R3}, in addition to regions \mbox{R4 and R5} (locations of these regions in the MUSE FOV are presented in Figure\,\ref{fig:results-optical-shock-on-chandra}), show significant emission from the coronal Fe lines \mbox{[\ion{Fe}{XIV}]~$\lambda5303$}, \mbox{[\ion{Fe}{XI}]~$\lambda7892$}, and \mbox{[\ion{Fe}{X}]~$\lambda6375$}, as well as lower-ionization lines such as \mbox{[\ion{Fe}{III}]} and \mbox{[\ion{Fe}{II}]} as shown in \mbox{Figures\,\ref{fig:results-spectra-from-regions_nw_w_ww4} and \ref{fig:appendix-results-spectra-from-r4-5-6}} and \mbox{Tables\,\ref{tab:appendix_fit_results_table_R1-R3} and \ref{tab:appendix_fit_results_table_R4-R5}}. In general, the radial velocities and FWHMs of the coronal Fe lines are higher (\mbox{$v_\mathrm{s}\lesssim|170|$\,km\,s$^{-1}$} and FWHMs \mbox{$\lesssim410$\,km\,s$^{-1}$}) than the corresponding values for \mbox{[\ion{Fe}{II}]} (\mbox{$v_\mathrm{s}\lesssim|130|$\,km\,s$^{-1}$} and FWHMs \mbox{$\lesssim220$\,km\,s$^{-1}$}), indicating that the coronal Fe lines (being broader and at higher radial velocities) come from faster shocks than the lower-ionization state Fe lines. This difference is particularly clear between \mbox{[\ion{Fe}{II}]~$\lambda5159$} and \mbox{[\ion{Fe}{XIV}]~$\lambda5303$}, whose proximity in wavelength minimize resolution-related effects. However, our results are subject to significant systematic uncertainties due to the wavelenght-dependent spectral resolution, which introduces a caveat especially to the FWHM results.

The Fe-line ratio \mbox{[\ion{Fe}{XIV}]~$\lambda5303$/[\ion{Fe}{XI}]~$\lambda7892$} is a shock-velocity diagnostic. Since both lines are detected in five regions (\mbox{R1--R5}, \mbox{Tables\,\ref{tab:appendix_fit_results_table_R1-R3} and \ref{tab:appendix_fit_results_table_R4-R5}}), we can estimate the blast-wave shock velocities in these locations. The computed ratio values are listed in Table\,\ref{tab:discussion_observed_ratio-density} with $1\sigma$ uncertainties. We then convert these Fe line ratio values into shock velocities, in which we rely on the blast-wave modeling by \citet{Vogt-2017}. This model by \citeauthor{Vogt-2017} is based on the \mbox{MAPPINGS V} code v10.3 and assumes SMC abundances. Since our analysis is limited to these two coronal forbidden Fe lines, the abundance differences between the MCs do not have a significant impact. To further characterise the surrounding diffuse ionized CSM/ISM, \citeauthor{Vogt-2017} adopt a base depletion factor of \mbox{$\log D_\mathrm{Fe} = -1.00$}, pre-shock hydrogen density of \mbox{$n_\mathrm{H} = 10$\,cm$^{-3}$}, and evolve the shock for 2000\,yr. The resulting model relating the Fe-line ratio to shock velocities is shown in Figure\,\ref{fig:discussion-fe-ratio-shock-vel}.

Now, we can estimate the blast-wave shock velocities in \mbox{R1--R5} based on their corresponding Fe-line ratio values. Table\,\ref{tab:discussion_observed_ratio-density} and Figure\,\ref{fig:discussion-fe-ratio-shock-vel} show that these shock velocities range from \mbox{$390^{+20}_{-10}$\,km\,s$^{-1}$} in R5 to \mbox{$440^{+30}_{-20}$\,km\,s$^{-1}$} in R1, while R2 and R3 are exhibiting intermediate velocities. The uncertainties of all these velocity estimates show significant overlap. In more detail, the upper edge of the $1\sigma$ uncertainty for R1 is limited by the available range presented by \citeauthor{Vogt-2017} and is likely of the same magnitude as the other two upper limits (i.e. \mbox{$\sim$ 20--30\,km\,s$^{-1}$}). Additionally, we are able to estimate a lower bound for the blast-wave shock velocity in R4 (\mbox{$450$\,km\,s$^{-1}$}), since the measured Fe line ratio exceeds the value range provided by \citeauthor{Vogt-2017}. However, the \mbox{[\ion{Fe}{XI}]~$\lambda7892$} signal is very low in this region (also exhibiting low signal to noise in general), which might affect the resulting velocity estimate.

\citet{Brantseg-2014} estimated blast-wave shock velocities based on X-ray observations, inferring the velocities from fitted post-shock plasma temperatures (e.g. \mbox{$T\sim$0.5--1\,keV} in the W). The velocity we obtain for R1 \mbox{($\sim440$\,km\,s$^{-1}$)} is consistent within $\sim2\sigma$ with the velocity \citeauthor{Brantseg-2014} estimated in the NW for a region (B) that overlaps with our R1. Similarly, the velocity of R3 \mbox{($\sim410$\,km\,s$^{-1}$)} is consistent within $\sim2\sigma$ with region C in \citet{Brantseg-2014}, though these regions do not strictly overlap. In X-rays, the shock velocity appears lower \mbox{($\sim400$\,km\,s$^{-1}$)} in the westernmost region of the forward shock. This region D in \citeauthor{Brantseg-2014} is partly outside our MUSE FOV but overlaps also with our R4, where we estimate the lower bound for the highest blast wave shock velocity. In the SW, the velocity estimate from the X-ray observations is comparable in magnitude to that \citeauthor{Brantseg-2014} obtain in the NW \mbox{($\sim500$\,km\,s$^{-1}$)}, being slightly higher what we obtain for our region R5. However, these estimates are all consistent within $\sim3\sigma$, likely reflecting the differences between the used extraction regions between the optical and X-ray estimates. Outside of our regions \mbox{R1--R5}, we do not detect significant emission from either of these coronal lines. Region A in the N, as reported by \citet{Brantseg-2014}, shows the highest velocity \mbox{($\sim700$\,km\,s$^{-1}$)} in the X-ray measurements.

As discussed in \cite{Lundqvist-2022}, the low-ionization optical lines cannot originate from the same shocked gas responsible for the X-ray and the optical coronal line emission. This can be seen from the expression for the cooling time of the shocked gas
\begin{equation}
t_{\rm cool} \approx 830~\left(\frac{n_{\rm filament}}{100~\rm{cm}^{-3}}\right)^{-1}~\left(\frac{v_{\rm s}}{300 \rm~{km~ s}^{-1}}\right)^{3.4}~{\rm yr},
\label{eq:discussion-cooling-time}
\end{equation}
which was obtained by \citet{Gro2006} for the shocked CSM of SN\,1987A in the shock speed interval \mbox{$100 \rm~{km~ s}^{-1} \lesssim V_{\rm s} \lesssim 600 \rm~{km~ s}^{-1}$}. Furthermore, in Equation\,\ref{eq:discussion-cooling-time}, $n_{\rm filament}$ is the density of a filament before it has been shocked. While Equation\,\ref{eq:discussion-cooling-time} was tailored for the He- and N-enriched CSM of SN\,1987A, it otherwise considers LMC abundances, consistent with those of SNR\,0540. In particular, \citet{Brantseg-2014} estimated blast-wave shock velocities and pre-shock densities
\mbox{$v_\mathrm{s}\sim510^{+30}_{-20}$\,km\,s$^{-1}$} and \mbox{$n_{\rm filament} \sim 1~\rm{cm}^{-3}$} for the X-ray emission from the region around R1, assuming compression by a factor of four. These conditions imply long cooling times (longer than the age of SNR\,0540), indicating that such shocks are not radiative. Our observed shocks speeds of \mbox{$\gtrsim 400 \rm~{km~ s}^{-1}$}, responsible for the coronal lines (and also the X-rays), fit well into this picture.

To reach the gas conditions (e.g. the inferred temperatures from [\ion{S}{iii}] or [\ion{O}{iii}], Figure\,\ref{fig:discussion-s-iii-ratio-shock-temp}) that can produce the lower-ionisation state emission, either slower shocks, denser filaments or a combination of both is required. The [\ion{Fe}{vii}] lines offer a way to probe these conditions, as the coronal Fe-emission has not yet had time to recombine to the ionisation state required for [\ion{Fe}{vii}] emission. In collisional equilibrium, the [\ion{Fe}{vii}] lines peak in relative abundance at \mbox{$\sim 2.5\times10^5$\,K} \citep{Lundqvist-2022}. \citeauthor{Lundqvist-2022} further report that this temperature corresponds to a shock velocity of \mbox{$\sim 100\,\rm{km~ s}^{-1}$}. Requiring the cooling times of these shocks to be short, \mbox{$\sim10$\%} of the age of SNR\,0540 i.e \mbox{120\,yr}, we can use Equation\,\ref{eq:discussion-cooling-time} to obtain an estimate for the pre-shock density: \mbox{$n_\mathrm{filament,O} = 50~\rm{cm}^{-3}$} (assuming a shock velocity of \mbox{$140\,\rm{km~ s}^{-1}$}). The passing shocks would then compress these densities to a post-shock density of \mbox{$\sim 5\times 10^3$\,cm$^{-3}$} (combined compression factor of $\sim100$) and the filament would cool to temperatures of \mbox{$\sim10^4$\,K}. These post-shock conditions appear to be consistent with the observed post-shock densities and temperatures we inferred from [\ion{S}{ii}] and [\ion{S}{iii}] emission (Table~\ref{tab:discussion_observed_ratio-density}).

Assuming dynamical pressure equilibrium i.e. that the kinetic energy per unit volume transferred to the filaments by the shocks \mbox{($n_\mathrm{filament}v_\mathrm{s}^2$)} is constant, we can estimate either shock velocities or pre-shock densities. Using the radiative shock conditions derived above for the optical low-ionisation emission \mbox{($n_\mathrm{filament,O}=50$\,cm$^{-1}$ and $v_\mathrm{s,O}=140$\,km\,s$^{-1}$)} and the pre-shock density from X-ray measurements (\mbox{$n_\mathrm{filament,X}=1$\,cm$^{-1}$}, \citealt{Brantseg-2014}), we obtain a corresponding shock velocity \mbox{$v_\mathrm{s,X}=10^3$\,km\,s$^{-1}$} responsible for the X-ray emission. This is somewhat higher, by a factor of \mbox{$\sim1.5$--2}, than the shock velocities reported by \citeauthor{Brantseg-2014}, and likely reflects the presence of a range of filament densities spanning $1-100\,\rm{cm}^{-3}$ or non-equilibrium conditions. Similarly, we can estimate the pre-shock filament density for the gas responsible for the coronal Fe emission by using the shock velocity estimate of \mbox{$\sim420$\,km\,s$^{-1}$}, which yields a pre-shock density of \mbox{$n_\mathrm{filament,Fe}\sim6$\,cm$^{-1}$} and a cooling time \mbox{$\sim4\times10^4$\,yr}.

\subsection{Origin of the Unidentified Emission Lines in R7}\label{subsec:discussion-R4}

\begin{figure*}
    \includegraphics[width=\linewidth]{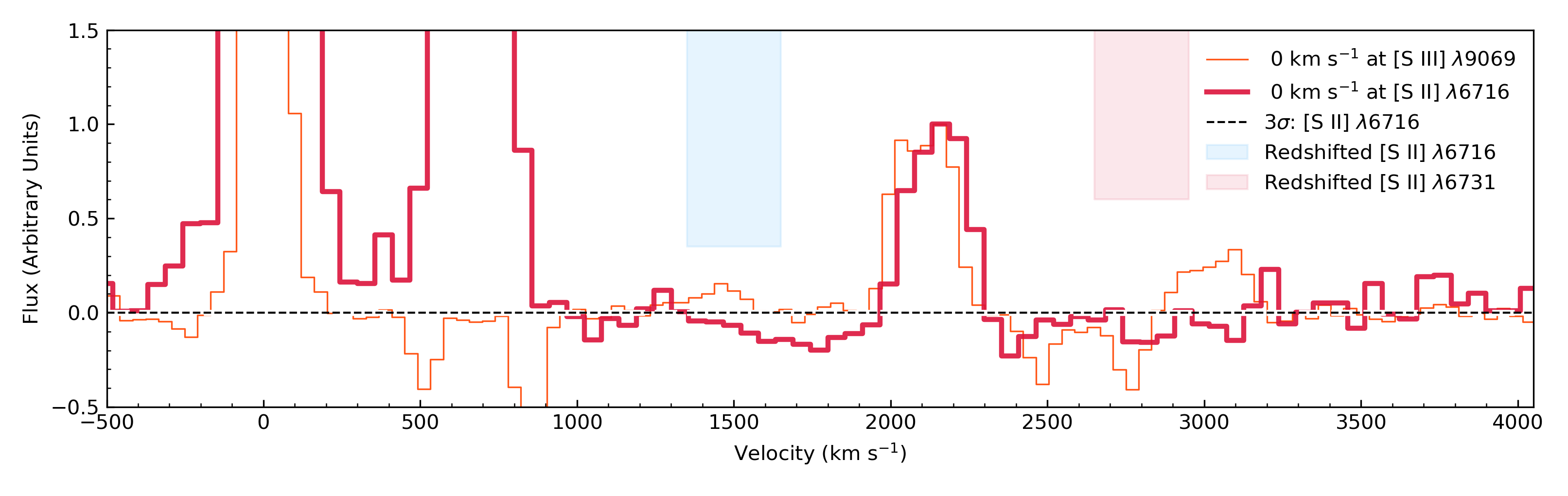}
    \caption{MUSE: Spectra from R7 with rest wavelengths at \mbox{[\ion{S}{II}]\,$\lambda$6716} (red) and \mbox{[\ion{S}{III}]\,$\lambda$9069} (orange). The features \mbox{uID-6765} and \mbox{uID-9132} are located at \mbox{$\sim$2000\,km\,s$^{-1}$} and \mbox{uID-9160} at \mbox{$\sim$3000\,km\,s$^{-1}$}. The fluxes are normalised with respect to the peaks of \mbox{uID-6765} and \mbox{uID-9132}. The shaded areas around \mbox{1500\,km\,s$^{-1}$} and \mbox{2750\,km\,s$^{-1}$} indicate the expected location and minimum peak hight of the doublet line, were \mbox{uID-6765} to be associated with the \mbox{[\ion{S}{II}]\,$\lambda\lambda$6716,6731} doublet.}
    \label{fig:discussion-if-mystery-lines-sulphur}
\end{figure*}

R7, located south of the PWN, differs from the other selected regions as it does not exhibit significant shock emission (see the location of R7 in Figure\,\ref{fig:results-optical-shock-on-chandra}, and the narrow emission lines in the R7 spectrum in Figure\,\ref{fig:results-spectra-from-regions_nw_w_ww4}). Instead, the spectrum from R7 shows unidentified emission lines \mbox{uID-5187}, {uID-6765}, {uID-9132}, and {uID-9160}. All of these features (excluding the blended \mbox{uID-5187}) have similar widths and are bright enough to rise above the minimum signal threshold, leading us to conclude that they represent genuine emission features rather than noise. These unidentified emission lines are also inconsistent with any typical emission lines originating from the CSM/ISM, prompting further investigation into alternative origins. The rest of this section explores three possible scenarios:

\begin{enumerate}[label=\textup{(\roman*)}, leftmargin=*, align=right, widest=iii]
    \item Cas A-like S-rich high-velocity ejecta clump,
    \item ejecta clump emitting in lines other than S,
    \item background source.
\end{enumerate}

As three of these unidentified emission lines, {uID-6765}, {uID-9132}, and {uID-9160}, are located near S-lines, it is natural to consider them originating from a high-velocity ejecta clump emitting in S (option i). These kinds of high-velocity ejecta clumps are abundant in Cas A (e.g. \citealt{Fesen-2016}). Figure\,\ref{fig:discussion-if-mystery-lines-sulphur} illustrates this scenario, where {uID-6765} could correspond to the \mbox{[\ion{S}{II}]\,$\lambda$6716} peak in the \mbox{[\ion{S}{II}]-doublet}, while {uID-9132} and {uID-9160} could be associated with \mbox{[\ion{S}{III}]\,$\lambda$9069}. In this interpretation, {uID-6765}, and {uID-9132} would overlap at \mbox{$\sim+2000$\,km\,s$^{-1}$}, while {uID-9160} would be located at \mbox{$\sim+3000$\,km\,s$^{-1}$}. 

If {uID-6765} is redshifted \mbox{[\ion{S}{II}]}-emission, we should also observe a clear counterpart for the other peak in the \mbox{[\ion{S}{II}]}-doublet. The shaded areas in Figure\,\ref{fig:discussion-if-mystery-lines-sulphur} mark the possible spectral ranges for this counterpart, depending on which peak in the \mbox{[\ion{S}{II}]}-doublet {uID-6765} belongs. The minimum peak hight of the counterpart is constrained by the \mbox{[\ion{S}{II}]\,$\lambda$6716 / [\ion{S}{II}]\,$\lambda$6731} ratio \citep{Osterbrock-2006}, as indicated in the figure. Notably, we detect no counterpart within these regions, not even above the minimum signal level. This suggests that {uID-6765} is unlikely to be associated with the {\mbox[\ion{S}{II}]\,$\lambda\lambda$6716,6731} doublet. It would also be unusual to find two \mbox{[\ion{S}{III}]\,$\lambda$9069}-rich ejecta clumps with such similar properties (Figures~\ref{fig:results-vel-spectra-unknown}~and~\ref{fig:results-vel-slices-unknown}), except that one is at \mbox{$\sim2000$} and the other at \mbox{$3000$\,km\,s$^{-1}$} in radial velocity.

In Cas A, \citet{Fesen-2016} find that the high velocity ejecta clumps are located out ahead of the forward shock and main shell ejecta, and exhibit transverse velocities of order \mbox{$10^4$\,km\,s$^{-1}$}. The radial velocities of these Cas A ejecta clumps are of the order of \mbox{$10^2$\,km\,s$^{-1}$} and the emission lines are generally narrower than other ejecta lines. In the case of SNR\,0540, the unidentified emission lines are also quite narrow (\mbox{$\sim200$\,km\,s$^{-1}$}, Figure\,\ref{fig:results-vel-spectra-unknown}), differing from the other (ejecta) lines such as H and O (Figure\,\ref{fig:results-spectra-from-regions_nw_w_ww4}) similar to Cas A. However, R7 is not located beyond the forward-shock shell, but rather near the H-blob (Figure\,\ref{fig:results-s-ii-ratio}), which is associated with ejecta \citepalias{L21}. Assuming free expansion, R7 has a transverse velocity of \mbox{~$2000$--$3000$\,km\,s$^{-1}$}, which would be comparable to the radial velocities of the unidentified lines if they originate from S. However, the absence of the second peak in the \mbox{[\ion{S}{II}]}-doublet, the two comparable \mbox{[\ion{S}{III}]\,$\lambda$9069} emitting ejecta clumps with different radial velocities, the proximity of R7 to the ejecta rather than beyond the forward shock, and the similarity between the unidentified lines' transverse and radial velocities -- despite the narrow line profiles -- suggest that these unidentified lines cannot plausibly originate from a S-rich high-velocity ejecta clump.

A second possible scenario (option ii) is that the unidentified lines originate from an ejecta clump emitting in lines other than S, such as Fe. The Fe-origin is motivated by the enhanced Fe-emission suggested in the southern part of SNR\,0540 by \citet{Park-2010}. Unfortunately, the most common Fe lines do not provide a good match within reasonable transverse velocities (up to \mbox{$\sim5000$\,km\,s$^{-1}$}). However, there is a vast number of weaker, less commonly observed Fe lines that we have not systematically checked and that could possibly produce such emission lines as {uID-6765}, {uID-9132}, and {uID-9160}. If any of these lesser-known Fe lines correspond to the unidentified emission in R7, it would suggest that this region contains ejecta synthesised in the deepest layers of the progenitor's core. This kind of an ejecta knot would then have been expelled out towards the SNR boundary, similar to what has been observed in Cas A \citep{Hughes-2000,Hwang-2003}. Confirming the Fe overabundance result from \citet{Park-2010} in the southern part of the remnant and the Fe origin for the unidentified lines in R7 would provide valuable constraints on the explosion and nucleosynthesis. It would also provide evidence for the location of the reverse shock in the southern parts of SNR\,0540, since such an Fe-rich ejecta knot would most likely be energised by interaction with the reverse shock.

However, there are two key issues with this interpretation. First, the region where \citet{Park-2010} detected potential Fe enhancement does not coincide with R7, but is instead located further out, with transverse velocities ranging from \mbox{$\sim4000$--6000\,km\,s$^{-1}$}. Second, since the strongest and most commonly observed Fe lines do not align with the wavelengths of the unidentified lines, it is unclear why weaker and more rarely observed Fe lines appear in SNR\,0540 while other, more prominent Fe lines remain undetected. 

Finally, in addition to a possible Fe-origin, we have also considered other ejecta lines that might explain the unidentified emission lines. For instance, {uID-5187} could potentially correspond to \mbox{[\ion{Ar}{III}]\,$\lambda$5192}, since we observe \mbox{[\ion{Ar}{III}]\,$\lambda$7136}. Given the similar line profiles of the unidentified emission lines, it is likely that they are all produced by the same emitting species (if they have an ejecta-origin). However, none of the typical emission line species provide a convincing match to all (or even the majority of) the unidentified lines.

It is also possible that the unidentified lines are produced by a background source with a high redshift (option iii). A spiral galaxy would be a plausible candidate, as it could account for both the spatial movement relative to the radial velocity of the emission lines (Figure\,\ref{fig:results-vel-slices-unknown}) and the observed \mbox{$\sim200$\,km\,s$^{-1}$} line widths (Figure\,\ref{fig:results-vel-spectra-unknown}, and for typical spiral galaxy optical spectra we refer to e.g. \citealt{Charlot-2001,Sanches-2012}). Assuming a redshift of $z=0.36$, the \mbox{[\ion{S}{II}]} doublet would be shifted to approximately the wavelengths of {uID-9132}, and {uID-9160}. H$\alpha$ would fall at \mbox{$\sim8925$\,Å}, a region affected by noise, potentially making it difficult to detect. At this redshift, 1\,arcsec corresponds to $\sim5$\,kpc, implying that this potential background object spans \mbox{$\sim10$\,kpc} in diameter based on its angular extent in Figure\,\ref{fig:results-vel-slices-unknown}, placing it within the size range of a dwarf galaxy. This interpretation is supported by the expected rotation velocities of such systems, $\sim100$\,km\,s$^{-1}$ \citep{Navarro-1996}, which would produce the integrated line widths of $\sim200$\,km\,s$^{-1}$, comparable to our observed values of the unidentified lines. However, a key issue with this scenario is the absence of H$\beta$ at $\sim6610$\,Å, despite the good signal-to-noise in this region of the spectrum. Furthermore, the apparent \mbox{[\ion{S}{II}]\,$\lambda\lambda$6716,6731} doublet ratio in R7 is $\sim4$ (Figure\,\ref{fig:results-vel-spectra-unknown}), significantly higher than a ratio of \mbox{$\lesssim1.4$} observed in typical gas temperatures and densities of star-forming regions \citep{Osterbrock-2006}. Combined with the lack of H-lines at a consistent redshift, these factors make a background galaxy origin unlikely.

\subsection{Overall Picture of Shock Emission in SNR\,0540}\label{subsec:discussion-overall_picture}

SNR\,0540, believed to result from a red supergiant that exploded as Type II SN \citep{Chevalier-2006}, has likely been evolving in a low-density cavity blown out by the progenitor's stellar wind \citep{Brantseg-2014}. According to \citeauthor{Brantseg-2014}, the blast wave has reached the clumpy, dense material at the western edge of this cavity, yielding bright emission in both X-rays and optical. In this scenario, the blast wave in the E, where emission is significantly dimmer, has likely not yet encountered comparable ambient densities. Overall, the material surrounding SNR\,0540 appears highly inhomogeneous. In the following, we relate our results to this broader picture of the remnant's interaction with its surroundings.

As reported in Section\,\ref{sec:results}, we detect clumpy optical shock emission (Figures\,\ref{fig:results-fit-results-shock} and \ref{fig:appendix-gaussian-fit-results}) within SNR\,0540, which we attribute to the blast wave encountering density enhancements in the surrounding medium. Such filaments of shocked ISM(/CSM) in SNR\,0540 have previously been observed in the optical by \citet{Lundqvist-2022}, who identified four such shocked ISM knots within SNR\,0540. Additional evidence for an inhomogeneous and clumpy ISM interacting with the passing blast wave comes from \citet{Brantseg-2014}, where variations in the radio rotation measure were observed in the western part of the blast-wave shell. This finding suggests that, as expected in a clumpy medium, the electron density and magnetic field direction fluctuate on relatively small spatial scales.

We also find that the blast-wave shock velocities in SNR\,0540 are low for such a young remnant, on the order of \mbox{$10^2$\,km\,s$^{-1}$} (Table\,\ref{tab:discussion_observed_ratio-density} and Figure\,\ref{fig:discussion-fe-ratio-shock-vel}). For comparison, the forward shock velocities in Cas\,A are an order of magnitude higher, \mbox{$10^3$\,km\,s$^{-1}$} \citep{Vink-2022}. For SNR\,0540, similar low shock velocities of the same magnitude \mbox{($10^2$\,km\,s$^{-1}$)} have also been reported by \citealt{Brantseg-2014} and \citet{Lundqvist-2022}, as well as for SNR\,1E\,0102.2-7219 in the SMC by \citet{Vogt-2017}. The presence of such slow shock velocities in young SNRs provides further evidence that the blast wave is interacting with dense ISM material. This conclusion is also supported by our density estimates (Table\,\ref{tab:discussion_observed_ratio-density}). 

Additionally, the X-ray observations (e.g. \citealt{Park-2010,Brantseg-2014}) reveal synchrotron emission from the hard arcs, seen in the SW and NE of the remnant (middle panel of Figure\,\ref{fig:intro} and the right panel of Figure\,\ref{fig:results-optical-shock-on-chandra}), requiring shock speeds of at least an order of magnitude greater than those we estimate from the optical coronal Fe emission (Table\,\ref{tab:discussion_observed_ratio-density} and Figure\,\ref{fig:discussion-fe-ratio-shock-vel}). Interestingly, the location of the synchrotron emission in the SW overlaps with regions where we observe slower, shocked [\ion{Fe}{XIV}]-emitting filaments (right panel of Figure\,\ref{fig:results-optical-shock-on-chandra}). This suggests that these shocks responsible for the slower coronal Fe emission have decelerated to their current velocities relatively recently, during the past few hundred years, as also noted by \citet{Brantseg-2014}.

SNR\,0540 shows pronounced spatial variation in brightness across all wavelengths. Bright emission from the NW to the SW is evident in X-rays (e.g. \citealt{Hwang-2001,Park-2010}), as well as in our optical results from \mbox{[\ion{Fe}{XIV}]\,$\lambda5303$} and \mbox{[\ion{S}{II}]\,$\lambda6731$}, which trace the bright X-ray blast wave in Figure\,\ref{fig:results-optical-shock-on-chandra}. Similar spatial variations are observed in the radio \citep{Manchester-1993,Brantseg-2014}. In contrast, the emission from the N, to E and to the S, is considerably fainter in X-rays (excluding the hard arc in the NE), but both the radio and optical observations reveal a brighter blob in the SE. This feature is best seen in \mbox{[\ion{S}{II}]} (top left panel of Figure\,\ref{fig:results-fit-results-shock}, overlapping with R9) but emits also in other lines as seen in Figure\,\ref{fig:appendix-results-spectra-from-r8-9-10} in the Appendix~\ref{appendix:r5-r8} (as well as in Figures\,\ref{fig:appendix-results-vel-spectra-bright-emission-lines} and \ref{fig:appendix-results-vel-spectra-coronal-fe-lines} and Table\,\ref{tab:appendix_fit_results_table_R6-R8-R9}). Notably, no significant X-ray emission is detected in this region. This suggests that the local temperatures in this region may be too low to produce detectable X-ray emission, yet still sufficient to generate bright optical and radio emission.

R7, the region in the S, remains enigmatic as discussed above (Section\,\ref{subsec:discussion-R4}). The unidentified lines found in this work, and the Fe-enhancement observed by \citet{Park-2010} (although not strictly overlapping in location) do not rule out the possibility that these features are Fe lines. If these unidentified emission lines from R7 can be associated with Fe, though unlikely, it would suggest that R7 is an Fe-rich ejecta clump heated by the reverse shock. This would then indicate that (at least part of) the reverse shock is much closer to the PWN than previously thought. \citet{Brantseg-2014} found possible evidence for the reverse shock in SNR~0540, particularly in the far W of the remnant (their region C, west from our R3). However, this region C is located much further out (with transverse velocities ranging \mbox{$\sim5000$--6000\,km\,s$^{-1}$}) compared to the transverse velocity of R7 (only \mbox{$\sim2000$--3000\,km\,s$^{-1}$}).

When it comes to the optical emission from the regions corresponding to C in \citeauthor{Brantseg-2014}, we cannot confirm any evidence for a reverse shock in this location. Most likely, the reverse shock, if existing, is still too close to the forward shock for us to resolve it with our current instruments. This scenario, also reported in e.g \citet{Hwang-2001}, would additionally support the interpretation that the remnant has encountered dense ambient material only relatively recently.

In this work, we have added SNR\,0540 to the population of SNRs whose optical shock emission has been investigated with integral-field spectroscopic data. \citet{Vogt-2017} used MUSE observations to investigate coronal Fe emission in SNR\,1E\,0102.2-7219 in the SMC. They discovered a thin, partial ring of \mbox{[\ion{Fe}{XIV}]\,$\lambda5303$} and \mbox{[\ion{Fe}{XI}]\,$\lambda7892$} encircling the fast-moving O-rich ejecta. \citeauthor{Vogt-2017} interpreted the brightest parts of this ring as shocks driven into dense ISM, and the fainter sections to trace the forward shock that is propagating into a lower-density gas. In SNR\,0540, unlike SNR~1E~0102.2-7219, no complete forward-shock ring is seen in the optical for SNR\,0540. Parts of this ring in the NE and S are faintly visible in soft X-rays (left panel of Figure\,\ref{fig:results-optical-shock-on-chandra}), but these regions are likely too faint for detection in the optical due to stronger background contamination. Optical coronal Fe emission has also been observed in the LMC SNRs N49 \citep{Dopita-2016} and N132D \citep{Dopita-2018}. Both of these works report a detection of \mbox{[\ion{Fe}{XIV}]\,$\lambda5303$} and \mbox{[\ion{Fe}{X}]\,$\lambda6374$} emission. This emission is interpreted to arise from shocks propagating with velocities of \mbox{$\sim350$\,km\,s$^{-1}$}, similar to the conditions described by \citet{Vogt-2017} and the shocks responsible for the coronal iron emission detected in this work.

\section{Summary and Conclusions} \label{sec:conclusion}

We have investigated the optical shock emission from SNR\,0540 using MUSE at the VLT. MUSE provides optical integral-spectroscopic data covering the spectral range \mbox{4650--9300 Å} with a \mbox{1 $\times$ 1 arcmin$^2$} FOV that covers nearly the entire remnant. After subtracting the background and continuum emission, we analysed the spatial and spectral properties of emission lines associated with shock-interacting regions. The main conclusions are summarised below. 

\begin{enumerate}[label=\textup{\arabic*.}, leftmargin=*, align=right, widest=4]
    \item We identify clumpy optical shocked CSM/ISM emission across the remnant. The brightest shock emission lines being the \mbox{[\ion{S}{II}]\,$\lambda\lambda6716,6731$} doublet (radial velocities \mbox{$v_\mathrm{r}\lesssim|130|$\,km\,s$^{-1}$} and FWHMs \mbox{$\lesssim220$\,km\,s$^{-1}$}) and the coronal line \mbox{[\ion{Fe}{XIV}]\,$\lambda5303$} (\mbox{$v_\mathrm{r}\lesssim|170|$\,km\,s$^{-1}$} and FWHMs \mbox{$\lesssim410$\,km\,s$^{-1}$}). These optical shock-emission lines appear to trace the blast-wave shell previously observed in X-rays and support the scenario that the blast wave is interacting with the surrounding CSM/ISM.
    \item We estimate post-shock electron densities by measuring the ratio \mbox{[\ion{S}{II}]\,$\lambda6716$/$\lambda6731$}. The densities have spatial variation, the highest densities (\mbox{$\sim10\times10^3$\,cm$^{-3}$}) residing in the bright knots in the NW and W and lowest in the SE (\mbox{$\sim3\times10^3$\,cm$^{-3}$}). We also find a lower bound indicating high post-shock densities in the SW (\mbox{$\sim100\times10^3$\,cm$^{-3}$}), although this result may be affected by the lower signal-to-noise ratio of the corresponding spectrum. Additionally, the signal in the N is not high enough to draw further conclusions, apart from the fact that the density appears to be significantly lower than in other regions, as is also suggested by previous studies.
    \item In addition, we estimate blast-wave shock velocities by measuring the ratio \mbox{[\ion{Fe}{XIV}]\,$\lambda5303$/[\ion{Fe}{XI}]\,$\lambda7892$}. Consistent with previous studies in X-rays, we found low shock velocities (\mbox{$\sim400$km\,s$^{-1}$}) for such a young remnant. This result, accompanied with the high density estimates, supports the scenario that the blast-wave of SNR\,0540 is interacting with the surrounding inhomogeneous ISM/CSM, particularly in the western regions. 
    \item We also estimate post-shock electron temperatures using the \mbox{[\ion{S}{III}]\,$\lambda9069$/$\lambda6312$} ratio and find values in the range \mbox{$\sim0.8$--$1.2\times10^4$\,K}. Combined with the post-shock electron densities derived from the \mbox{[\ion{S}{II}]}-doublet ratio, and that the high-ionisation state emission, such as \mbox{[\ion{Fe}{XIV}]}, has not yet had time to cool and recombine into lower ionisation states like the observed \mbox{[\ion{Fe}{VII}]}, these results support a scenario in which the observed \mbox{[\ion{Fe}{VII}]} originates from cooling gas clumps and that the filaments most likely exhibit a range of pre-shock densities spanning \mbox{1 and 100\,cm$^{-3}$}.
    \item In a region, \mbox{$\sim$2000--3000\,km\,s$^{-1}$} south (in transverse velocity) from the PSR, we detect unidentified emission lines at approximate wavelengths 5187, 6765, 9132, and 9160\,Å. We explore potential origins for these features, but they do not obviously align with any scenario. Given the previous Fe enhancement observed in the south (\citealt{Park-2010}, although not directly overlapping with this region) and the numerous possible Fe-lines in this wavelength range, we cannot entirely exclude the possibility that these unidentified lines originate from Fe. However, several issues complicate this interpretation. Other potential origins, such as a sulphur-rich Cas-A-like ejecta clump, or a background source, also face challenges. If the unidentified emission lines originate indeed from Fe, it would imply that the region in the S is an Fe-rich ejecta clump heated by the reverse shock. This would suggest that the reverse shock in SNR\,0540 would have propagated much closer to the central PWN than previously thought.
\end{enumerate}

There are several diagnostic line ratios outside our spectral coverage that could offer further insights into the temperatures, densities, and shock velocities in SNR\,0540. For instance, the line ratios \mbox{[\ion{O}{III}]\,$\lambda5007$/$\lambda4363$} and \mbox{\ion{He}{II}\,$\lambda4686$/\ion{He}{I}\,$\lambda5876$} are sensitive to shock velocity, with the former also tracing electron temperatures \citep{Dopita-2016,Dopita-2018}. In addition, other temperature-sensitive emission line ratios, such as \mbox{[\ion{O}{II}]\,$\lambda\lambda$7319--7331/$\lambda\lambda$3726,3729} and \mbox{[\ion{S}{II}]\,$\lambda\lambda$4069,4076/$\lambda\lambda$6716,6731}, would provide further information (e.g., \citealt{Groningsson-2008}). 

Future observations in the infrared with JWST could also provide valuable information on the shock dynamics and chemical composition of SNR\,0540. Observations of key regions such as the bright shock region in the NW, a region in the far W, where \citet{Brantseg-2014} found possible signs of the reverse shock, and the enigmatic region in the south, where we observe unidentified emission lines, would be particularly interesting. 

\section*{Acknowledgments}

This work was supported by the Knut \& Alice Wallenberg Foundation.
JDL acknowledges support from a UK Research and Innovation Future Leaders Fellowship (MR/T020784/1).

The following software made this work possible: ASTROPY \citep{astropy22,astropy18,astropy13}, ASTROQUERY \citep{astroquery19}, MATPLOTLIB \citep{matplotlib-2007}, PHOTUTILS \citep{photutils24}, SCIPY \citep{scipy}, STATSMODELS \citep{statsmodels-2010}, SAOImage DS9 \citep{DS9}.

\section*{Data Availability}
 
The data underlying this article are all publicly available. The MUSE data can be accessed from \url{http://archive.eso.org/cms.html} (ESO programme 0102.D-0769), and the Chandra data are available at \url{https://cxc.harvard.edu/cda/} (ObsID 5549, 7270, and 7271).

\bibliographystyle{mnras}
\bibliography{snr0540-shocks} 

\appendix

\section{Additional Spatial and Spectral Properties of Shock Emission in SNR\,0540} \label{appendix:r5-r8}

\subsection{Spatial properties of Additional Emission Lines} \label{subsec:appendix-additional-gaussian-results}

\begin{figure*}
    \includegraphics[width=\textwidth]{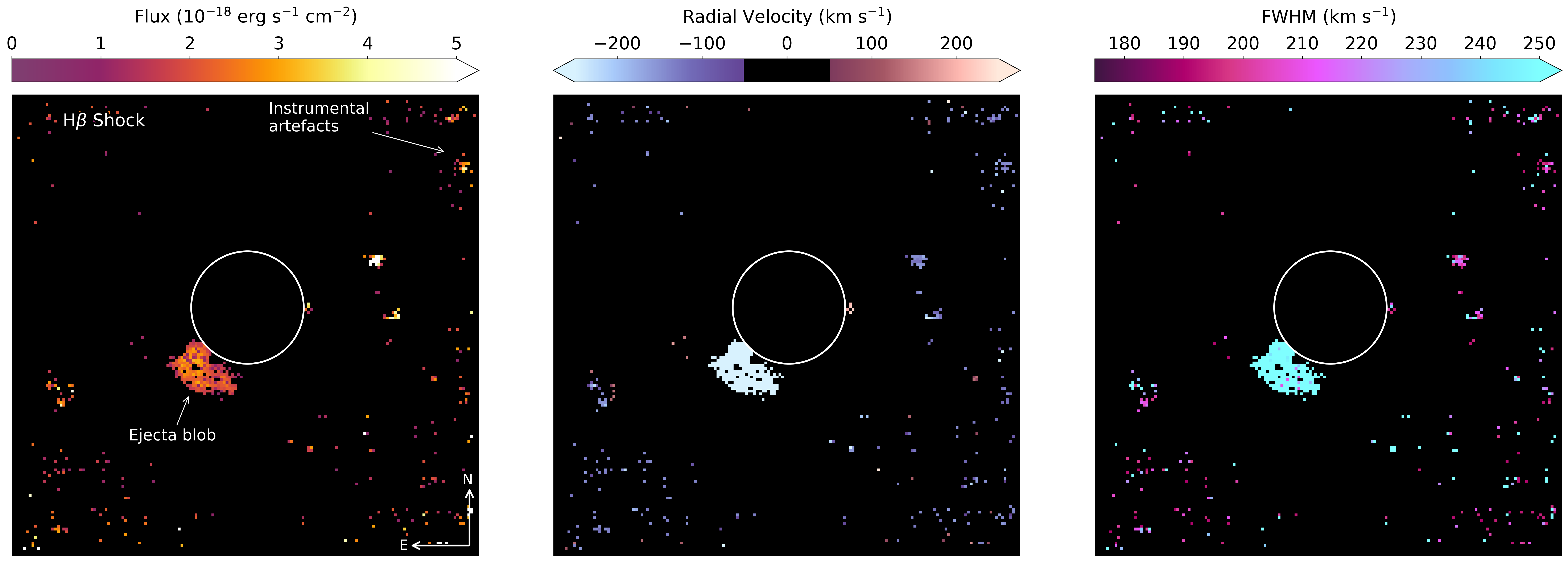}
    \includegraphics[width=\textwidth]{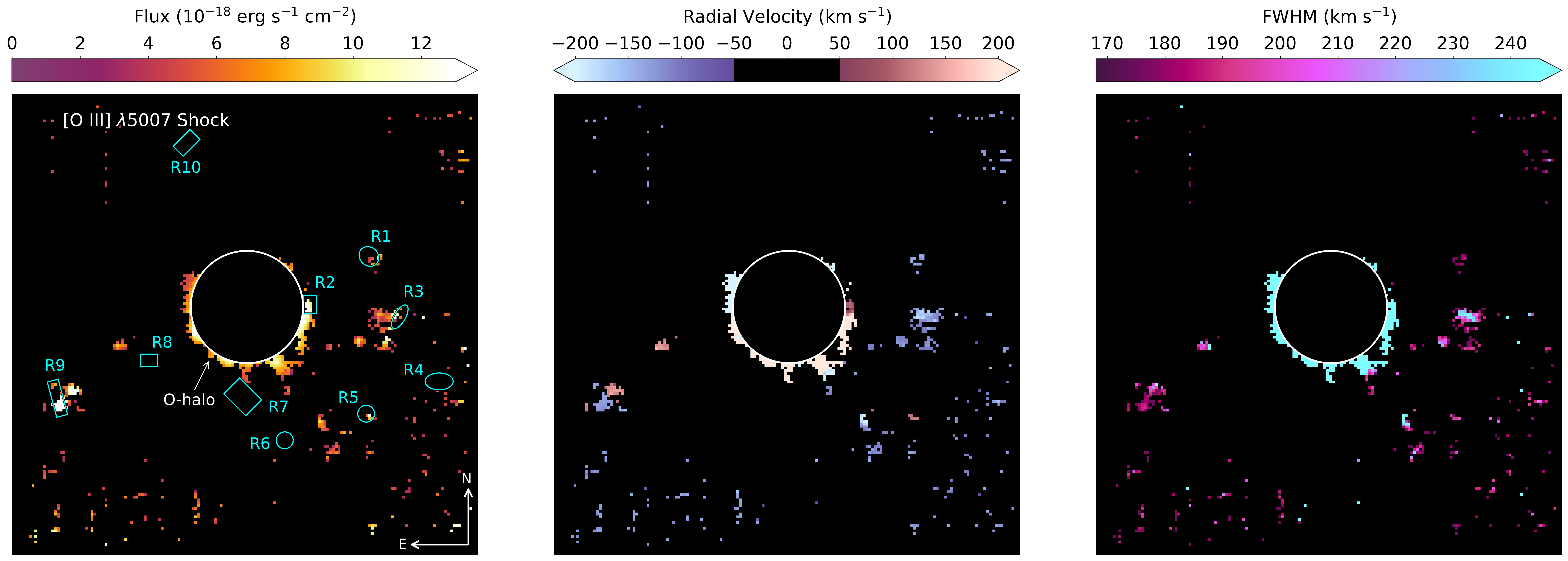}
    \includegraphics[width=\textwidth]{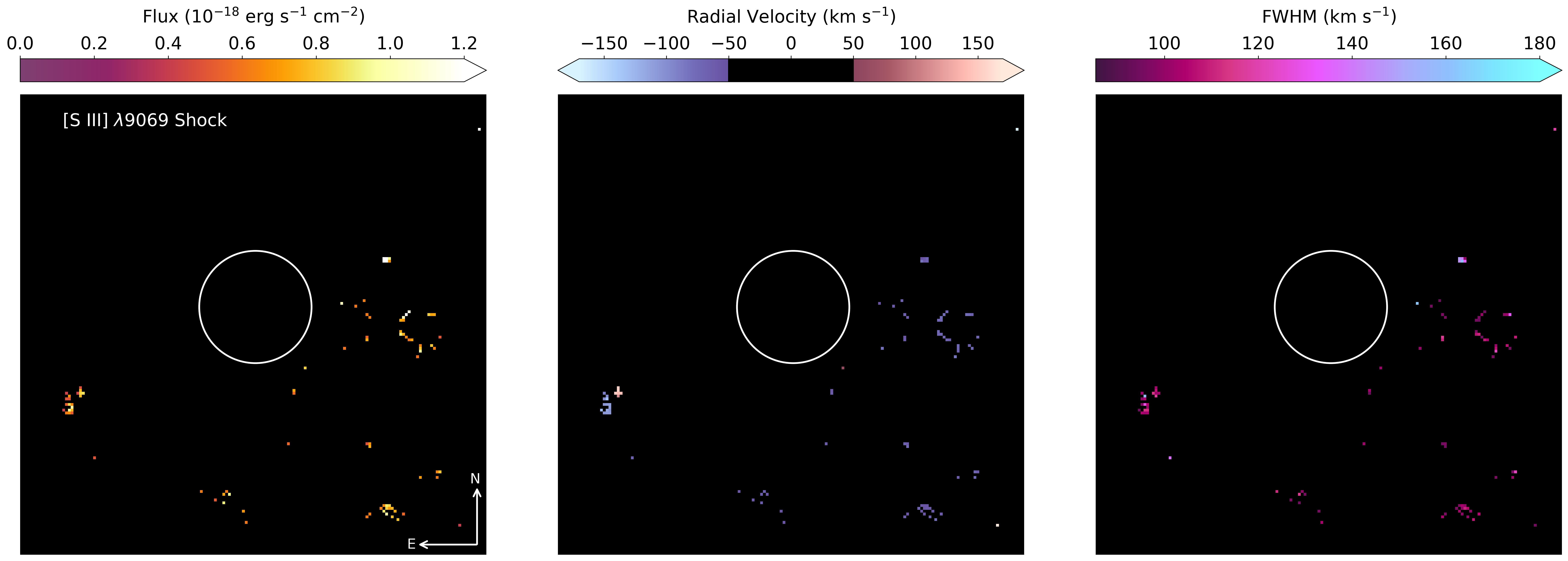}
    \caption{MUSE: Shock component fit results for additional emission lines. The PWN region within radius \mbox{8\,arcsec} (denoted with a white circle) is masked. For all panels the spatial binning is $2\times2$ compared to the original MUSE spatial sampling. The pixels with radial velocity within \mbox{$\pm50$\,km\,s$^{-1}$} are masked due to limitations set by the spectral resolution (or in case of \mbox{[\ion{S}{III}]\,$\lambda$9069} the low signal to noise). Note that the colour scales differ between the rows. \textit{Top Row:} \mbox{H$\beta$} shock component. \textit{Middle Row:} \mbox{[\ion{O}{III}]\,$\lambda$5007} shock component. A signal threshold of \mbox{$5\sigma$} is used here to ensure that the most significant shock emission is depicted, see text. The cyan ellipses and rectangles denote different extraction regions \mbox{R1--R10} described in Section\,\ref{sec:results}. \textit{Bottom Row:} \mbox{[\ion{S}{III}]\,$\lambda$9069} shock component.
    \label{fig:appendix-gaussian-fit-results}}
\end{figure*}

For completeness, we include here the shock emission analysis results for \mbox{H$\beta$}, \mbox{[\ion{O}{III}]\,$\lambda5007$}, and \mbox{[\ion{S}{III}]\,$\lambda$9069} in the top, middle, and bottom row of Figure\,\ref{fig:appendix-gaussian-fit-results}, respectively. In case of \mbox{H$\beta$} and \mbox{[\ion{O}{III}]\,$\lambda5007$}, the fitting algorithm (outlined in Section\,\ref{sec:methods}) captures the brightest shock emission in the NW, W, and SE similarly as for \mbox{[\ion{S}{II}]\,$\lambda$6731} in the top row of Figure\,\ref{fig:results-fit-results-shock}. Due to the lower spectral resolution of MUSE near \mbox{H$\beta$} and \mbox{[\ion{O}{III}]\,$\lambda5007$} compared to \mbox{[\ion{S}{II}]\,$\lambda$6731}, it is more challenging to separate the much weaker shock emission from the tail of the strong unshocked emission, which is reflected in Figure\,\ref{fig:appendix-gaussian-fit-results} as smaller and fewer shock emission knots. In the case of \mbox{[\ion{O}{III}]\,$\lambda5007$}, the emission line profiles across the entire FOV are complex due to contributions from the O-halo, strong ISM, and the presence of an \mbox{[\ion{Fe}{III}]\,$\lambda4986$} line between the doublet components. To ensure we capture at least the most prominent knots of the \mbox{[\ion{O}{III}]\,$\lambda5007$} shock emission, we increased the signal threshold in the fitting algorithm for \mbox{[\ion{O}{III}]\,$\lambda5007$} from $3\sigma$ to $5\sigma$.

We note that the dominant feature in the SE of the PWN in the \mbox{H$\beta$} results of Figure\,\ref{fig:appendix-gaussian-fit-results} is not shocked ISM but rather the H-emitting ejecta blob discovered in \citetalias{L21}. This feature is saturated in radial velocity and FWHM. In addition, some instrumental artefacts remain at the edges of the FOV. Similarly, in the \mbox{[\ion{O}{III}]\,$\lambda5007$} results, parts of the previously observed O-halo (\citealt{Morse-2006}; \citetalias{L21}) remains visible around the masked PWN. This emission component is also highly saturated in all the different panels in the middle row of Figure\,\ref{fig:appendix-gaussian-fit-results}. Notable is that the extraction region R2 seems to overlap with this halo (most clearly seen in the radial velocity panel) in line-of-sight (as discussed in relation to Figure\,\ref{fig:results-vel-spectra-bright-emission-lines}).

Even though the spectral resolution is much higher near \mbox{[\ion{S}{III}]\,$\lambda$9069} compared to all the aforementioned emission lines, the MUSE spectra exhibit low signal to noise in this spectral region (as discussed in Section\,\ref{sec:methods}). The effect of this limitation can be seen in the bottom row of Figure\,\ref{fig:appendix-gaussian-fit-results}, where even smaller knots of \mbox{[\ion{S}{III}]\,$\lambda$9069} shock emission is observed. Nevertheless, the same properties in the flux, radial velocity and FWHM are apparent compared to the aforementioned shock emission results for \mbox{H$\beta$}, \mbox{[\ion{O}{III}]\,$\lambda5007$}, and \mbox{[\ion{S}{II}]\,$\lambda$6731} in Section\,\ref{sec:results}.

\subsection{Spectral Properties of R4--R6 and R8--R10}

\begin{figure*}
    \includegraphics[width=\textwidth]{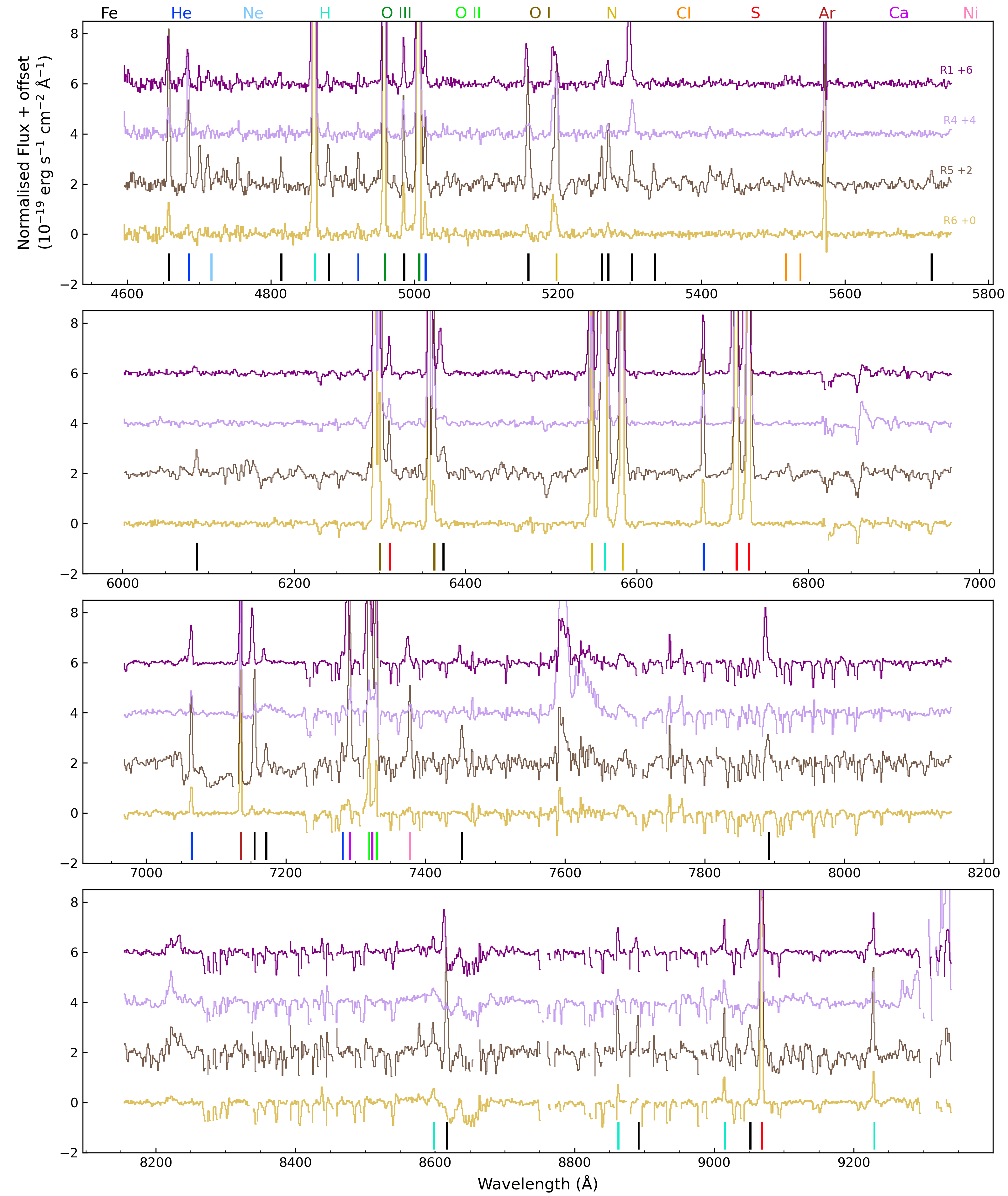}
    \caption{Background-subtracted MUSE spectra from shock-dominated regions in the R1 (purple), R4 (lavender), R5 (brown), and R6 (yellow). The fluxes are normalised with the number of pixels in the extraction aperture (locations and sizes of the apertures are shown in Figures\,\ref{fig:results-optical-shock-on-chandra} and \ref{fig:appendix-gaussian-fit-results}) and an offset is added for visual clarity. Additionally fluxes below -1 (in corresponding units) are masked for visual clarity before adding an offset. Short vertical lines indicate the exact wavelengths of the brightest emission lines, with different colors corresponding to different atomic or ionic species. The brightest emission lines are also listed in Table\,\ref{tab:appendix_observed_emission_lines}.
    \label{fig:appendix-results-spectra-from-r4-5-6}}
\end{figure*}

\begin{figure*}
    \includegraphics[width=\textwidth]{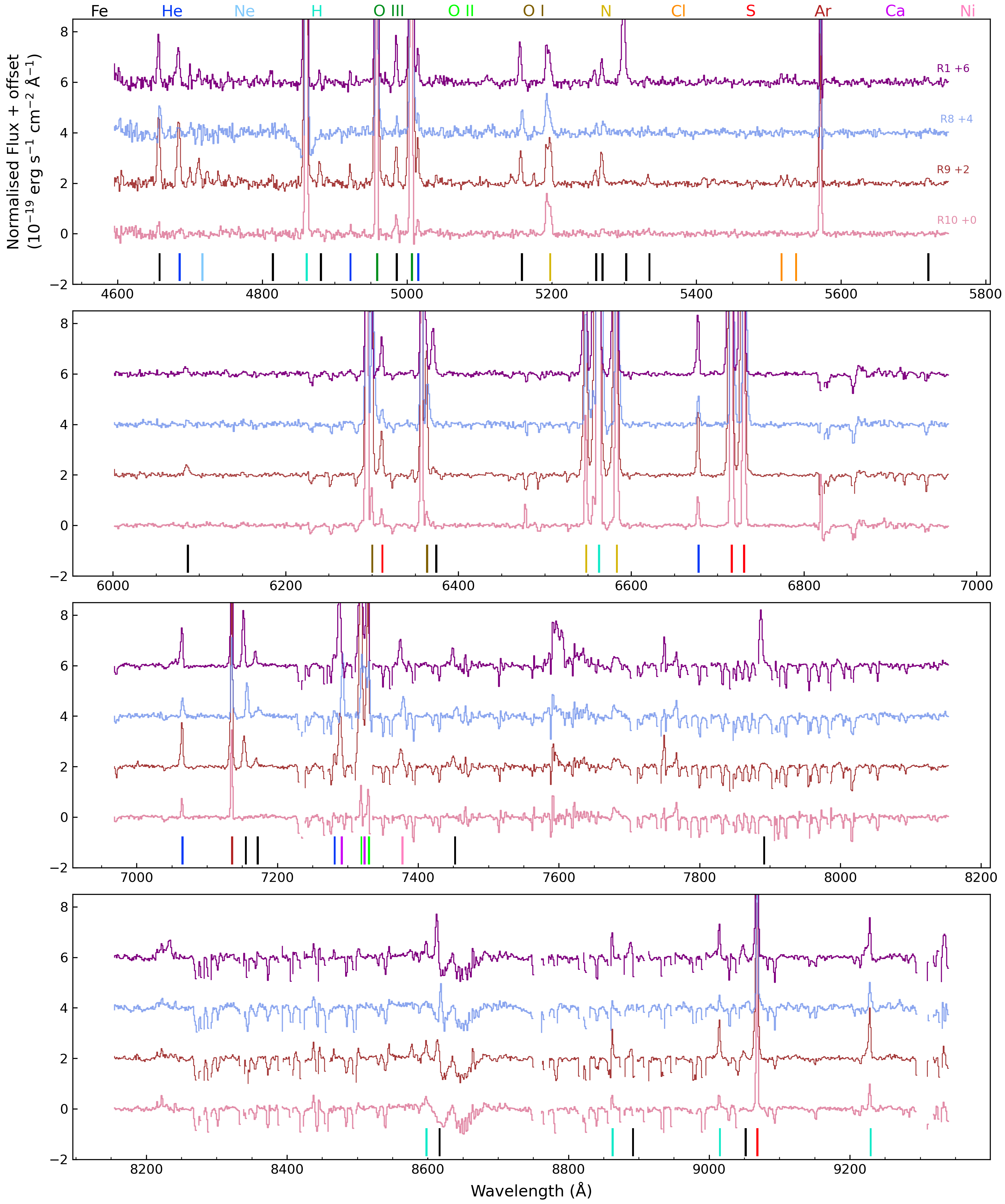}
    \caption{Background-subtracted MUSE spectra from shock-dominated regions in the R1 (purple), R8 (light blue), R9 (dark red), and R10 (pink). The fluxes are normalised with the number of pixels in the extraction aperture (locations and sizes of the apertures are shown in Figure\,\ref{fig:results-optical-shock-on-chandra} and \ref{fig:appendix-gaussian-fit-results}) and an offset is added for visual clarity. Additionally fluxes below -1 (in corresponding units) are masked for visual clarity before adding an offset. Short vertical lines indicate the exact wavelengths of the brightest emission lines, with different colors corresponding to different atomic or ionic species. The brightest emission lines are also listed in Table\,\ref{tab:appendix_observed_emission_lines}.
    \label{fig:appendix-results-spectra-from-r8-9-10}}
\end{figure*}

For completeness, we include here the rest of the regions shown in Figure\,\ref{fig:results-optical-shock-on-chandra} i.e. spectra extracted from regions \mbox{R4--R6} and \mbox{R8--R10} (Figures\,\ref{fig:appendix-results-spectra-from-r4-5-6} and \ref{fig:appendix-results-spectra-from-r8-9-10}, respectively). The identified brightest emission lines that exceed three times the standard deviation measured from corresponding background region fluxes are marked to the same figure and also listed in Table\,\ref{tab:appendix_observed_emission_lines}. 

Table\,\ref{tab:appendix_fit_results_table_R1-R3} contains shock component results from the fit algorithm (presented in Section\,\ref{sec:methods}) from \mbox{R1--R3}, and \mbox{Tables\,\ref{tab:appendix_fit_results_table_R4-R5} and \ref{tab:appendix_fit_results_table_R6-R8-R9}} the corresponding results from \mbox{R4 and R5}, and from \mbox{R6, R8 and R9}, respectively (R7 does not contain significant shock emission). We found that the shock emission signal from R10 is too low to produce significant fit results and therefore no values are listed.

\begin{figure*}
    \includegraphics[width=254pt]{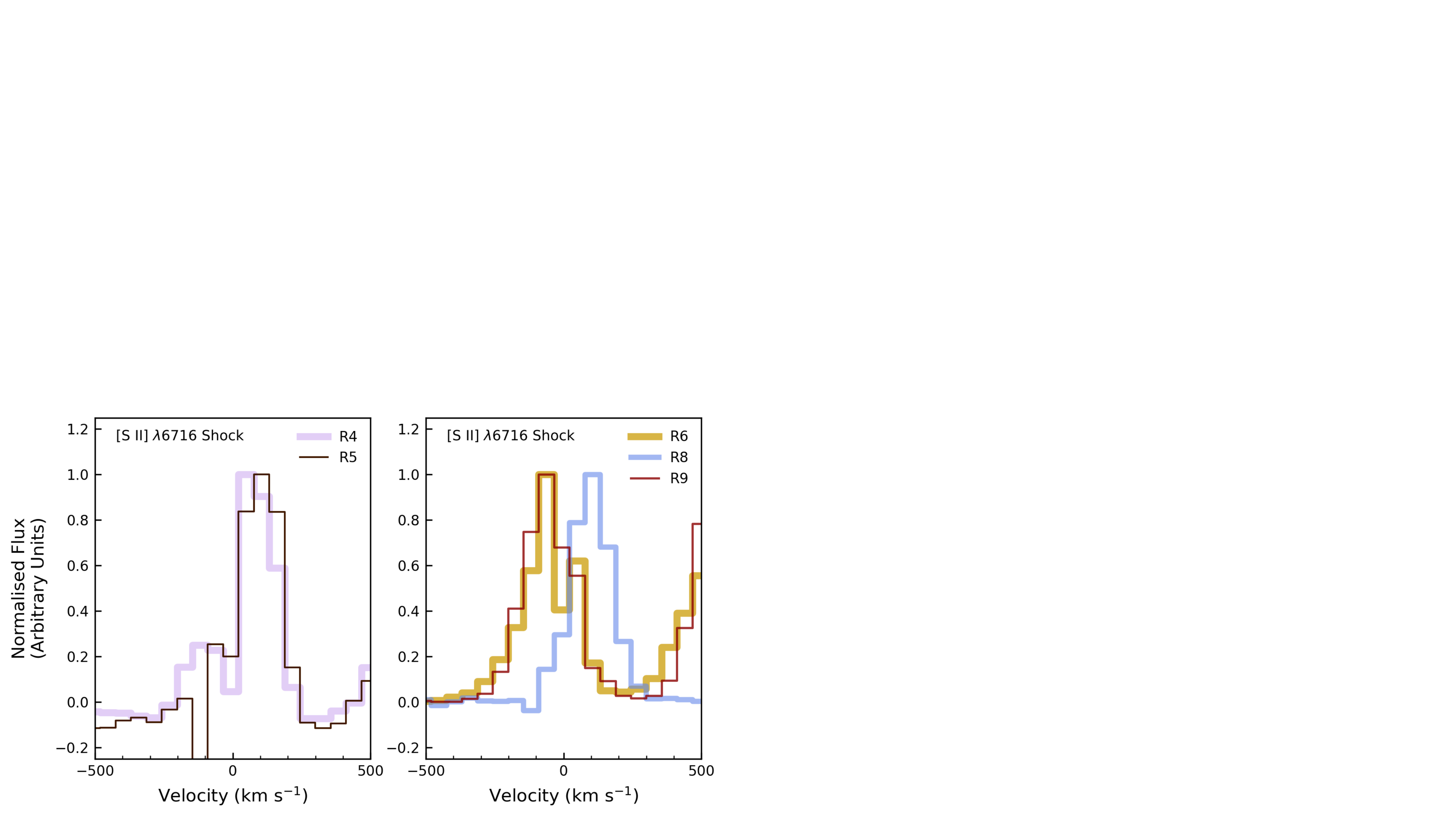}
    \caption{MUSE: Shock component velocity spectra of \mbox{[\ion{S}{II}]\,$\lambda6716$} from regions R4 (lavender), R5 (dark brown), R6 (yellow), R8 (light blue) and R9 (dark red, locations of these regions can be found in Figure\,\ref{fig:results-optical-shock-on-chandra} and \ref{fig:appendix-gaussian-fit-results}). A strong unshocked ISM component has been subtracted from spectra before the fluxes are normalised with respect to the hight of the shock peak. Velocity spectra are only presented from regions where a significant shock component is detected (all fit results from \mbox{R4 and R5}, as well as from \mbox{R6, R8, and R9} can be found in \mbox{Tables\,\ref{tab:appendix_fit_results_table_R4-R5} and \ref{tab:appendix_fit_results_table_R6-R8-R9}}, respectively).
    \label{fig:appendix-results-vel-spectra-bright-emission-lines}}
\end{figure*}

We also include the \mbox{[\ion{S}{II}]\,$\lambda$6716} shock emission velocity profiles from regions where significant shock component is detected in Figure\,\ref{fig:appendix-results-vel-spectra-bright-emission-lines} for completeness. Similarly as for Figure\,\ref{fig:results-vel-spectra-bright-emission-lines}, the strong unshocked ISM component has been subtracted according to the two-component fit results. The rest of the shock component fit results for regions \mbox{R4--R6} and \mbox{R8--R10} are listed in \mbox{Tables\,\ref{tab:appendix_fit_results_table_R4-R5} and \ref{tab:appendix_fit_results_table_R6-R8-R9}}.

\begin{figure*}
    \includegraphics[width=\textwidth]{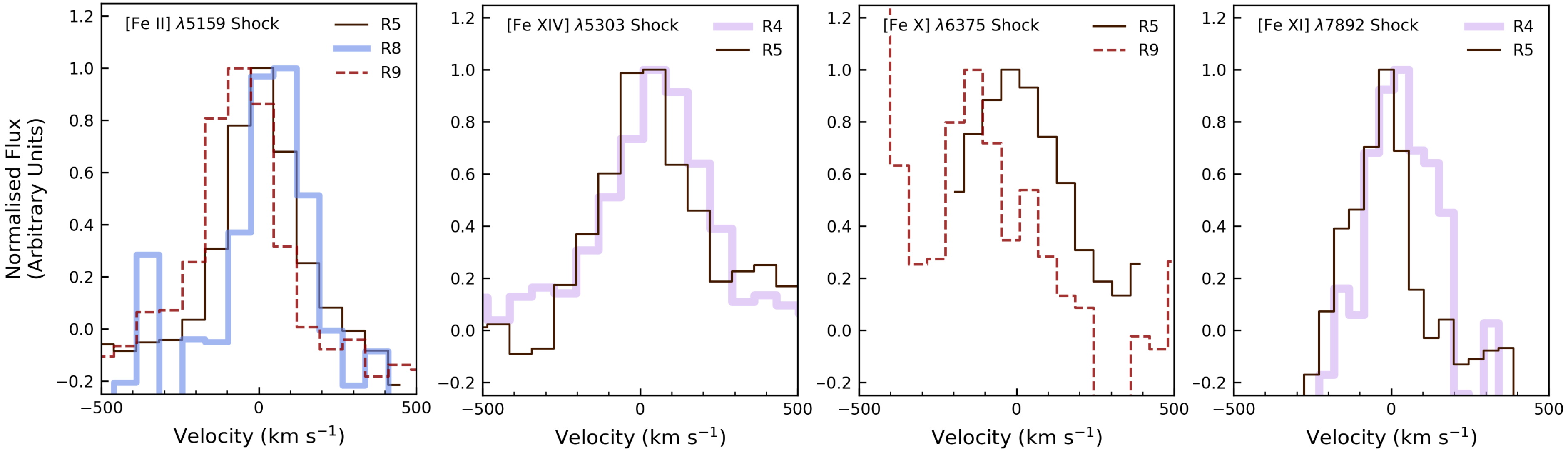}
    \caption{MUSE: Shock component velocity spectra of coronal Fe lines from regions R4 (lavender), R5 (dark brown), R8 (light blue), and R9 (dashed dark red, locations of these regions can be found in Figure\,\ref{fig:results-optical-shock-on-chandra} and \ref{fig:appendix-gaussian-fit-results}). The fluxes are normalised with respect to the hight of the shock peak. Velocity spectra are only presented from regions where a significant shock component is detected (all fit results from \mbox{R4 and R5}, as well as from \mbox{R6, R8, and R9} can be found in \mbox{Tables\,\ref{tab:appendix_fit_results_table_R4-R5} and \ref{tab:appendix_fit_results_table_R6-R8-R9}}, respectively).
    \label{fig:appendix-results-vel-spectra-coronal-fe-lines}}
\end{figure*}

Following Figure\,\ref{fig:results-vel-spectra-iron}, where velocity profiles are presented for Fe lines from regions \mbox{R1--R3}, we construct Fe-line velocity profiles for the rest of the regions (in Figure\,\ref{fig:appendix-results-vel-spectra-coronal-fe-lines}) where significant Fe emission is detected. The complete Fe-line fit results for regions \mbox{R4--R6} and \mbox{R8--R10} are listed in \mbox{Tables\,\ref{tab:appendix_fit_results_table_R4-R5} and \ref{tab:appendix_fit_results_table_R6-R8-R9}}.

\subsection{Testing the Extinction Correction with the [\ion{Fe}{VII}]-line ratio}

The Fe lines \mbox{[\ion{Fe}{VII}]\,$\lambda5721$} and \mbox{[\ion{Fe}{VII}]\,$\lambda6087$} share the same upper level, meaning their intensity ratio \mbox{[\ion{Fe}{VII}]\,$\lambda5721$/$\lambda6087$} is expected to be constant \citep{Young-2005}. After correcting for extinction, this ratio is predicted to be 0.656 \citep{Young-2005}. By measuring the observed line ratio from SNR\,0540, we can test the reliability of the extinction correction we use in this work, which is based on the color excess derived from the Balmer decrement measurement in \citetalias{Tenhu-2024}.

We detect significant emission from \mbox{[\ion{Fe}{VII}]\,$\lambda5721$} and \mbox{[\ion{Fe}{VII}]\,$\lambda6087$} in four regions \mbox{R1--R3} and R9 (Tables\,\ref{tab:appendix_fit_results_table_R1-R3} and \ref{tab:appendix_fit_results_table_R6-R8-R9}). The measured \mbox{[\ion{Fe}{VII}]\,$\lambda5721$/$\lambda6087$} ratios are $0.7\pm0.4$, $0.5\pm0.2$ $0.8\pm0.2$ and $0.6\pm0.1$ for R1, R2, R3, and R9, respectively. All values lie within $1\sigma$ of the theoretical ratio 0.656, supporting the extinction correction ${E(B-V)=0.27}$\,mag estimated in \citetalias{Tenhu-2024}.

\begin{table}
    \centering
    \begin{threeparttable}
    \caption{Observed bright emission lines from SNR\,0540.}
    \begin{tabular}{ll}
\toprule
           ID &                                    $\lambda$  (Å) \\
\midrule
          H I &    4861.3, 6562.8, 8598.3, 8862.8, 9014.9, 9229.0 \\
         He I &                            5015.7, 7065.2, 7281.4 \\
        He II &                                    4685.7, 6678.2 \\
    $[$N I$]$ &                                    5179.9, 5200.3 \\
   $[$N II$]$ &                                    6548.1, 6583.5 \\
    $[$O I$]$ &                                    6300.3, 6363.8 \\
   $[$O II$]$ &                                    7319.9, 7330.2 \\
  $[$O III$]$ &                                    4958.9, 5006.8 \\
  $[$Ne IV$]$ &                                            4714.2 \\
   $[$S II$]$ &                                    6716.4, 6730.8 \\
  $[$S III$]$ &                                    6312.1, 9068.6 \\
 $[$Cl III$]$ &                                    5517.7, 5537.9 \\
 $[$Ar III$]$ &                                            7135.8 \\
  $[$Ca II$]$ &                                    7291.5, 7323.9 \\
  $[$Fe II$]$ &  4814.54, 5158.8, 5261.6, 5333.6, 5721.3, 7155.2, \\
              &            7172.0, 7452.5, 8617.0, 8891.9, 9051.9 \\
 $[$Fe III$]$ &                      4658.1, 4881.0, 4986, 5270.4 \\
  $[$Fe VI$]$ &                                            5335.2 \\
 $[$Fe VII$]$ &                                    5720.7, 6087.0 \\
   $[$Fe X$]$ &                                            6374.5 \\
  $[$Fe XI$]$ &                                            7891.8 \\
 $[$Fe XIV$]$ &                                            5303.1 \\
  $[$Ni II$]$ &                                            7377.8 \\
\bottomrule
\end{tabular}
    \label{tab:appendix_observed_emission_lines}
    \end{threeparttable}
\end{table}

\begin{landscape}
\begin{table}
    \centering
    \begin{threeparttable}
    \caption{Shock component fit results for R1-R3.}
    \begin{tabular}{lcc|ccc|ccc|ccc}
\toprule
            ID &  $\lambda$ &  $R$ &   $S^{\mathrm{R1}}$ & $v_{\mathrm{r}}^{\mathrm{R1}}$ & FWHM$^{\mathrm{R1}}$ &  $S^{\mathrm{R2}}$ & $v_{\mathrm{r}}^{\mathrm{R2}}$ & FWHM$^{\mathrm{R2}}$ &  $S^{\mathrm{R3}}$ & $v_{\mathrm{r}}^{\mathrm{R3}}$ & FWHM$^{\mathrm{R3}}$ \\

           &  (Å) &   & (erg\,s$^{-1}$\,cm$^{-2}$ & (km\,s$^{-1}$) &  (km\,s$^{-1}$) & (erg\,s$^{-1}$\,cm$^{-2}$ & (km\,s$^{-1}$) &  (km\,s$^{-1}$) & (erg\,s$^{-1}$\,cm$^{-2}$ & (km\,s$^{-1}$) &  (km\,s$^{-1}$) \\

           &   &   & arcsec$^{-2}$) &  &   & arcsec$^{-2}$) &  &   & arcsec$^{-2}$) &  &  \\
\midrule
    H$\beta^\mathrm{a}$ &    4861.29 &  190 &    1754.5 $\pm$ 0.2 &                   -184 $\pm$ 6 &    189.59 &       450 $\pm$ 20 &                    222 $\pm$ 4 &    189.59 &        403 $\pm$ 4 &                   -218 $\pm$ 2 &    189.59 \\
 $[$O III$]^\mathrm{a}$ &    4958.91 &  173 &  1060.64 $\pm$ 0.06 &               -136.8 $\pm$ 0.4 &    180.80 $\pm$ 0.03 &        956 $\pm$ 8 &                    168 $\pm$ 8 &    500.00 $\pm$ 0.02 $^\mathrm{c}$ &        983 $\pm$ 5 &                   -142 $\pm$ 7 &      213.4 $\pm$ 0.3 \\
 $[$O III$]^\mathrm{a}$ &    5006.84 &  173 &    3181.9 $\pm$ 0.2 &               -136.8 $\pm$ 0.4 &    180.80 $\pm$ 0.03 &      2870 $\pm$ 30 &                    168 $\pm$ 8 &    500.00 $\pm$ 0.02 $^\mathrm{c}$ &      2950 $\pm$ 20 &                   -142 $\pm$ 7 &      213.4 $\pm$ 0.3 \\
   $[$Fe II$]$ &    5158.79 &  161 &          25 $\pm$ 2 &                   -126 $\pm$ 8 &      195.7 $\pm$ 0.3 &         30 $\pm$ 3 &                     68 $\pm$ 9 &      214.8 $\pm$ 0.3 &         15 $\pm$ 4 &                   -50 $\pm$ 30 &      161.2 $\pm$ 0.8 \\
  $[$Fe XIV$]$ &    5303.06 &  155 &       1420 $\pm$ 50 &                   -168 $\pm$ 4 &    235.62 $\pm$ 0.08 &       170 $\pm$ 30 &                   160 $\pm$ 20 &      283.3 $\pm$ 0.4 &       230 $\pm$ 40 &                   -30 $\pm$ 30 &      413.4 $\pm$ 0.4 \\
  $[$Fe VII$]$ &    5720.7 &  140 &       5 $\pm$ 2 &              -250 $\pm$ 30 &    139 &       10 $\pm$ 2 &   20 $\pm$ 30 &      233.3 $\pm$ 0.6 &       8 $\pm$ 1 &        16 $\pm$ 20 &      225.8 $\pm$ 0.4 \\
  $[$Fe VII$]$ &    6087.0 &  129 &       7 $\pm$ 2 &              -110 $\pm$ 40 &    500.0 $\pm$ 0.4 $^\mathrm{c}$ &       17 $\pm$ 2 &   21 $\pm$ 20 &      253.0 $\pm$ 0.3 &       10 $\pm$ 2 &        -20 $\pm$ 30 &      284.2 $\pm$ 0.6 \\
  $[$S III$]^\mathrm{a,b}$ &    6312.1 &  123 &    17.8 $\pm$ 0.3 &                -105 $\pm$ 3 &  255.02 $\pm$ 0.02 &       12.0 $\pm$ 0.3 &                 73 $\pm$ 4 &  123 $\pm$ 0.05 &      32.3 $\pm$ 0.3 &              -131.8 $\pm$ 0.9 &  222.69 $\pm$ 0.02 \\
    $[$Fe X$]$ &    6374.52 &  121 &          30 $\pm$ 5 &                  -160 $\pm$ 20 &      207.7 $\pm$ 0.4 &      8.9 $\pm$ 0.2 &                     72 $\pm$ 3 &    291.81 $\pm$ 0.05 &     10.4 $\pm$ 0.2 &                    -80 $\pm$ 2 &    349.08 $\pm$ 0.05 \\
  $[$S II$]^\mathrm{a}$ &    6716.44 &  123 &    2211.0 $\pm$ 0.2 &                -90.6 $\pm$ 0.5 &  205.563 $\pm$ 0.008 &       2027 $\pm$ 9 &                 86.5 $\pm$ 0.8 &  217.130 $\pm$ 0.009 &      1550 $\pm$ 20 &              -100.85 $\pm$ 0.2 &  133.882 $\pm$ 0.009 \\
  $[$S II$]^\mathrm{a}$ &    6730.81 &  123 &        3626 $\pm$ 5 &                -90.6 $\pm$ 0.5 &  205.563 $\pm$ 0.008 &      3410 $\pm$ 20 &                 86.5 $\pm$ 0.8 &  217.130 $\pm$ 0.009 &   2067.0 $\pm$ 0.3 &               -100.8 $\pm$ 0.2 &  133.882 $\pm$ 0.009 \\
   $[$Fe XI$]$ &    7891.86 &   95 &          36 $\pm$ 7 &                  -160 $\pm$ 20 &      156.0 $\pm$ 0.5 &          7 $\pm$ 3 &                    70 $\pm$ 40 &          200 $\pm$ 2 &          8 $\pm$ 3 &                   -50 $\pm$ 20 &      156.4 $\pm$ 0.8 \\
 $[$S III$]^\mathrm{a}$ &    9068.60 &   92 &        200 $\pm$ 60 &                  -100 $\pm$ 20 &           92 $\pm$ 1 &       180 $\pm$ 70 &                    50 $\pm$ 40 &      277.8 $\pm$ 0.7 &       650 $\pm$ 60 &                    -73 $\pm$ 4 &     92.12 $\pm$ 0.07 \\
\bottomrule
\end{tabular}
 
    \label{tab:appendix_fit_results_table_R1-R3}
    \begin{tablenotes}
        \footnotesize
        \item Note: a) Two-component fit; shock emission component deblended from the unshocked ISM component. Some emission lines have a FWHM less than the achieved spectral resolution, in which case only an upper bound is reported. \\
        b) \mbox{[\ion{S}{III}]\,$\lambda6312$} line is deblended from the strong [\ion{O}{I}]\,$\lambda6300$ line for R2, see text. \\
        c) The upper bound set in the fitting algorithm for the allowed FWHM values is \mbox{$500$\,km\,s$^{-1}$}. \\
    \end{tablenotes}
    \end{threeparttable}
\end{table}
\begin{table}
    \centering
    \begin{threeparttable}
    \caption{Shock component fit results for R4 and R5.}
    \begin{tabular}{lcc|ccc|ccc}
\toprule
            ID &  $\lambda$ &  $R$ & $S^{\mathrm{R4}}$ & $v_{\mathrm{r}}^{\mathrm{R4}}$ & FWHM$^{\mathrm{R4}}$ & $S^{\mathrm{R5}}$ & $v_{\mathrm{r}}^{\mathrm{R5}}$ & FWHM$^{\mathrm{R5}}$ \\

            &  (Å) &   & (erg\,s$^{-1}$\,cm$^{-2}$ & (km\,s$^{-1}$) &  (km\,s$^{-1}$) & (erg\,s$^{-1}$\,cm$^{-2}$ & (km\,s$^{-1}$) &  (km\,s$^{-1}$)\\

           &   &   & arcsec$^{-2}$) &  &   & arcsec$^{-2}$) &  &   \\
\midrule
     H$\beta^\mathrm{a}$ &    4861.29 &  190 &              -- &              -- &                -- &             -- &           -- &               --  \\
  $[$O III$]^\mathrm{a}$ &    4958.91 &  173 &              -- &              -- &                -- &             -- &           -- &               --  \\
  $[$O III$]^\mathrm{a}$ &    5006.84 &  173 &              -- &              -- &                -- &             -- &           -- &               --  \\
             $[$Fe II$]$ &    5158.79 &  161 &              -- &              -- &                -- &     70 $\pm$ 4 &    5 $\pm$ 7 &   212.2 $\pm$ 0.2 \\
            $[$Fe XIV$]$ &    5303.06 &  155 &    340 $\pm$ 20 &     60 $\pm$ 10 &   353.1 $\pm$ 0.2 &   360 $\pm$ 40 &  20 $\pm$ 20 &   315.8 $\pm$ 0.3 \\
            $[$Fe VII$]$ &     5720.7 &  140 &              -- &              -- &                -- &     14 $\pm$ 3 &  40 $\pm$ 20 &   261.9 $\pm$ 0.6 \\
            $[$Fe VII$]$ &     6087.0 &  129 &              -- &              -- &                -- &             -- &           -- &               --  \\
$[$S III$]^\mathrm{a}$ &     6312.1 &  123 &              -- &              -- &                -- &             -- &           -- &               --  \\
              $[$Fe X$]$ &    6374.52 &  121 &              -- &              -- &                -- &     25 $\pm$ 2 & -20 $\pm$ 20 &   320.0 $\pm$ 0.3 \\
   $[$S II$]^\mathrm{a}$ &    6716.44 &  123 & 444.2 $\pm$ 0.2 &  84.9 $\pm$ 0.7 & 150.68 $\pm$ 0.04 &  1280 $\pm$ 70 &   93 $\pm$ 5 & 147.07 $\pm$ 0.06 \\
   $[$S II$]^\mathrm{a}$ &    6730.81 &  123 &     829 $\pm$ 5 &  84.9 $\pm$ 0.7 & 150.68 $\pm$ 0.04 & 2900 $\pm$ 200 &   93 $\pm$ 5 & 147.07 $\pm$ 0.06 \\
             $[$Fe XI$]$ &    7891.86 &   95 &       4 $\pm$ 3 &     30 $\pm$ 50 &       193 $\pm$ 2 &     18 $\pm$ 5 & -20 $\pm$ 20 &  120.0 $\pm$ 0.07 \\
  $[$S III$]^\mathrm{a}$ &    9068.60 &   92 &              -- &              -- &                -- &             -- &           -- &               --  \\
\bottomrule
\end{tabular}
    \label{tab:appendix_fit_results_table_R4-R5}
    \begin{tablenotes}
        \footnotesize
        \item Note: a) Two-component fit; shock emission component deblended from the unshocked ISM component. \\
    \end{tablenotes}
    \end{threeparttable}
\end{table}
\end{landscape}

\begin{landscape}
\begin{table}
    \centering
    \begin{threeparttable}
    \caption{Shock component fit results for R6, R8, and R9.}
    \begin{tabular}{lcc|ccc|ccc|ccc}
\toprule
            ID &  $\lambda$ &  $R$ & $S^{\mathrm{R6}}$ & $v_{\mathrm{r}}^{\mathrm{R6}}$ & FWHM$^{\mathrm{R6}}$ & $S^{\mathrm{R8}}$ & $v_{\mathrm{r}}^{\mathrm{R8}}$ & FWHM$^{\mathrm{R8}}$ & $S^{\mathrm{R9}}$ & $v_{\mathrm{r}}^{\mathrm{R9}}$ & FWHM$^{\mathrm{R9}}$ \\

            &  (Å) &   & (erg\,s$^{-1}$\,cm$^{-2}$ & (km\,s$^{-1}$) &  (km\,s$^{-1}$) & (erg\,s$^{-1}$\,cm$^{-2}$ & (km\,s$^{-1}$) &  (km\,s$^{-1}$) & (erg\,s$^{-1}$\,cm$^{-2}$ & (km\,s$^{-1}$) &  (km\,s$^{-1}$) \\

           &   &   & arcsec$^{-2}$) &  &   & arcsec$^{-2}$) &  &   & arcsec$^{-2}$) &  &  \\
\midrule
    H$\beta^\mathrm{a}$ &    4861.29 &  190 &                 -- &                              -- &                    -- &       351 $\pm$ 9 &                    144 $\pm$ 3 &    189.59 &     1590 $\pm$ 60 &                  -170 $\pm$ 20 &    189.59 \\
 $[$O III$]^\mathrm{a}$ &    4958.91 &  173 &                 -- &                              -- &                    -- &       304 $\pm$ 4 &                   150 $\pm$ 20 &      258.2 $\pm$ 0.2 &     4890 $\pm$ 30 &                  -130 $0\pm$ 2 &    172.60 \\
 $[$O III$]^\mathrm{a}$ &    5006.84 &  173 &                 -- &                              -- &                    -- &      910 $\pm$ 10 &                   150 $\pm$ 20 &      258.2 $\pm$ 0.2 &    14680 $\pm$ 80 &                   -130 $\pm$ 2 &    172.60 \\
   $[$Fe II$]$ &    5158.79 &  161 &                 -- &                              -- &                    -- &        16 $\pm$ 4 &                    50 $\pm$ 20 &      176.8 $\pm$ 0.6 &        18 $\pm$ 2 &                   -60 $\pm$ 20 &      216.3 $\pm$ 0.3 \\
  $[$Fe XIV$]$ &    5303.06 &  155 &                 -- &                              -- &                    -- &                 -- &                              -- &                    -- &                 -- &                              -- &                    -- \\
  $[$Fe VII$]$ &    5720.7 &  140 &       -- &               -- &    -- &       -- &   -- &    -- &       6.0 $\pm$ 0.9 &        10 $\pm$ 20 &      283.1 $\pm$ 0.4 \\
  $[$Fe VII$]$ &    6087.0 &  129 &       -- &               -- &    -- &       -- &   -- &    -- &       10.6 $\pm$ 0.8 &        -40 $\pm$ 10 &      293.7 $\pm$ 0.2 \\  
  $[$S III$]^\mathrm{a}$ &    6312.1 &  123 &    -- &                -- &  -- &       -- &                 -- &  -- &      43.4 $\pm$ 0.5 &              -90 $\pm$ 2 &  134.08 $\pm$ 0.05 \\
    $[$Fe X$]$ &    6374.52 &  121 &                 -- &                              -- &                    -- &                 -- &                              -- &                    -- &    5.09 $\pm$ 0.2 &                  -120 $\pm$ 20 &    389.60 $\pm$ 0.06 \\
  $[$S II$]^\mathrm{a}$ &    6716.44 &  123 &   742.5 $\pm$ 0.3 &                    -61 $\pm$ 7 &    248.09 $\pm$ 0.08 &  1411.0 $\pm$ 0.3 &                 96.9 $\pm$ 0.8 &    167.92 $\pm$ 0.02 &  3213.4 $\pm$ 0.2 &                    -61 $\pm$ 2 &  217.056 $\pm$ 0.008 \\
  $[$S II$]^\mathrm{a}$ &    6730.81 &  123 &      900 $\pm$ 50 &                    -61 $\pm$ 7 &    248.09 $\pm$ 0.08 &     2210 $\pm$ 30 &                 96.9 $\pm$ 0.8 &    167.92 $\pm$ 0.02 &      3958 $\pm$ 8 &                    -61 $\pm$ 2 &  217.056 $\pm$ 0.008 \\
   $[$Fe XI$]$ &    7891.86 &   95 &                 -- &                              -- &                    -- &                 -- &                              -- &                    -- &                 -- &                              -- &                    -- \\
 $[$S III$]^\mathrm{a}$ &    9068.60 &   92 &       195 $\pm$ 2 &                    -67 $\pm$ 6 &     92.12 $\pm$ 0.07 &        77 $\pm$ 3 &                    165 $\pm$ 4 &     92.12 $\pm$ 0.07 &      390 $\pm$ 20 &                   -107 $\pm$ 3 &      112.3 $\pm$ 0.3 \\
\bottomrule
\end{tabular}
    \label{tab:appendix_fit_results_table_R6-R8-R9}
    \begin{tablenotes}
        \footnotesize
        \item Note: a) Two-component fit; shock emission component deblended from the unshocked ISM component. Some emission lines have a FWHM less than the achieved spectral resolution, in which case only an upper bound is reported. \\
    \end{tablenotes}
    \end{threeparttable}
\end{table}
\end{landscape}

\bsp	
\label{lastpage}
\end{document}